\documentclass{article}

\usepackage{arxiv}

\usepackage[utf8]{inputenc} % allow utf-8 input
\usepackage[T1]{fontenc}    % use 8-bit T1 fonts
\usepackage{hyperref}       % hyperlinks
\usepackage{url}            % simple URL typesetting
\usepackage{booktabs}       % professional-quality tables
\usepackage{amsmath}
\usepackage{amsfonts}       % blackboard math symbols
\usepackage{amssymb}
\usepackage{amsthm}
\usepackage{algorithmic}
\usepackage{algorithm}
\usepackage{nicefrac}       % compact symbols for 1/2, etc.
\usepackage{microtype}      % microtypography
\usepackage{lipsum}		% Can be removed after putting your text content
\usepackage{graphicx,xcolor}
\usepackage[numbers]{natbib}
\usepackage{doi}
\usepackage{cleveref}
\usepackage{subcaption}
\usepackage{svg}
\usepackage{csquotes}
\usepackage{enumerate}
\usepackage{cancel}
\usepackage{todonotes}
\usepackage{lineno}

\usepackage{localmacros}

\title{Riemannian geometry for efficient analysis of protein dynamics data}

\author{ 
Willem Diepeveen
\\
	Faculty of Mathematics\\
	University of Cambridge\\
	Cambridge, UK \\
	\texttt{wd292@cam.ac.uk} \\
	\And
 Carlos Esteve-Yag\"ue
 \\
	Faculty of Mathematics\\
	University of Cambridge\\
	Cambridge, UK \\
	\texttt{ce423@cam.ac.uk} \\
    \And
 Jan Lellmann
 \\
	Institute of Mathematics and Image Computing\\
	University of L\"ubeck\\
	L\"ubeck, Germany \\
	\texttt{lellmann@mic.uni-luebeck.de} \\
    \And
 Ozan \"Oktem
 \\
	Department of Mathematics\\
	KTH–Royal Institute of Technology\\
	Stockholm, Sweden \\
	\texttt{ozan@kth.se} \\
    \And
 Carola-Bibiane Sch\"onlieb
 \\
	Faculty of Mathematics\\
	University of Cambridge\\
	Cambridge, UK \\
	\texttt{cbs31@cam.ac.uk} \\
}

\hypersetup{
pdfauthor={Willem Diepeveen, Carlos Esteve-Yag\"ue, Jan Lellmann, Ozan \"Oktem, Carola-Bibiane Sch\"onlieb},
pdfkeywords={First keyword, Second keyword, More},
}

\begin{document}
\maketitle

\begin{abstract}
    An increasingly common mathematical viewpoint is that protein dynamics data sets reside in a non-linear subspace of low conformational energy. Ideal data analysis tools for such data sets should therefore account for such non-linear geometry. The Riemannian geometry setting can be suitable for a variety of reasons. First, it comes with a rich mathematical structure to account for a wide range of non-linear geometries that can be modelled after an energy landscape. Second, many standard data analysis tools initially developed for data in Euclidean space can also be generalised to data on a Riemannian manifold. In the context of protein dynamics, a conceptual challenge comes from the lack of a suitable smooth manifold and the lack of guidelines for constructing a smooth Riemannian structure based on an energy landscape. In addition, computational feasibility in computing geodesics and related mappings poses a major challenge. This work considers these challenges. The first part of the paper develops a novel local approximation technique for computing geodesics and related mappings on Riemannian manifolds in a computationally feasible manner. The second part constructs a smooth manifold of point clouds modulo rigid body group actions and a Riemannian structure that is based on an energy landscape for protein conformations. The resulting energy landscape-based Riemannian geometry is tested on several data analysis tasks relevant for protein dynamics data. It performs exceptionally well on coarse-grained molecular dynamics simulated data. In particular, the geodesics with given start- and end-points approximately recover corresponding molecular dynamics trajectories for proteins that undergo relatively ordered transitions with medium sized deformations. The Riemannian protein geometry also gives physically realistic summary statistics and retrieves the underlying dimension even for large-sized deformations within seconds on a laptop.

\end{abstract}

% keywords can be removed


% 49Q10: Optimization of shapes other than minimal surfaces
% 53C22  	Geodesics in global differential geometry
% 53Z10  	Applications of differential geometry to biology
% 53Z50  	Applications of differential geometry to data and computer science
% 65D18 Numerical aspects of computer graphics, image analysis, and computational geometry
% 92-04 Software, source code, etc. for problems pertain- ing to biology
% 92-08 Computational methods for problems pertaining to biology
% 92-10 Mathematical modeling or simulation for prob- lems pertaining to biology

\blfootnote{Our code for the main algorithms is available at \href{https://github.com/wdiepeveen/Riemannian-geometry-for-efficient-analysis-of-protein-dynamics-data}{https://github.com/wdiepeveen/Riemannian-geometry-for-efficient-analysis-of-protein-dynamics-data}. }

\section{Introduction}
\label{sec:intro-protein-geometry}

Protein dynamics
% conformation 
data are becoming available at an ever-increasing rate
% already ubiquitous and the rate at which these data become available is not expected to slow down over the next couple of years. These surges in data availability are not only 
due to developments in molecular dynamics simulation \cite{kolloff2022machine,rydzewski2023manifold} -- generating more trajectories of large molecular assemblies across longer time scales
% generating more high-quality trajectories
--, and also due to developments on the experimental side through recent advances in the processing of heterogeneous cryogenic electron microscopy data \cite{jamali2023graph,zhong2021cryodrgn} -- enabling reconstruction of several discrete conformations or continuous large-scale motions of proteins and complexes. With this
% all these 
data, proper analysis tools are becoming increasingly important to extract information, e.g., interpolation, extrapolation, computing meaningful summary statistics, and dimension reduction. Consequently, the development of such tools has been an active area of research for several decades. In particular, suitable interpolation between discrete conformations has given insight into conformational transitions without running full-scale molecular dynamics simulations \cite{zheng2017survey}. Extrapolation has enabled one to explore different regions of conformation space \cite{kurkcuoglu2016clustenm}. Finally, a good notion of mean conformation and of low-rank approximation has retrieved the most important large-scale motions \cite{amadei1993essential} in continuous trajectory data sets.

% 
% [with all this data ... we need good tools, e.g., proper interpolation, computing meaningful summary statistics, dimension reduction, or even extrapolation/ prediction]
% [ - having good tools for these tasks has received much attention over the last decades...]
% [In particular, \cite{zheng2017survey}] % survey interpolating protein states
% [But also the by now widely used \cite{amadei1993essential}] % pca for protein motion at the mean of the data set
% [or for exploring protein space \cite{kurkcuoglu2016clustenm}] % explore conformation space through addition

The construction of such
% the above-mentioned 
tools relies strongly on the protein energy landscape, and even more so on \emph{normal mode analysis} (NMA), which is a technique to describe the flexible states accessible to a protein from
% about 
an equilibrium conformation through linearisation of the energy landscape. This idea was explored in the seminal works by Go et al.~\cite{go1983dynamics} and by Brooks and Karplus \cite{brooks1983harmonic,brooks1985normal} in the early 1980s. The idea was taken further by Tirion \cite{tirion1996large}, who approximates the physical energy by a Hookean potential to resolve the issue of possible negative eigenvalues of the Hessian of the physical energy. The latter approximation has many extensions \cite{bastolla2014computing}, and has been highly successful in 
% many more applications in several fields of 
structural biology \cite{lopez2016new,sorzano2019survey}.

% [protein movement here]
% % most of the tools above rely on the energy landscape of proteins ...

Despite the success of the NMA framework, there is 
a major conceptual disadvantage
% flaw 
when it is used for analysis of protein conformations:
% . That is, 
NMA is local and linear by construction and it has been observed numerous times that the quality of the approximations degrades for large-scale deformations \cite{mahajan2017jumping}. 
% In other words, the above-described tools suffer from the same locality and linearity issue as well. 
To resolve the problem of locality and linearity, ideal energy landscape-based data analysis tools should be able to interpolate and extrapolate over \emph{energy-minimising non-linear paths}, compute \emph{non-linear means on such energy-minimising paths} and compute \emph{low-rank approximations over curved subspaces spanned by energy-minimising paths}. Deep learning has recently emerged as one way of modelling with non-linearity under physics-based constraints, see for example
% e.g., as in 
\cite{ramaswamy2021deep}. However, in addition to requiring substantial
% besides 
training time, such methods focus on physically correct interpolation within the -- protein-specific -- training data set. Therefore, 
% So 
they cannot be expected to generalize well to proteins that are structurally very different.
% when analysing data that has not been in the training data, which is bound to happen when new data is coming in or when exploring new parts of conformation space. 

% On top of that, training a deep learning model tends to be expensive.

% but in the current state it suffers from two issues: (i) we need a lot of data (ii) interpolation can be managed this way, but if there is new data coming in that is not in the training set or if we want to extrapolate to a region that was not in the data set, we cannot expect these methods to perform well.]

% \wdp{TODO: finish above paragraph}

% motivation for using Riemannian geometry
% [for small deformations, the above tools are alright, but will degrade in quality for large-scale deformations]
% \cite{mahajan2017jumping} % talks about degrading of NMA

% [tools should be able to interpolate over proper non-liniear paths, compute non-linear means and compute low-rank approximations over non-linear spaces.]
% [DL is a recent way of doing this, but it seems a bit steep to train a neural network for every new protein \cite{ramaswamy2021deep} ]

Instead, the framework of \emph{Riemannian geometry} \cite{carmo1992riemannian,sakai1996riemannian} could be a more suitable choice. Here, interpolation can be performed
% done 
over non-linear \emph{geodesics} 
% -- 
or related higher order interpolation schemes \cite{bergmann2018bezier},
% --
extrapolation can be done using the \emph{Riemannian exponential mapping}, the data mean is naturally generalized to the \emph{Riemannian barycentre} \cite{karcher1977riemannian}, and low rank approximation knows several extensions that find the most important geodesics through a data set using the \emph{Riemannian logarithmic mapping} \cite{diepeveen2023curvature,fletcher2004principal}\footnote{On top of that, having Riemannian geometry opens more doors. For one, because many popular methods rely solely on a notion of distance \cite{glielmo2021unsupervised} -- and thus can be generalized to Riemannian manifolds.}. Riemannian geometry also allows 
% one 
to specify
% choose 
the type of non-linearity, i.e., there is a choice \emph{which curves are length-minimising} -- which then naturally implies
% tells 
\emph{the distance between any two points}. In particular, there is the possibility to take physics into account in building
% and potential to build 
a custom \emph{geometry for proteins} through modelling the Riemannian manifold structure after the energy landscape.
% -- [in such a way that it possibly also overcomes the locality issue of NMA]. 
Then, ideally, length-minimising curves over which we perform
% do 
data analysis are energy-minimising and the distance between two conformations corresponds to the energy needed for the transition of one into the other.
In this work, we set out to construct such a geometry for proteins that will enable more natural data analysis of protein dynamics data. In particular, our goal is threefold: (i) construct a smooth manifold of protein conformations (ii) with a suitable Riemannian structure that models the conformation energy landscape, (iii) which allows computationally feasible analysis of protein dynamics data.

% - versitile manifold
% - good Riemannian structure based on energy
% - computationally feasible for data analysis tasks

% \todo[inline]{For the figure, it would be easiest to make a 2D grid with multiple lines that would represent geodesics. We can construct these lines through Bezier curves I guess.}

\subsection{Related work}
\label{sec:intro-protein-geometry-related-work}
As a geometry for proteins comes in three parts, we will discuss the related work in a similar fashion.

% [As we will see in the following, there are no clear answers for the three goals above -- to the best of our knowledge, there is no work that attempts to do this.] In the following, we will consider various options for constructing a manifold, constructing geometry and how computational feasibility comes into play subsequently and see what different challenges come with them.

\paragraph{Smooth manifolds of protein conformations.}
Before constructing a suitable and computationally feasible \emph{Riemannian manifold}, one needs a \emph{smooth manifold} \cite{boothby2003introduction,lee2013smooth} whose elements represent various conformations of a fixed protein. The type of Riemannian structure we aim for -- one that is based on the energy landscape -- comes with constraints. In particular, since the energy boils down to a sum of interaction potentials that only depend on the Euclidean distance between 
pairs of atoms
% each two atoms 
in the protein, two conformations can only be distinguished up to rigid-body transformations. This observation leaves
% gives us 
two choices for constructing a smooth manifold that allows for such a symmetry. We either choose an already \emph{invariant parametrisation} that is also a smooth manifold or choose to identify each protein conformation -- in a suitable subset of protein conformations -- with all of its rigid-body transformations, i.e., we construct a \emph{quotient manifold}. As we will see below, several challenges need to be overcome before using either approach.

% these manifolds are constructed so that they are invariant under rigid body motion, which is a symmetry of the interaction energy and hence a require into building smooth manifolds.

A classical parametrisation of a protein is to model the macromolecule as a chain of peptides, each one living in a plane, 
% and 
where each plane is determined by two conformation angles per peptide, e.g., see \cite{penner2016moduli} in which the extension to so-called \emph{fat graphs} to encode hydrogen bonds is also discussed. This parametrisation is a smooth manifold and is invariant under rotations and translations, but not 
% to 
reflections. In other words, it is not a proper invariant parametrisation manifold. In addition, to the best of our knowledge, an invariant parametrisation manifold has not arisen in the literature so far.
% Such a problem is likely to arise for any other invariant parametrisation manifold that is not a point cloud.
% In addition, and more practically, building an energy landscape-based Riemannian structure on any invariant parametrisation manifold that is not a point cloud will be challenging due to the conversion . That is, because such parametrisations are indirect since the interaction potential acts on the atoms and not on the modeled edges connecting them. This will also be valid for any other invariant parametrisation manifold that is not a point cloud.

% However, more problematic is that, if we want to build a Riemannian structure that is based on energy, this parametrisation is very indirect since the interaction potential acts on the atoms and not on the modeled edges connecting them. The latter challenge will also be valid for any other invariant parametrisation manifold that is not a point cloud.

Constructing quotient manifolds has been explored extensively in the shape manifold literature (see \cite{younes2010shapes} for recent accounts of the theory). In particular, the modelling of shapes in a given embedding space can roughly be broken down into point clouds, curves, diffeomorphisms and surfaces \cite{younes2012spaces} -- with suitable symmetries quotiented out. Out of these four options, the first three --  point clouds, curves, diffeomorphisms -- seem most suitable for modelling protein conformations. However, as we will see, each of them comes with a challenge.

Starting with \emph{point clouds} -- being the classical choice \cite{brooks1983harmonic,brooks1985normal,go1983dynamics} for protein modelling under the energy landscape constraint -- a prime example of such a shape space is the seminal work by Kendall \cite{kendall1984shape}. Here, point clouds are centered and normalized with rotations quotiented out. The resulting quotient space is a smooth manifold for 2-dimensional point clouds, but it is not a manifold anymore in dimension 3 or higher. Kendall's construction is not just an unlucky one. The issue is caused by quotienting out the rotations, which will also happen if we consider point clouds without the normalization -- although this seems to be ignored in practice \cite{laga2018survey}. On top of that, quotienting out reflections as well will bring even more complications. 
% In other words, if we want to use point clouds modulo rigid body motions for proteins (in 3D), [we will need to be careful with construction]. 
% Curves and diffeomorphisms do not have the construction issue, but a new challenge arises as it is less natural to construct an energy-based Riemannian structure on there. 

In the case of \emph{curves,} there is also a challenge in the construction of a manifold. However, more importantly there are bound to be issues with the Riemannian constraint as it is not trivial how to extend the energy landscape to a continuum of interacting parts of the curve. In particular, the interaction potential blows up if two atoms become arbitrarily close, which is conceptually challenging to tailor to a continuous curve.

For \emph{diffeomorphisms,} a similar issue also arises if the diffeomorphism class is infinite dimensional. However, even for the finite-dimensional case -- e.g., when applied to the conformation angle framework discussed above -- there is yet another more conceptual problem. That is, any Riemannian structure on the diffeomorphism action will by construction be independent of the conformation it acts on -- due to the typical homogeneity assumption. Subsequently, this renders such an approach unrealistic unless we only want to deform one template conformation, which is not our goal.

\paragraph{Riemannian geometry for protein conformations.}
Having an invariant parametrisation manifold or quotient manifold, the approaches one can follow to construct Riemannian geometry are similar. 
% So without loss of generality, we can focus on quotient manifolds for the simple reason that shape space literature has covered most conventional ways. The types of Riemannian structure is again strongly dependent on the underlying shape space. 
For point clouds and curves, the canonical construction is the norm \cite{laga2018survey}, i.e., constructing a \emph{metric tensor field} \cite{carmo1992riemannian,sakai1996riemannian}, which would also be the default in the conformation angle framework. For diffeomorphisms, there are more options, e.g., considering different reproducing kernel Hilbert spaces for the diffeomorphic flow vector fields~\cite{niethammer2019metric} or considering extended frameworks such as metamorphosis~\cite{trouve2005metamorphoses}. 

In any case,
% Either way though, 
the focus of construction has historically been on attaining well-posedness rather than on imposing physical laws or biological mechanisms. To the best of our knowledge, there is
% So unsurprisingly, there is -- to the best of our knowledge -- 
no readily available energy landscape-based Riemannian geometry for protein conformations.
% Recently there have been attempts to physics and biology-inspired Riemannian metrics as summarized in \cite{charon2023shape}. 
% And although there is to the best of our knowledge no systematic way to impose physics for point clouds and curves...
More recently, there has been a surge in work on bringing physics and biology into the diffeomorphism framework \cite{charon2023shape}. In particular, through adding a regulariser within a hybrid approach \cite{younes2018hybrid}, or through growth models \cite{goriely2017mathematics} where external actions can also be taken into account. Although the diffeomorphism framework still suffers from the same conceptual problem as discussed above, it should be mentioned that the ideas underlying these approaches -- inheriting the mathematical rigour from established theory, but also encoding physics or biology -- is something to strive for.

\paragraph{Computational feasibility for data analysis.}
% main point here is that the computational feasibility often is limited by the choice of riemannian structure
When analysing data on a Riemannian manifold, we need access to several standard manifold mappings. In particular, from the discussion above we have seen, that for basic data processing tasks we will at least need the ability to compute geodesics, the exponential mapping and the logarithmic mapping -- or good approximations thereof.

In the canonical form of Riemannian geometry, i.e., having just a metric tensor field, closed-form expressions for the above-mentioned mappings can realistically only be expected when having special structure from the specific choice of metric \cite{younes2008metric}. In other words, the necessary manifold mappings are not readily available, but need to be approximated -- typically by solving the geodesic equation under specific boundary conditions \cite{herzog2023manifold}. Such an approach to accessing these mappings is very slow in practice and can be numerically unstable -- as in the conformation angle framework \cite{bastolla2019can} --, which renders it unfeasible for the analysis of large high-dimensional data sets. 

Computationally feasible approximations for all of these mappings, i.e., without solving the geodesic equation, require extra structure as well \cite{rumpf2015variational}. Alternatively, individual mappings can be learned by neural networks to a certain extent \cite{yang2017quicksilver}. At a moment the special structure often comes down to using ambient linear structure, which is not readily translatable to quotient manifolds or invariant representation manifolds. In the case of learning the mappings we will need to be very careful with taking the appropriate symmetries into account, but will nonetheless inevitably require very careful curation of the training data to ensure generalisation
% suffer from the aforementioned out-of-training-distribution problem
-- although it should be said that this could be justified if we need more than the basic mappings, e.g., their differentials.

\subsection{Contributions}
\label{sec:intro-protein-geometry-contributions}
% we solve the point cloud problem -- so that is becomes easier to use ideas from course graining

In the above, we identified three open problems for building a computationally feasible Riemannian protein geometry: (i) there is typically -- without extra structure -- a trade-off between accurate modelling and computational feasibility, 
% but even with this sorted 
(ii) there is no readily available manifold for protein conformations, and (iii) existing Riemannian structures do not cater to protein energy landscapes. Naturally, our contribution of this work is also threefold:

\paragraph{Computationally feasible Riemannian geometry.}
For attaining computational feasibility we propose the notion of \emph{separation}, as a relaxation of the Riemannian distance, for additional structure. We show that separations are local approximations of the Riemannian distance (\cref{thm:separation-and-distance}), that yield a closed-form approximation of the logarithmic mapping (\cref{cor:log-approx-separation}), and that enable -- under mild conditions -- the construction of well-posed optimisation problems to approximate geodesics and the exponential mapping in a provably efficient fashion (\cref{thm:well-posed-ness-separation-geodesics,thm:well-posed-ness-separation-exp}).

\paragraph{Reverse-engineering Riemannian geometry for point cloud conformations.}
From the options outlined above, the point cloud is the most natural space to build geometry upon, and is additionally most in line with what is being used in practice. We resolve the challenge of constructing a quotient manifold out of point clouds by passing to an appropriate subset before taking the quotient (\cref{thm:quotient-manifold-Rdnd-mod-Ed}) and provide guidelines for reverse-engineering a metric tensor field and a separation from a suitable family of metrics (\cref{thm:class-of-separations-on-pointcloud-manifold}). We note that these results are still generally applicable for designing geometry on point clouds, so can also be used beyond the protein dynamics data application.

\begin{figure}[h!]
    \centering
    \begin{subfigure}{0.21\linewidth}
    \makebox[10pt]{\raisebox{40pt}{\rotatebox[origin=c]{90}{\small \shortstack{Molecular \\ dynamics} }}}
    \hfill
        \includegraphics[width=0.85714\linewidth]{"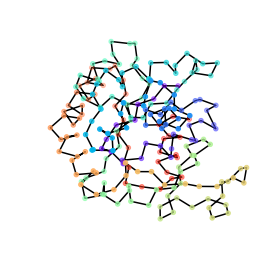"}
        \\
        \makebox[10pt]{\raisebox{40pt}{\rotatebox[origin=c]{90}{\small \shortstack{$\separation^\GyRaParam$-geodesic from \\ closed to open} }}}
        \hfill
        \includegraphics[width=0.85714\linewidth]{"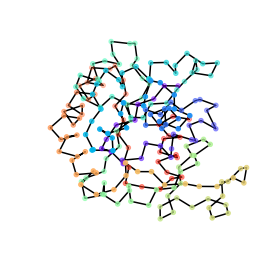"}
        \\
        \makebox[10pt]{\raisebox{40pt}{\rotatebox[origin=c]{90}{\small \shortstack{Rank 1 approx. at \\ the $\separation^\GyRaParam$-barycentre} }}}
        \hfill
        \includegraphics[width=0.85714\linewidth]{"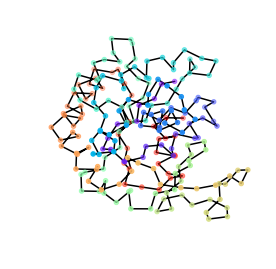"}
        \caption{$t=0$}
    \end{subfigure}
    \hfill
    \begin{subfigure}{0.18\linewidth}
        \includegraphics[width=\linewidth]{"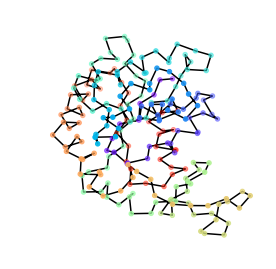"}
        \\
        \includegraphics[width=\linewidth]{"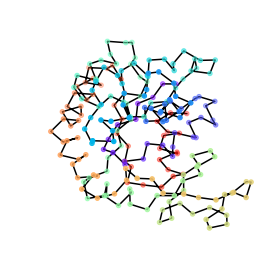"}
        \\
        \includegraphics[width=\linewidth]{"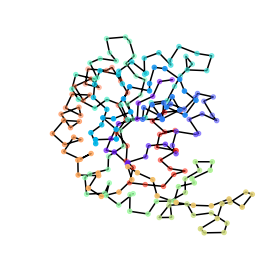"}
        \caption{$t=0.25$}
    \end{subfigure}
    \hfill
    \begin{subfigure}{0.18\linewidth}
        \includegraphics[width=\linewidth]{"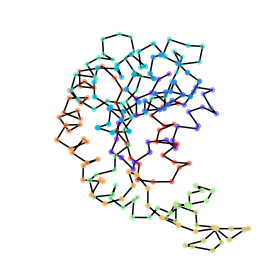"}
        \\
        \includegraphics[width=\linewidth]{"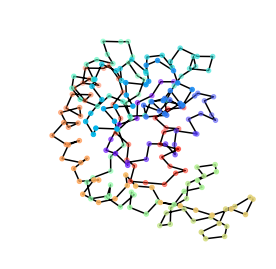"}
        \\
        \includegraphics[width=\linewidth]{"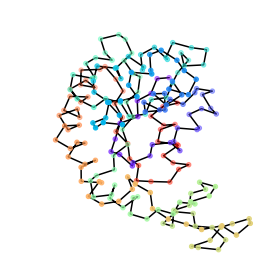"}
        \caption{$t=0.5$}
    \end{subfigure}
    \hfill
    \begin{subfigure}{0.18\linewidth}
        \includegraphics[width=\linewidth]{"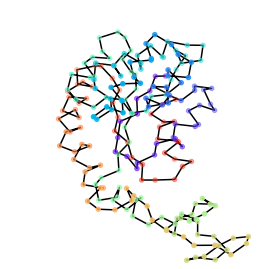"}
        \\
        \includegraphics[width=\linewidth]{"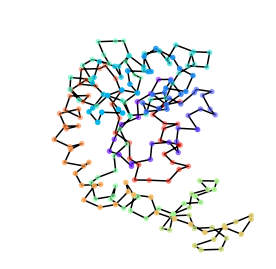"}
        \\
        \includegraphics[width=\linewidth]{"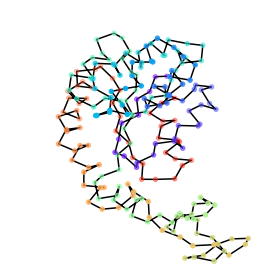"}
        \caption{$t=0.75$}
    \end{subfigure}
    \hfill
    \begin{subfigure}{0.18\linewidth}
        \includegraphics[width=\linewidth]{"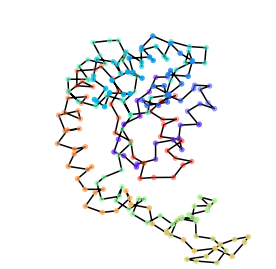"}
        \\
        \includegraphics[width=\linewidth]{"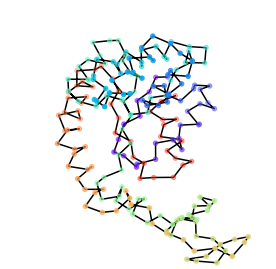"}
        \\
        \includegraphics[width=\linewidth]{"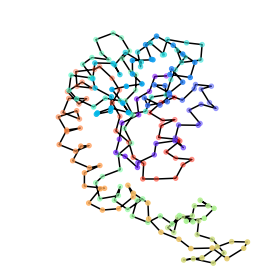"}
        \caption{$t=1$}
    \end{subfigure}
    \caption{Several snapshots of the $\mathrm{C}_\alpha$ atoms of the adenylate kinase protein under a closed ($t=0$) to open ($t=1$) transition. The top row is generated from a molecular dynamics simulation, the middle row is a $\separation^\GyRaParam$-geodesic between the end points and the bottom row a $\separation^\GyRaParam$-geodesic obtained from rank 1 approximation of the data at the $\separation^\GyRaParam$-barycentre. The begin-to-end $\separation^\GyRaParam$-geodesic captures the large-scale motion exceedingly well, having only small errors at the lower green and yellow part of the protein, and the $\separation^\GyRaParam$-geodesic obtained from the rank 1 approximation captures the data almost perfectly.}
    \label{fig:md-geo-low-rank-4ake}
\end{figure}

\paragraph{Riemannian geometry for efficient analysis of protein dynamics data.}
Finally, we 
% show how the developed theory helps us to construct Riemannian geometry for protein conformations. 
% In particular, we show that given a metric satisfying mild assumptions, we can reverse-engineer a metric tensor field that generates the metric as a separation (\cref{thm:class-of-separations-on-pointcloud-manifold}). This then allows us to use all the efficient approximations of manifold mappings. This makes life easier since it is typically much more straightforward to model an approximate energy landscape -- that is also a metric --, than to construct a metric tensor field directly. 
% Subsequently, we 
propose a specific metric and show in numerical results that the Riemannian manifold reverse-engineered from this metric is suitable for protein dynamics data processing through several examples with molecular dynamics trajectories of the adenylate kinase protein and the SARS-CoV-2 helicase nsp 13 protein. 
Our most encouraging findings are, that on a coarse-grained level the Riemannian manifold structure approximately recovers complete adenylate kinase trajectories from just a begin-and endpoint (\cref{fig:md-geo-low-rank-4ake}), 
% -- although not for SARS-CoV-2 helicase nsp 13 due to the larger deformation size and higher level of disorder --, 
gives physically realistic summary statistics for both proteins and retrieves the underlying dimension despite large conformational changes -- all within mere seconds on a laptop.
% geodesics and barycentres leave adjacent peptide distance invariant and that low-rank decompositions correctly retrieves the intrinsic dimension of the data -- despite the large conformational change. That is, a single geodesics approximates the molecular dynamics trajectory exceedingly well.
In particular, we demonstrate that our approach can break down the former 636-dimensional conformations to just a non-linear 1-dimensional space and break down the latter 1764-dimensional conformations to a non-linear 7-dimensional space without a significant loss in approximation accuracy.

\subsection{Outline}
\label{sec:intro-protein-geometry-outline} 
This article is structured as follows. \Cref{sec:prelim-protein-geometry} covers basic notation from differential and Riemannian geometry. In \cref{sec:separation-protein-geometry} we define the notion of separation on general Riemannian manifolds and show how this object enables the construction of provably computationally feasible approximations of geodesics, the exponential mapping and the logarithmic mapping. \Cref{sec:riemannian-protein-geometry} proposes a point cloud-based differential geometry of protein conformations and general guidelines for additionally constructing protein geometry. \Cref{sec:riemannian-protein-geometry-separation} considers a concrete example with clear ties to normal mode analysis, but arguably overcomes the open problem of locality and linearity. In \cref{sec:data-analysis-protein-geometry} we argue for deviating from established standards for more advanced data analysis tools and provide guidelines for doing so in the case of low rank approximation. The proposed Riemannian protein geometry is then tested numerically in \cref{sec:numerics-protein-geometry} through several data analysis tasks with molecular dynamics simulation data of the adenylate kinase protein and the SARS-CoV-2 helicase nsp 13 protein. Finally, we summarize our findings in \cref{sec:conclusions-protein-geometry}. The main article describes the general ideas and major results. All the proofs are carried out in the appendix.

\section{Notation}
\label{sec:prelim-protein-geometry}

Here we present some basic notations from differential and Riemannian geometry, see \cite{boothby2003introduction,carmo1992riemannian,lee2013smooth,sakai1996riemannian} for details. 
% \dn{What does this sentence mean exactly? Maybe instead say something like "Here we present some basics and notation from differential and Riemannian geometry, see \cite{}... for more details." ?} \wdp{fixed}

Let $\manifold$ be a smooth manifold. We write $C^\infty(\manifold)$ for the space of smooth functions over $\manifold$. The \emph{tangent space} at $\mPoint \in \manifold$, which is defined as the space of all \emph{derivations} at $\mPoint$, is denoted by $\tangent_\mPoint \manifold$ and for \emph{tangent vectors} we write $\mTVector_\mPoint \in \tangent_\mPoint \manifold$. For the \emph{tangent bundle} we write $\tangent\manifold$ and smooth vector fields, which are defined as \emph{smooth sections} of the tangent bundle, are written as $\vectorfield(\manifold) \subset \tangent\manifold$. If $\manifold$ is a \emph{quotient manifold}, i.e., $\manifold := \manifoldB/\group$ for some  smooth manifold $\manifoldB$ and \emph{Lie group} $\group$, we denote its elements with square brackets $[\mPoint]\in \manifoldB/\group$. Besides a tangent space, quotient manifolds are equipped with a \emph{vertical} and \emph{horizontal space} for $\mPoint\in [\mPoint]$. The former is defined as $\vertical_{\mPoint} \manifoldB :=\ker (D_{\mPoint}\pi)$, where $D_{\mPoint}\pi: \tangent_\mPoint \manifoldB \to \tangent_{\pi(\mPoint)}\manifoldB/\group$ is the differential of the \emph{canonical projection} mapping $\pi:\manifoldB \to \manifoldB/\group$, and the latter is the linear subspace $\horizontal_{\mPoint} \manifoldB \subset \tangent_\mPoint \manifoldB$ such that $\horizontal_{\mPoint} \manifoldB \cap \vertical_{\mPoint} \manifoldB = \{0\}$ and $\horizontal_{\mPoint} \manifoldB \oplus\vertical_{\mPoint} \manifoldB = \tangent_\mPoint \manifoldB$. To distinguish horizontal vectors and tangent vectors, we write $\mTVector_{\diamond\mPoint} \in \horizontal_{\mPoint} \manifoldB$ and $\mTVector_{[\mPoint]} \in \tangent_{[\mPoint]} \manifoldB/\group$. 
% Furthermore, additional structure on manifolds and derived notions thereof are typically defined through tensor fields of the form $T~:~\vectorfield(\manifold)^n \to \vectorfield(\manifold)^m$ or $T~:~\vectorfield(\manifold)^n \to C^\infty(\manifold)$. A tensor field can -- by definition of being $C^\infty(\manifold)$-modules -- be restricted to a tangent space at each point $\mPoint \in \manifold$, for which we use the notation $T_{\mPoint}: \tangent_\mPoint \manifold^n \to \tangent_\mPoint \manifold^m$ or $T_{\mPoint}: \tangent_\mPoint \manifold^n \to \Real$. 

A smooth manifold $\manifold$ becomes a \emph{Riemannian manifold} if it is equipped with a smoothly varying \emph{metric tensor field} $(\,\cdot\,, \,\cdot\,) \colon \vectorfield(\manifold) \times \vectorfield(\manifold) \to C^\infty(\manifold)$. This tensor field induces a \emph{(Riemannian) metric} $\distance_{\manifold} \colon \manifold\times\manifold\to\Real$. The metric tensor can also be used to construct a unique affine connection, the \emph{Levi-Civita connection}, that is denoted by $\nabla_{(\,\cdot\,)}(\,\cdot\,) : \vectorfield(\manifold) \times \vectorfield(\manifold) \to \vectorfield(\manifold)$. 
This connection is in turn the cornerstone of a myriad of manifold mappings.
One is the notion of a \emph{geodesic}, which for two points $\mPoint,\mPointB \in \manifold$ is defined as a curve $\geodesic_{\mPoint,\mPointB} \colon [0,1] \to \manifold$ with minimal length that connects $\mPoint$ with $\mPointB$. 
% This notion is well-defined if the manifold is \emph{geodesically connected}, i.e., any two points $\mPoint,\mPointB\in\manifold$ can be connected with a geodesic that is contained in $\manifold$.
Another closely related notion is the curve $t \mapsto \geodesic_{\mPoint,\mTVector_\mPoint}(t)$  for a geodesic starting from $\mPoint\in\manifold$ with velocity $\dot{\geodesic}_{\mPoint,\mTVector_\mPoint} (0) = \mTVector_\mPoint \in \tangent_\mPoint\manifold$. This can be used to define the \emph{exponential map} $\exp_\mPoint \colon \mathcal{D}_\mPoint \to \manifold$ as 
\[ 
\exp_\mPoint(\mTVector_\mPoint) := \geodesic_{\mPoint,\mTVector_\mPoint}(1)
\quad\text{where $\mathcal{D}_\mPoint \subset \tangent_\mPoint\manifold$ is the set on which $\geodesic_{\mPoint,\mTVector_\mPoint}(1)$ is defined.} 
\]
% The manifold $\manifold$ is said to be \emph{complete} whenever $\mathcal{D}_p = \mathcal{T}_{\mPoint}\manifold$.
Furthermore, the \emph{logarithmic map} $\log_\mPoint \colon \exp(\mathcal{D}'_\mPoint ) \to \mathcal{D}'_\mPoint$ is defined as the inverse of $\exp_\mPoint$, so it is well-defined on  $\mathcal{D}'_{\mPoint} \subset \mathcal{D}_{\mPoint}$ where $\exp_\mPoint$ is a diffeomorphism. 
Moreover, the \emph{Riemannian gradient} of a smooth function $F \colon \manifold\to \Real$ denotes the unique vector field $\Grad F \in \vectorfield(\manifold)$ such that 
\[
    (\Grad F,
      \mTVector )_\mPoint :=  \mTVector_\mPoint F := D_{\mPoint} F(\cdot) [\mTVector_\mPoint] ,
    \quad\text{holds for any $\mTVector \in \vectorfield(\manifold)$ and $\mPoint\in\manifold$,}
\]
where $D_{\mPoint}F(\cdot): \tangent_\mPoint \manifold \to \Real$ denotes the differential of $F$.

% \todo[inline]{Maybe also discuss the notions from optimisation. e.g. retraction, geodesically convex.}

% \todo[inline]{grad as Riemannian gradient in here and Hess, differential}
\section{Computationally feasible Riemannian geometry}
\label{sec:separation-protein-geometry}

In this section we focus on constructing efficient approximations of several manifold mappings. We restrict ourselves to the main results and refer the reader to \cref{app:proofs-separation-protein-geometry} for auxiliary lemmas and the proofs.

% [proofs in \cref{app:proofs-separation-protein-geometry}]
% \subsection{Relaxation of the Riemannian distance and an approximate logarithmic mapping}
% \label{sec:separation-protein-geometry-relaxation}

Realistically, efficient approximation of manifold mappings requires additional structure as argued in \cref{sec:intro-protein-geometry-related-work}. Our approach is similar to the \emph{variational time discretisation} approach by Rumpf et al. \cite{rumpf2015variational}, which has shown to be successful for several applications \cite{heeren2016splines,heeren2012time,wirth2011continuum}. The key difference is that we use \emph{fully intrinsic notions} that do not rely on Euclidean Taylor approximations or ambient Banach space structure. This allows for several natural approximation schemes that are distinct from the ones proposed in \cite{rumpf2015variational} and are more general. Our theoretical framework relies on the notion of \emph{separation}.

% In \cite{rumpf2015variational} a second-order approximation of the Riemannian distance was assumed. However, checking [this] was done extrinsically through Euclidean Taylor approximations. Below, we will assume similar structure -- that of the separation -- and show that we attain a second-order approximation of the distance in an intrinsic fashion.

% maybe also cite rumpf2015 here and say that assuming 3rd-order approximations for distance is fairly standard, but not done in an intrinsic way yet

\begin{definition}[separation]
\label{def:separation}
    Let $(\manifold, (\cdot, \cdot))$ be a Riemannian manifold. A mapping $\separation:\manifold\times \manifold\to \Real$ is called a \emph{separation with respect to the Riemannian metric tensor field $(\cdot, \cdot)$} if it satisfies the following properties:
    \begin{enumerate}[(i)]
        \item $\separation$ is a metric on $\manifold$,
        \item for any $\mPoint\in \manifold$, there exists a neighbourhood of $\mPoint$ in which the mapping $\separation(\mPoint, \cdot)^2$ is smooth,
        \item for any $\mTVector_{\mPoint} \in \tangent_\mPoint \manifold$ at any $\mPoint\in \manifold$ the identity $\nabla_{\mTVector_{\mPoint}} \Grad \separation(\mPoint, \cdot)^2 = 2 \mTVector_{\mPoint} \in \tangent_\mPoint \manifold$ holds.
    \end{enumerate}
\end{definition}

As we will see below, separations can be used to approximate basic Riemannian manifold mappings. This works primarily because they are designed to approximate the Riemannian distance, as shown in the following theorem.

% The main reason for their success comes from the fact that they are approximations of the Riemannian distance.

\begin{theorem}[Relation between a separation and $\distance_{\manifold}$]
\label{thm:separation-and-distance}
    Let $(\manifold,  (\cdot, \cdot))$ be a Riemannian manifold and $\distance_{\manifold}:\manifold\times \manifold\to \Real$ be the distance on $\manifold$ generated by $(\cdot, \cdot)$. Furthermore, let $\separation:\manifold\times \manifold\to \Real$ be separation on $\manifold$ with respect to $(\cdot, \cdot)$. 
    % [We can characterize the relation between $\distance_\manifold$ and $\separation$ in the following way]

    Then, for $\mPoint\in \manifold$,
    \begin{equation}
        \separation(\mPoint, \mPointB)^2 = \distance_{\manifold}(\mPoint, \mPointB)^2 + \mathcal{O}(\distance_{\manifold}(\mPoint, \mPointB)^3),\quad \text{as $\mPointB \to \mPoint$,}
        \label{eq:thm-relation-sepa-dist-i}
    \end{equation}
    and the approximation \cref{eq:thm-relation-sepa-dist-i} becomes exact, i.e., for any $\mPointB$ in a neighbourhood of $\mPoint\in \manifold$,
    \begin{equation}
        \separation(\mPoint, \mPointB)^2 = \distance_{\manifold}(\mPoint, \mPointB)^2
        \label{eq:thm-relation-sepa-dist-ii}
    \end{equation}
     if and only if additionally $\Grad \separation(\mPoint, \cdot)^2\mid_{\mPointC} = - 2 \log_{\mPointC}\mPoint$ for any $\mPointC\in \geodesic_{\mPointB,\mPoint}$.
\end{theorem}

We note that property \cref{eq:thm-relation-sepa-dist-i} is very useful in practice and also underlies the convergence of the variational time discretisation approximations in \cite{rumpf2015variational}. However, in approximating the manifold mappings we take a different route. For one, a separation directly gives us a first-order approximation of the Riemannian logarithm -- rather than solving a discrete geodesic problem first. That is, defining the \emph{$\separation$-logarithmic map} $\log_{\mPoint}^{\separation}: \manifold \to \tangent_\mPoint \manifold$ as
\begin{equation}
    \log_{\mPoint}^{\separation} (\mPointB) := -\frac{1}{2} \Grad \separation(\mPointB, \cdot)^2\mid_{\mPoint} \in \tangent_\mPoint \manifold,
\end{equation}
we have the following result.

\begin{corollary}
\label{cor:log-approx-separation}
Under the assumptions in \cref{thm:separation-and-distance}, the $\separation$-logarithmic map satisfies for any $\mPoint\in \manifold$,
    \begin{equation}
    \|\log_{\mPoint}^{\separation} \mPointB  - \log_{\mPoint}\mPointB\|_{\mPoint} \in \mathcal{O}(\distance_\manifold(\mPoint, \mPointB)^2), \quad \text{as $\mPointB \to \mPoint$}.
\end{equation}
\end{corollary}

For the geodesics we will also resort to a different approximation scheme. In particular, we construct the $\separation$-geodesic $\geodesic_{\mPoint, \mPointB}^{\separation} : [0,1] \to \manifold$ as
\begin{equation}
    \geodesic_{\mPoint, \mPointB}^{\separation}(t) \in \argmin_{\mPointC \in \manifold} \frac{1 - t}{2}\separation (\mPoint, \mPointC)^2 + \frac{t}{2} \separation (\mPointC, \mPointB)^2.
    \label{eq:separation-geodesics}
\end{equation}
Note that even if we could choose $\separation$ to be the exact Riemannian metric $\distance_\manifold$, the optimisation problem \cref{eq:separation-geodesics} does not necessarily have minimisers. So we have to be careful here. On the flip side, if $\separation := \distance_\manifold$ and a solution exists, the problem is locally \emph{strongly geodesically convex}, i.e., strongly convex along geodesics \cite{boumal2023introduction}, and geodesics $\geodesic_{\mPoint, \mPointB}(t)$ at time $t$ are minimisers, which at least gives consistency. 

Ideally, a separation would inherit such properties. As it turns out, existence is guaranteed under an additional metric completeness assumption, whereas geodesic convexity is inherited directly.

\begin{theorem}
\label{thm:well-posed-ness-separation-geodesics}
     Let $(\manifold, (\cdot, \cdot))$ be a Riemannian manifold, and consider any metric $\separation: \manifold \times \manifold \to \Real$ on $\manifold$ and the minimisation problem \cref{eq:separation-geodesics} for arbitrarily fixed $t \in [0,1]$.
     \begin{enumerate}[(i)]
         \item If $\separation$ is a complete metric, a minimiser exists.
         \item If $\separation$ is a separation with respect to $(\cdot, \cdot)$, the problem is locally strongly geodesically convex for $\mPoint$ and $\mPointB$ close enough.
     \end{enumerate}
\end{theorem}

Although property (i) in \cref{thm:well-posed-ness-separation-geodesics} is important from a mathematical point of view, property (ii) has more practical consequences. In particular, as a consequence we enjoy linear convergence when solving the problem with simple Riemannian gradient descent \cite{zhang2016first}\footnote{Unit step size is sufficient due to local Lipschitz gradients by assumption (iii) in \cref{def:separation}.}. Since the exponential mapping is not available, we assume a \emph{retraction}, i.e., a first-order approximation of the exponential mapping, instead, without affecting the convergence rate \cite{absil2009optimization}. In other words, having a complete separation yields a more convenient and lower-dimensional optimisation problem than computing discrete geodesics \cite{rumpf2015variational}. 

Finally, for approximating the exponential mapping beyond first-order we are using a slight modification of the recursive scheme used in \cite{rumpf2015variational}. Assuming a retraction $\operatorname{retr}_{\mPoint}: \tangent_\mPoint \manifold\to \manifold$, we construct the $\separation$-exponential mapping $\exp^{\separation}_{\mPoint}: \tangent_\mPoint \manifold\to \manifold$ as
\begin{equation}
    \exp^{\separation}_{\mPoint} (\mTVector_{\mPoint}) := \mPoint^M \in \manifold,
\end{equation}
where
\begin{equation}
    \mPoint^\sumIndC := \geodesic_{\mPoint^{\sumIndC-2}, \mPoint^{\sumIndC-1}}^{\separation}(2), \quad \mPoint^0 = \mPoint, \quad \text{and} \quad \mPoint^1 = \operatorname{retr}_{\mPoint} \Bigl(\frac{1}{M}\mTVector_{\mPoint}\Bigr), \quad \text{for some $M\in \Natural$.}
    \label{eq:w-exp-recursive-scheme}
\end{equation}

Note that $t=2$ in \cref{eq:w-exp-recursive-scheme} lies outside of the admissible region in \cref{thm:well-posed-ness-separation-geodesics}. However, this is amendable and we obtain a similar result. So once again, we obtain a provably efficient approximation.

\begin{theorem}
\label{thm:well-posed-ness-separation-exp}
     Let $(\manifold, (\cdot, \cdot))$ be a Riemannian manifold, and consider any metric $\separation: \manifold \times \manifold \to \Real$ on $\manifold$ and the minimisation problem \cref{eq:separation-geodesics} for $t = 2$.
     \begin{enumerate}[(i)]
         \item If $\separation$ is a complete metric, a minimiser exists.
         \item If $\separation$ is a separation with respect to $(\cdot, \cdot)$, the problem is locally strongly geodesically convex for $\mPoint$ and $\mPointB$ close enough.
     \end{enumerate}
\end{theorem}

% \todo[inline]{Rephrase remark below}
% \begin{remark}
%    Note that existence of geodesics and the exponential mapping is normally shown through proving geodesic completeness combined with the Hopf-Rinow Theorem, e.g., as in \cite{herzog2023manifold}. Our assumptions are weaker. They also hold for separations on non-connected manifolds. So naturally we have to be careful, because even though minimisers exist, separation geodesics are not necessarily continuous curves without the locality assumption. 
    % [However, beyond scope of this work.]
% \end{remark}
\section{Reverse-engineering Riemannian geometry for point cloud conformations}
\label{sec:riemannian-protein-geometry}
The goal in this section is to \emph{set up guidelines} for constructing Riemannian geometry for proteins that comes with a separation on this space, which then enables access to all computationally feasible manifold mapping approximations proposed in \cref{sec:separation-protein-geometry}. The key idea is to \emph{construct a metric first and reverse-engineer a metric tensor field on which the original metric is a separation}. For full detail, we refer the reader to the proofs in \cref{app:proofs-quotient-manifold}.

% \subsection{A smooth quotient manifold of point cloud conformations (1 page)}
% \label{sec:quotient-manifold-protein-geometry}
Before considering Riemannian geometry, we first need a smooth manifold.
As motivated in \cref{sec:intro-protein-geometry-related-work}, we will start from a point cloud-based manifold and take a suitable quotient.
% In \cref{sec:intro-protein-geometry-related-work} we have seen that there is no readily available manifold that allows for the appropriate (rigid body) symmetries required for an energy landscape-based Riemannian structure, but have also argued that a point cloud-based quotient manifold is the most suitable representation, if we can overcome challenges in construction. To alleviate the construction problem -- i.e., actually obtaining a quotient manifold rather than a quotient space --, 
We start with the sets
%construct a smooth quotient manifold for modeling proteins conformations from the sets
\begin{equation}
    \PCmanifoldA := \{\eMat \in \Real^{\dimInd\times \proteinLen} \mid \eMat = (\ePoint_1, \ldots, \ePoint_\proteinLen)  \text{ s.t. } \ePoint_\sumIndA\neq \ePoint_\sumIndB \in \Real^\dimInd\},
    \label{eq:Rdnx}
\end{equation}
and
\begin{equation}
    \PCmanifoldB := \{\eMat \in \Real^{\dimInd\times \proteinLen} \mid \eMat - \frac{1}{\proteinLen}\eMat \mathbf{1}_{\proteinLen} \mathbf{1}_{\proteinLen}^{\top} \in \Real^{\dimInd\times \proteinLen}_\dimInd \},
    \label{eq:Rdndx}
\end{equation}
where $\mathbf{1}_{\proteinLen}:= (1, \ldots, 1)^\top \in \Real^{\proteinLen}$ and $\Real^{\dimInd\times \proteinLen}_\dimInd$ is the set of matrices of size $\dimInd \times \proteinLen$ and rank $\dimInd$.
We note that $\PCmanifoldA$ models the physical constraint that two atoms or course-grained pseudo-atoms cannot overlap, whereas $\PCmanifoldB$ models that point clouds do not live in an affine subspace and is chosen so that the quotient space defined in \cref{thm:quotient-manifold-Rdnd-mod-Ed} will actually be a smooth manifold. 

Our first step is combining the two sets through intersection, which gives us a smooth manifold.

% \subsection{A smooth manifold of point clouds}
% \label{sec:quotient-manifold-protein-geometry-point-clouds}

\begin{proposition}
\label{thm:point-cloud-manifold}
The set
\begin{equation}
    %\Real^{\dimInd\times \proteinLen}_{\dimInd, \star, *} 
    \PCmanifold := \PCmanifoldB \cap \PCmanifoldA 
    %= \{\eMat \in \Real^{\dimInd\times \proteinLen} \mid \eMat = (\ePoint_1, \ldots, \ePoint_\proteinLen) \text{ s.t. } \ePoint_\sumIndA\neq \ePoint_\sumIndB \in \Real^\dimInd \text{ and } \eMat - \frac{1}{\proteinLen}\eMat\mathbf{1}_{\proteinLen} \mathbf{1}_{\proteinLen}^{\top} \in \Real^{\dimInd\times \proteinLen}_\dimInd \}
    \label{eq:point-cloud-manifold}
\end{equation}
is a smooth $(\proteinLen\cdot \dimInd)$-dimensional manifold if $\proteinLen \geq \dimInd +1$.
\end{proposition}

% \subsection{Point clouds modulo Euclidean group action}
% \label{sec:quotient-manifold-protein-geometry-group}

Next, we consider left group actions on $\PCmanifold$. More generally, if $\group$ is a Lie group and $\manifold$ is a smooth manifold, a \emph{left group action} of $\group$ on $\manifold$ is a map $\leftGroupAction: \group \times \manifold \rightarrow \manifold$ that satisfies
\begin{equation}
\begin{array}{cll}
\leftGroupAction(\idPoint, \mPoint) = \mPoint, & \text { for all } \mPoint \in \manifold, & \text{(identity)}\\
\leftGroupAction(\gPoint_1, \leftGroupAction(\gPoint_2, \mPoint)) = \leftGroupAction(\gPoint_1 \cdot \gPoint_2, \mPoint), & \text { for all } \gPoint_1, \gPoint_2 \in \group \text { and } \mPoint \in \manifold. & \text{(compatibility)}
\end{array}
\label{eq:identity-compatibility-lie-group}
\end{equation}

In the following, we will consider the Euclidean group $\mathbb{E}(\dimInd)$. We represent elements of this Lie group in the canonical way as $(\EdOrthoref, \EdTrans)\in \mathbb{E}(\dimInd)$, where $\EdOrthoref \in \mathbb{O}(\dimInd)$ is a $\dimInd\times \dimInd$ orthogonal matrix and $\EdTrans \in \Real^\dimInd$ is a (translation) vector. We can extend the left group action of $\mathbb{E}(\dimInd)$ on $\Real^\dimInd$ to an action on $\PCmanifold$. That is, we define $\leftGroupAction: \mathbb{E}(\dimInd) \times \PCmanifold \to \PCmanifold$ as 
\begin{equation}
    \leftGroupAction((\EdOrthoref, \EdTrans), \eMat) := \EdOrthoref \eMat + \EdTrans \mathbf{1}_{\proteinLen}^{\top}, \quad (\EdOrthoref, \EdTrans)\in \mathbb{E}(\dimInd), \; \eMat\in \PCmanifold.
    \label{prop:left-group-action-Rdnd}
\end{equation} 

The left group action \cref{prop:left-group-action-Rdnd} on $\PCmanifold$ allows us to construct the quotient space $\PCmanifold/\mathbb{E}(\dimInd)$. By construction this quotient space is also a smooth manifold.
\begin{theorem}
\label{thm:quotient-manifold-Rdnd-mod-Ed}
The quotient space $\PCmanifold /\mathbb{E}(\dimInd)$ is a smooth ($\proteinLen\cdot\dimInd - (\dimInd +1)\cdot\dimInd/2$)-dimensional manifold if $\proteinLen \geq \dimInd +1$.
\end{theorem}

% \subsection{Reverse-engineering a separation (2 pages)}
% \label{sec:riemannian-protein-geometry-reverse-engineering}

% [proofs in \cref{app:proofs-riemannian-protein-geometry-reverse-engineering}]

% Do we want to cite -- because they also assume the same, but do it in an extrinsic setting, which we cannot do now
 % also uses the trick to do euclidean hessian to to taylor approximation -- does not hold for general manifolds
% \cite{rumpf2014geometry}
% \cite{berkels2013discrete}

Now that we have a smooth quotient manifold, the next step is to construct a Riemannian structure that is computationally feasible, yet adapted for protein conformations. The following result gives clear guidelines for doing that, i.e., it shows that having a suitable metric on the quotient space $\PCmanifold /\mathbb{E}(\dimInd)$ always allows us to find a Riemannian manifold on which that metric is a separation. 

% [we talked about symmetry and that a Riemannian structure has to satisfy symmetry] [This means that the Riemannian distance must satisfy ...] 
% [As it turns out, having a metric will always allow us to find a Riemannian manifold on which that metric is a separation]

\begin{theorem}
\label{thm:class-of-separations-on-pointcloud-manifold}
    Consider a metric $\separation: \PCmanifold /\mathbb{E}(\dimInd) \times \PCmanifold /\mathbb{E}(\dimInd) \to \Real$ on $\PCmanifold /\mathbb{E}(\dimInd)$ for $\proteinLen \geq \dimInd+1$ of the form
    \begin{equation}
        \separation([\eMat], [\eMatB]) = \tilde{\separation}(\eMat, \eMatB), \quad \eMat\in [\eMat], \eMatB\in [\eMatB],
        \label{eq:thm-lass-of-separations-on-pointcloud-manifold-w-form}
    \end{equation}
    where $\tilde{\separation}: \PCmanifold \times  \PCmanifold\to \Real$ is a mapping that is invariant under $\mathbb{E}(\dimInd)$ action in both arguments. Additionally, assume that $\tilde{\separation}(\eMat,\cdot)^2$ is smooth in a neighbourhood of $\eMat$ for any $\eMat\in \PCmanifold$ and that the Euclidean Hessian $D_{\eMat} \nabla \tilde{\separation}(\eMat, \cdot)^2 : \Real^{\dimInd\times \proteinLen} \to \Real^{\dimInd\times \proteinLen}$ is positive definite when restricted to the horizontal space $\horizontal_{\eMat} \PCmanifold$ and has kernel $\ker(D_{\eMat} \nabla \tilde{\separation}(\eMat, \cdot)^2) = \vertical_{\eMat} \PCmanifold$ for all $\eMat \in \PCmanifold$.
    
    Then, the bi-linear form $(\cdot, \cdot): \vectorfield (\PCmanifold /\mathbb{E}(\dimInd)) \times \vectorfield (\PCmanifold /\mathbb{E}(\dimInd)) \to C^\infty(\PCmanifold /\mathbb{E}(\dimInd))$ given by 
    \begin{equation}
        (\mTVector, \mTVectorB)_{[\eMat]} := \frac{1}{2} \Bigl(D_{\eMat} \nabla \tilde{\separation}(\eMat, \cdot)^2 [\mTVector_{\diamond \eMat}],  \mTVectorB_{\diamond \eMat}\Bigr)_2, \quad \eMat \in [\eMat], \text{ where $(\cdot, \cdot)_2$ is the $\ell^2$ inner product,}
    \end{equation}
     defines a metric tensor field on $\PCmanifold /\mathbb{E}(\dimInd)$, and $\separation$ is a separation under $(\cdot, \cdot)$.
\end{theorem}

% The above result gives us a guideline for designing efficient Riemannian geometry for protein conformations [and all the differential geometry is done]. In particular, we can construct a metric that models after a protein conformation landscape and find

\section{A Riemannian geometry and separation for protein dynamics data}
\label{sec:riemannian-protein-geometry-separation}

Now we are ready to choose a Riemannian geometry for efficient analysis of protein conformations through \cref{thm:class-of-separations-on-pointcloud-manifold}. This amounts to choosing a metric suited for quantifying similarity between protein conformations and reverse-engineering a metric tensor field and a separation. The
% In this section we will carefully construct a 
metric will have
% that has 
strong \emph{similarities to normal mode analysis} discussed in \cref{sec:intro-protein-geometry}, but is arguably \emph{less likely to suffer from locality and linearity issues} discussed in \cref{sec:intro-protein-geometry-related-work}. 
% In other words, the proposed geometry is consistent with the literature on deforming proteins. 
% In particular, one consequence of our non-local metric is that separation geodesics will approximately preserve bond-lengths between atoms or pseudo-atoms. 
We restrict ourselves once again to stating the main results and refer the reader to \cref{app:proofs-chosen-Riemannian} for the details. Furthermore, all statements will be given for general $\dimInd$-dimensional point clouds -- even though we only need the case $\dimInd=3$ for proteins.
% Additionally, in our Riemannian framework we enjoy more benefits and options than previously possible, e.g., preservation of bond-lengths over geodesics, we have an exponential [No longer need two states to explore what is in between, but one state and a velocity vector is sufficient] and logarithmic mapping [which tells in what direction one walks to get to a certain point, which is useful in low rank approximation]

Invoking \cref{thm:class-of-separations-on-pointcloud-manifold} requires selecting
% we need 
a metric of the form \cref{eq:thm-lass-of-separations-on-pointcloud-manifold-w-form}. Passing to a \emph{Euclidean distance matrix} \cite{dokmanic2015euclidean} representation of point clouds allows us to construct a whole family of $\mathbb{E}(\dimInd)$ invariant metrics.
% -- that are also invariant with respect to $\mathbb{E}(\dimInd)$.

\begin{proposition}
\label{thm:candidate-metric}
Let $\distance_{>0}: \Real_{>0} \times \Real_{>0} \to \Real$ be any metric over the positive real numbers.

Then, the function $\separation:\PCmanifold /\mathbb{E}(\dimInd) \times \PCmanifold /\mathbb{E}(\dimInd) \to \Real$ given by
\begin{equation}
    \separation([\eMat], [\eMatB]):= \sqrt{\sum_{\sumIndA} \sum_{\sumIndB > \sumIndA} \distance_{>0}(\|\ePoint_\sumIndA - \ePoint_\sumIndB\|_2, \|\ePointB_\sumIndA - \ePointB_\sumIndB\|_2)^2}, \quad \eMat \in [\eMat], \eMatB \in [\eMatB],
    \label{eq:candidate-metric-thm}
\end{equation}
defines a metric over $\PCmanifold /\mathbb{E}(\dimInd)$ if $\proteinLen \geq \dimInd+1$.
\end{proposition}

The expression in \cref{eq:candidate-metric-thm} has a similar form to the classical quadratic -- Hookean -- potential proposed in \cite{tirion1996large} when interpreting 
the $[\eMatB]$ entry
% one of the entries 
as a reference protein conformation. However, there are two important differences: in \cite{tirion1996large} (i) the interactions are restricted to atoms within some cut-off radius -- violating the metric axioms --, and (ii) the quadratic potential stipulates that the distance -- or energy needed -- to transition into a state $\eMatC \in \Real^{\dimInd\times \proteinLen}\setminus \PCmanifoldA$ is \emph{finite and small}, i.e., to a state with $\ePointC_\sumIndA = \ePointC_\sumIndB$ for some $\sumIndA\neq \sumIndB$. The latter difference tells us that there are many non-physical states close by, which is arguably a key source of the locality problem. In reality it would take infinite energy to move two atoms to the same location, which means that these states should be infinitely far away\footnote{Note that we are asking here for a complete metric, which already gives us existence of separation-geodesics \cref{eq:separation-geodesics} and also has physical meaning now.}.

% [Maybe mentionn here already that the classical choice would be to choose d geq0 to be the hookean potential (however, this does not define a metric due to the cut-off, and is non-smooth due to the cut-off). Also, if it were fixed, the then to be metric is not complete. So we might have issues with existence. Although it should be said that interpolation schemes based on such an idea work reasonably well in practice.]

Instead, in the following we will make another choice for $\distance_{>0}$. That is, we choose the complete metric $\distance_{>0}: \Real_{>0} \times \Real_{>0} \to \Real$ over the positive real numbers given by
\begin{equation}
    \distance_{>0}(a, b) := \Bigl|\log\Bigl(\frac{a}{b}\Bigr)\Bigr|.
    \label{eq:candidate-metric-component}
\end{equation}
Intuitively, this metric inserted into \cref{eq:candidate-metric-thm} tells us that \emph{two point-clouds are close together if their respective pair-wise distances are in the same order of magnitude}. Such a metric directly solves the locality problem, since local interactions are fairly strong -- and blow up when moving towards a state $\eMatC$ with $\ePointC_\sumIndA = \ePointC_\sumIndB$ --, but does not need an explicit cut-off as long-range interactions are relatively weak. We note that such a metric is truly more in line with well-known interaction potentials such as the famous Lennard-Jones potential than the quadratic potential from \cite{tirion1996large}. With non-locality accounted for, non-linearity is naturally taken care of through availability of $\separation$-geodesics, the $\separation$-exponential mapping and the $\separation$-logarithmic mapping, if we can ensure completeness of the metric. 
% Finally, and more heuristically, such a metric is also expected to model large-scale deformations better than the quadratic potential as it is more similar to well-known interaction potentials such as the famous Lennard-Jones potential.

% [A comparison between the Lennard-Jones potential and our proxy-potential \cref{eq:candidate-metric-component}].

\begin{remark}
\label{rem:nice-interpolation}
    Note that we also expect from a Riemannian manifold with the metric \cref{eq:candidate-metric-thm} under \cref{eq:candidate-metric-component}
    %such a metric 
    as separation to preserve small invariant distances, e.g., protein bond lengths, over $\separation$-geodesics.
    % , which is yet more motivation why this choice is suitable for protein conformation data analysis.
    We expect the same from $\separation$-based Riemannian barycentres \cite{karcher1977riemannian}.
\end{remark}

% Now propose a metric and compare this to the Lennard-Jones potential.

% There is one minor issue, that is, the metric becomes degenerate as one travels over the manifold towards the [boundary of the fixed rank constraint.] Indeed, given a point in $\Real^{\dimInd \times \proteinLen}_{*} \setminus \Real^{\dimInd \times \proteinLen}_{\dimInd, \star}$, the distance towards this point is finite from any other point in the quotient manifold. So if there were to exist a Riemannian metric tensor field that would generate a metric [], this metric would not be \emph{complete}. We get the same issues as outlined in \cref{rem:non-completeness-positive-metric}.

Despite the metric \cref{eq:candidate-metric-component} being complete on the positive real numbers, the metric \cref{eq:candidate-metric-thm} under \cref{eq:candidate-metric-component} is not complete on $\PCmanifold /\mathbb{E}(\dimInd)$ yet as the distance -- or energy needed -- to transition into a state
% similar phenomena as above can occur for states 
$\eMatC \in \Real^{\dimInd\times \proteinLen}\setminus \PCmanifoldB$ is again finite and small, i.e., to a state where all points lie on an affine subspace. 
A simple way to resolve this is to add an extra term that blows up towards such states. In particular, we propose the full metric $\separation^\GyRaParam:\PCmanifold /\mathbb{E}(\dimInd) \times \PCmanifold /\mathbb{E}(\dimInd) \to \Real$ given by
\begin{equation}
    \separation^\GyRaParam ([\eMat], [\eMatB]):= \sqrt{\sum_{\sumIndA} \sum_{\sumIndB > \sumIndA} \Biggl(\log \biggl(\frac{\|\ePoint_\sumIndA - \ePoint_\sumIndB\|_2}{\|\ePointB_\sumIndA - \ePointB_\sumIndB\|_2} \biggr)\Biggr)^2 + \GyRaParam \Biggl(\log \biggl(\frac{\det\bigl(\sum_{\sumIndA} (\ePoint_\sumIndA - \frac{1}{\proteinLen} \eMat \mathbf{1}_\proteinLen) \otimes (\ePoint_\sumIndA - \frac{1}{\proteinLen} \eMat \mathbf{1}_\proteinLen)\bigr)}{\det\bigl(\sum_{\sumIndA} (\ePointB_\sumIndA - \frac{1}{\proteinLen} \eMatB \mathbf{1}_\proteinLen) \otimes (\ePointB_\sumIndA - \frac{1}{\proteinLen} \eMatB \mathbf{1}_\proteinLen)\bigr)} \biggr)\Biggr)^2} , \quad \eMat \in [\eMat], \eMatB \in [\eMatB],
    \label{eq:candidate-metric-full}
\end{equation}
where $\GyRaParam>0$ and $\mathbf{a} \otimes \mathbf{b} := \mathbf{a}\mathbf{b}^\top \in \Real^{\dimInd \times \dimInd}$ for $\mathbf{a}, \mathbf{b} \in \Real^{\dimInd}$. The above expression still gives us a metric.

\begin{corollary}
    The mapping $\separation^\GyRaParam:\PCmanifold /\mathbb{E}(\dimInd) \times \PCmanifold /\mathbb{E}(\dimInd) \to \Real$ in \cref{eq:candidate-metric-full} is a complete metric on $\PCmanifold /\mathbb{E}(\dimInd)$ if $\proteinLen \geq \dimInd+1$.
\end{corollary}

Intuitively, the additional term tells us that \emph{two conformations are close if their distribution of mass is similar}. Indeed, this can easily be seen through realizing that $\trace(\sum_{\sumIndA} (\ePoint_\sumIndA - \frac{1}{\proteinLen} \eMat \mathbf{1}_\proteinLen) \otimes (\ePoint_\sumIndA - \frac{1}{\proteinLen} \eMat \mathbf{1}_\proteinLen))$ is the square of the \emph{radius of gyration} -- a well-known quantity for describing whether proteins are in an open or closed state. In other words, this additional term is unlikely to jeopardize the constructed protein geometry -- especially for $\GyRaParam$ small so that the expected properties discussed in \cref{rem:nice-interpolation} are not disturbed too much.

% \begin{remark}
%     Note that $\trace((\eMat - \frac{1}{\proteinLen} \eMat \mathbf{1}_\proteinLen\mathbf{1}_\proteinLen^\top) (\eMat - \frac{1}{\proteinLen} \eMat \mathbf{1}_\proteinLen\mathbf{1}_\proteinLen^\top)^\top ))$ is a well-known quantity in comparing molecules, i.e., the radius of gyration -- or its square.
% \end{remark}

Finally, we have a metric and we are ready to construct a Riemannian manifold through invoking \cref{thm:class-of-separations-on-pointcloud-manifold}. The following result states that our candidate metric \cref{eq:candidate-metric-full} is a separation to a metric tensor field consisting of two terms, the leading first term being similar to the Hessian of the Hookean potential \cite{tirion1996large} with the difference that the pair-wise distances in the denominators are squared now\footnote{This effectively gives us smooth decay of the interaction strength rather than having to add an abrupt cut-off and is expected to give normal mode-like behaviour.}. Overall, this theorem tells us that the following \emph{Riemannian manifold enables computationally feasible data analysis of protein dynamics data along normal mode-like proxy energy-minimising non-linear paths without suffering from locality through the mappings from \cref{sec:separation-protein-geometry}}. 
% So overall, we really do inherit the best practices from normal mode analysis, but arguably do not suffer from basic locality due to the choice of metric and from linearity due to the availability of manifold mappings.

\begin{theorem}
\label{thm:protein-geometry-tensor-separation}
Let $(\pwdMetricTensor_{\diamond\eMat} + \GyRaParam \MetricTensorCorr_{\diamond\eMat}):  \Real^{\dimInd\times \proteinLen} \to \Real^{\dimInd\times \proteinLen}$ for $\GyRaParam>0$ and $\eMat:= (\ePoint_1, \ldots, \ePoint_\proteinLen)\in \PCmanifold$ with $\proteinLen \geq \dimInd+1$ be defined as
\begin{equation}
    (\pwdMetricTensor_{\diamond\eMat} + \GyRaParam \MetricTensorCorr_{\diamond\eMat}) (\mTVector_{\diamond \eMat}) := \sum_{\sumIndB=1}^\proteinLen \biggl(\Bigl((\pwdMetricTensor_{\diamond\eMat})_{1\sumIndB} + \GyRaParam (\MetricTensorCorr_{\diamond\eMat})_{1\sumIndB} \Bigr) \mTVectorCompon_\sumIndB, \ldots, \Bigl((\pwdMetricTensor_{\diamond\eMat})_{\proteinLen\sumIndB} + \GyRaParam (\MetricTensorCorr_{\diamond\eMat})_{\proteinLen\sumIndB} \Bigr) \mTVectorCompon_\sumIndB \biggr), \quad \mTVector_{\diamond \eMat} := (\mTVectorCompon_1, \ldots, \mTVectorCompon_\proteinLen),
\end{equation}
% $\tilde{\separation}^\GyRaParam: \Real^{\dimInd\times \proteinLen}_{\dimInd, \star, *}\times  \Real^{\dimInd\times \proteinLen}_{\dimInd, \star, *}\to \Real$ be a fuction given by the right hand side of \cref{eq:candidate-metric-full}
where each $(\pwdMetricTensor_{\diamond\eMat})_{\sumIndA\sumIndB} \in \Real^{\dimInd\times \dimInd}$ is defined as
% $\pwdMetricTensor_{\diamond\eMat}: \Real^{\dimInd\times \proteinLen} \to \Real^{\dimInd\times \proteinLen}$ 
% be defined as the linear map so that its restriction to the $\sumIndB$th input and $\sumIndA$th output column is
% defined through $\dimInd\times\dimInd$ blocks
\begin{equation}
    (\pwdMetricTensor_{\diamond\eMat})_{\sumIndA\sumIndB} := \left\{\begin{matrix}
 \sum_{\sumIndC\neq \sumIndA}  \frac{\ePoint_\sumIndA - \ePoint_\sumIndC}{\|\ePoint_\sumIndA - \ePoint_\sumIndC\|_2^2} \otimes \frac{\ePoint_\sumIndA - \ePoint_\sumIndC}{\|\ePoint_\sumIndA - \ePoint_\sumIndC\|_2^2}, & \text{for } \sumIndA = \sumIndB, \\
 -   \frac{\ePoint_\sumIndA - \ePoint_\sumIndB}{\|\ePoint_\sumIndA - \ePoint_\sumIndB\|_2^2} \otimes \frac{\ePoint_\sumIndA - \ePoint_\sumIndB}{\|\ePoint_\sumIndA - \ePoint_\sumIndB\|_2^2}, &  \text{for } \sumIndA\neq \sumIndB,
\end{matrix}\right. % (\pwdMetricTensor_{\diamond\eMat})_{\sumIndA\sumIndA}:=  \sum_{\sumIndB\neq \sumIndA}  \frac{\ePoint_\sumIndA - \ePoint_\sumIndB}{\|\ePoint_\sumIndA - \ePoint_\sumIndB\|_2^2} \otimes \frac{\ePoint_\sumIndA - \ePoint_\sumIndB}{\|\ePoint_\sumIndA - \ePoint_\sumIndB\|_2^2}, \quad \text{and}\quad  (\pwdMetricTensor_{\diamond\eMat})_{\sumIndA\sumIndB}:= -   \frac{\ePoint_\sumIndA - \ePoint_\sumIndB}{\|\ePoint_\sumIndA - \ePoint_\sumIndB\|_2^2} \otimes \frac{\ePoint_\sumIndA - \ePoint_\sumIndB}{\|\ePoint_\sumIndA - \ePoint_\sumIndB\|_2^2}, \quad \text{for } \sumIndA\neq \sumIndB,
\end{equation}
% and
% \begin{equation}
%     (\pwdMetricTensor_{\diamond\ePoint})_{\sumIndA\sumIndB}:= - 4  \frac{\ePoint_\sumIndA - \ePoint_\sumIndB}{\|\ePoint_\sumIndA - \ePoint_\sumIndB\|^2} \otimes \frac{\ePoint_\sumIndA - \ePoint_\sumIndB}{\|\ePoint_\sumIndA - \ePoint_\sumIndB\|^2}, \quad \sumIndA\neq \sumIndB.
% \end{equation}
and each $(\MetricTensorCorr_{\diamond\eMat})_{\sumIndA\sumIndB} \in \Real^{\dimInd\times \dimInd}$ is defined as
% $\MetricTensorCorr_{\diamond\eMat} : \Real^{\dimInd\times \proteinLen} \to \Real^{\dimInd\times \proteinLen}$ be defined similarly as
% through $\dimInd\times\dimInd$ blocks
\begin{equation}
    (\MetricTensorCorr_{\diamond\eMat})_{\sumIndA\sumIndB}:= 4 \ePointC_{\eMat,\sumIndA} \otimes \ePointC_{\eMat,\sumIndB}, \quad \text{for } \ePointC_{\eMat,\sumIndA} := \Bigl((\eMat - \frac{1}{\proteinLen} \eMat \mathbf{1}_\proteinLen\mathbf{1}_\proteinLen^\top) (\eMat - \frac{1}{\proteinLen} \eMat \mathbf{1}_\proteinLen\mathbf{1}_\proteinLen^\top)^\top \Bigr)^{-1} (\ePoint_\sumIndA - \frac{1}{\proteinLen} \eMat \mathbf{1}_\proteinLen).
\end{equation}
%\begin{equation}
 %   (\MetricTensorCorr_{\diamond\eMat})_{\sumIndA\sumIndB}:= 4 \Bigl((\eMat - \frac{1}{\proteinLen} \eMat \mathbf{1}_\proteinLen\mathbf{1}_\proteinLen^\top) (\eMat - \frac{1}{\proteinLen} \eMat \mathbf{1}_\proteinLen\mathbf{1}_\proteinLen^\top)^\top \Bigr)^{-1} (\ePoint_\sumIndA - \frac{1}{\proteinLen} \eMat \mathbf{1}_\proteinLen) \otimes \Bigl((\eMat - \frac{1}{\proteinLen} \eMat \mathbf{1}_\proteinLen\mathbf{1}_\proteinLen^\top) (\eMat - \frac{1}{\proteinLen} \eMat \mathbf{1}_\proteinLen\mathbf{1}_\proteinLen^\top)^\top \Bigr)^{-1} (\ePoint_\sumIndB - \frac{1}{\proteinLen} \eMat \mathbf{1}_\proteinLen).
%\end{equation}

Then, the bi-linear form $(\cdot,\cdot)^\GyRaParam: \vectorfield (\PCmanifold /\mathbb{E}(\dimInd)) \times \vectorfield (\PCmanifold /\mathbb{E}(\dimInd)) \to C^\infty(\PCmanifold /\mathbb{E}(\dimInd))\to \Real$ given by
\begin{equation}
    (\mTVector,\mTVectorB)^\GyRaParam_{[\eMat]}:= \Bigl((\pwdMetricTensor_{\diamond\eMat} + \GyRaParam \MetricTensorCorr_{\diamond\eMat}) (\mTVector_{\diamond \eMat}), \mTVectorB_{\diamond \eMat}\Bigr)_2, \quad \eMat \in [\eMat],
    \label{thm:candidate-metric-tensor-field}
\end{equation}
%where 
% $(\pwdMetricTensor_{\diamond\eMat} + \GyRaParam \MetricTensorCorr_{\diamond\eMat}):  \Real^{\dimInd\times \proteinLen} \to \Real^{\dimInd\times \proteinLen}$ defined as
% \begin{equation}
%     (\pwdMetricTensor_{\diamond\eMat} + \GyRaParam \MetricTensorCorr_{\diamond\eMat}) (\mTVector_{\diamond \eMat}) := \sum_{\sumIndB=1}^\proteinLen \biggl(\Bigl((\pwdMetricTensor_{\diamond\eMat})_{1\sumIndB} + \GyRaParam (\MetricTensorCorr_{\diamond\eMat})_{1\sumIndB} \Bigr) \mTVectorCompon_\sumIndB, \ldots, \Bigl((\pwdMetricTensor_{\diamond\eMat})_{\proteinLen\sumIndB} + \GyRaParam (\MetricTensorCorr_{\diamond\eMat})_{\proteinLen\sumIndB} \Bigr) \mTVectorCompon_\sumIndB \biggr), \quad \mTVector_{\diamond \eMat} := (\mTVectorCompon_1, \ldots, \mTVectorCompon_\proteinLen),
% \end{equation}
defines a metric tensor field on $\PCmanifold /\mathbb{E}(\dimInd)$ and the mapping $\separation^\GyRaParam:\PCmanifold /\mathbb{E}(\dimInd) \times \PCmanifold /\mathbb{E}(\dimInd) \to \Real$ in \cref{eq:candidate-metric-full} is a separation under $(\cdot, \cdot)^\GyRaParam$.
\end{theorem}

% \begin{remark}
%    As a final sanity check, note that the $\pwdMetricTensor_{\diamond\eMat}$ term in the metric tensor is similar to the Hessian of the Hookean potential \cite{tirion1996large} with the difference that the pair-wise distances in the denominators are squared now, which effectively gives us smooth decay of the interaction strength rather than having to add an abrupt cut-off. So overall, we really do inherit the best practices from normal mode analysis, but arguably do not suffer from basic locality due to the choice of metric and from linearity due to the availability of manifold mappings.
% \end{remark}

% [Note that the tensor A looks very much like the normal mode analysis framework. In the latter, the Hookean potential [Cite Tirion] does not decay, but are cut-off. Here it happens automatically in a smooth way. ]

% \cite{tekpinar2010predicting} % this one would be better in the

% \todo[inline]{Consider whether we want to merge this with the previous subsection now}

% \todo[inline]{Mention that we use addition as a retraction and notice that we really do need a proper exponential mapping if we want to walk far over the manifold as we know from NMA literature.}

\section{Riemannian geometry and analysis of protein dynamics data}
\label{sec:data-analysis-protein-geometry}
Before moving on to numerical experiments, we discuss why we should deviate from established standards for some more advanced data analysis tools. We do this in a case study considering low rank approximation. First, let us restate our complete approach using the results above. We represent protein conformations as elements of the quotient manifold $\PCmanifold /\mathbb{E}(\dimInd)$ (\cref{thm:quotient-manifold-Rdnd-mod-Ed}). On this manifold, we construct a physically motivated metric $\separation^{\GyRaParam}$ \cref{eq:candidate-metric-full} that we can efficiently evaluate, and show that this metric is a separation if we equip our manifold with the Riemannian structure in \cref{thm:protein-geometry-tensor-separation}. Using this Riemannian structure, we can use the separation $\separation^{\GyRaParam}$ to efficiently interpolate (\cref{thm:well-posed-ness-separation-geodesics}) and extrapolate (\cref{thm:well-posed-ness-separation-exp}) along non-linear proxy energy landscape-minimizing curves on the manifold of protein conformations.
% So far, our separation-based framework for Riemannian protein geometry (\cref{thm:protein-geometry-tensor-separation}) allows us to interpolate between points ($\separation^{\GyRaParam}$-geodesics), extrapolate into a given direction ($\separation^{\GyRaParam}$-exponential mapping) and retrieve such a direction from any point ($\separation^{\GyRaParam}$-logarithmic mapping) -- all over non-linear proxy energy landscape-minimizing curves. 

In the case of low-rank approximation, our goal is to find non-linear geodesic subspaces that capture protein dynamics data best. 
% However, consider what it means to capture data in the ``best'' way for this case. 
In works such as \cite{diepeveen2023curvature,fletcher2004principal}, the best rank-$r$ approximation is defined as having minimal error in the Riemannian distance. In our case the metric is modeled after the energy landscape for the various conformations of a given protein. Hence, in our Riemannian geometry there could be an energetically small conformational change that corresponds to a large change in terms of Euclidean distance. This is undesirable. So conceptually, we might want to move towards a low rank approximation that gives the lowest error in the Euclidean distance instead -- that is after re-centering and registering both conformations in a least-squares sense \cite{arun1987least} --, which is more in line with current practices of protein conformation comparison using the \emph{root-mean-square deviation} (RMSD).

% In our case where distance is modeled after the energy landscape, the issue arises that there could still be a large deformations in terms of Euclidean distance -- by design of our Riemannian geometry --, which is not what we are after

% consider separation based equivalents of several standard data analysis tools and discuss why data analysis of protein conformations is somewhat different to how we would normally analyse data] [based on the standard mappings we can build more advanced things. We will discuss low rank approximation at the barycentre, as we will see here very clearly how we need yet another generalization than the already available ones.]

% [Mention that in practice we care about different error metrics than are given by the Riemannian structure -- we only care about the mappings here]

Low RMSD can be achieved through a change of inner product. First, we need -- similarly to \cite{diepeveen2023curvature,fletcher2004principal} -- a point from which we compute the $\separation^\GyRaParam$-logarithmic mappings to all points in a data set $\{[\eMat_\sumIndC]\}_{\sumIndC=1}^\numData \subset \PCmanifold /\mathbb{E}(\dimInd)$ of $\numData\in \Natural$ points. For that, we will use -- in line with previous work -- the $\separation^\GyRaParam$-barycentre, which is defined as\footnote{Existence and local strong geodesic convexity can be shown in a similar fashion as in \cref{thm:well-posed-ness-separation-geodesics,thm:well-posed-ness-separation-exp}.}
\begin{equation}
    [\bar{\eMat}] \in \argmin_{[\eMat] \in \PCmanifold /\mathbb{E}(\dimInd)} B_{\separation^\GyRaParam} ([\eMat]),
    \label{eq:4ake-bary}
\end{equation}
where $B_{\separation^\GyRaParam} : \PCmanifold /\mathbb{E}(\dimInd) \to \Real$ is given by
\begin{equation}
    B_{\separation^\GyRaParam} ([\eMat]) := \frac{1}{2\numData} \sum_{\sumIndC = 1}^\numData \separation^\GyRaParam ([\eMat], [\eMat_{\sumIndC}])^2.
\end{equation}
Having the $\separation^\GyRaParam$-barycentre, we collect -- again in line with previous work -- the tangent vectors $\{\log^{\separation^\GyRaParam}_{[\bar{\eMat}]} [\eMat_\sumIndC]\}_{\sumIndC =1}^\numData \subset \tangent_{[\bar{\eMat}]} \PCmanifold /\mathbb{E}(\dimInd)$, but compute the Gram matrix $\Gramm \in \Real^{\numData \times \numData}$ with respect to the $\ell^2$-inner product on any horizontal space $\horizontal_{\bar{\eMat}} \PCmanifold$ for $\bar{\eMat}\in [\bar{\eMat}]$
\begin{equation}
    \Gramm_{\sumIndC,\sumIndD} := ((\log^{\separation^\GyRaParam}_{[\bar{\eMat}]} [\eMat_\sumIndC])_{\diamond \bar{\eMat}}, (\log^{\separation^\GyRaParam}_{[\bar{\eMat}]} [\eMat_\sumIndD])_{\diamond \bar{\eMat}})_2.
\end{equation}
The eigendecomposition of the Gram matrix $\Gramm = \mathbf{U} \Lambda \mathbf{U}^\top$, where $\mathbf{U} := (\mathbf{u}_1, \ldots, \mathbf{u}_\numData)\in \Real^{\numData \times \numData}$ orthonormal and $\Lambda := \diag (\lambda_1,\ldots, \lambda_\numData) \in \Real^{\numData \times \numData}$ with $\lambda_1 \geq \ldots \geq \lambda_\numData\geq 0$, then allows us to compute a rank-$r$ approximation $(\Xi_{[\bar{\eMat}]}^r)_\sumIndC$ of $\log^{\separation^\GyRaParam}_{[\bar{\eMat}]} [\eMat_\sumIndC]$ given by
\begin{equation}
    ((\Xi_{[\bar{\eMat}]}^r)_\sumIndC)_{\diamond \bar{\eMat}}:= \sum_{\sumIndC'=1}^r (\mathbf{u}_{\sumIndC'}, (\log^{\separation^\GyRaParam}_{[\bar{\eMat}]} [\eMat_\sumIndC])_{\diamond \bar{\eMat}})_2 \mathbf{u}_{\sumIndC'} \in \horizontal_{\bar{\eMat}} \PCmanifold.
    \label{eq:optimal-approximation-log-protein}
\end{equation}
The main rationale for this choice of inner product is that it achieves a low RMSD since
\begin{multline}
    \operatorname{RMSD}([\eMat_\sumIndC], \exp^{\separation^\GyRaParam}_{\bar{[\eMat]}}((\Xi_{[\bar{\eMat}]}^r)_\sumIndC)) := \frac{1}{\sqrt{\proteinLen}} \|[\eMat_\sumIndC]_{\diamond \eMat^0} - \exp^{\separation^\GyRaParam}_{\bar{[\eMat]}}((\Xi_{[\bar{\eMat}]}^r)_\sumIndC)_{\diamond \eMat^0}\|_2 \\
    = \frac{1}{\sqrt{\proteinLen}} \|\exp^{\separation^\GyRaParam}_{\bar{[\eMat]}}(\log^{\separation^\GyRaParam}_{[\bar{\eMat}]} [\eMat_\sumIndC])_{\diamond \eMat^0} - \exp^{\separation^\GyRaParam}_{\bar{[\eMat]}}((\Xi_{[\bar{\eMat}]}^r)_\sumIndC)_{\diamond \eMat^0}\|_2 
    \approx \frac{1}{\sqrt{\proteinLen}} \|(\log^{\separation^\GyRaParam}_{[\bar{\eMat}]} [\eMat_\sumIndC])_{\diamond \bar{\eMat}} -  ((\Xi_{[\bar{\eMat}]}^r)_\sumIndC)_{\diamond \bar{\eMat}}\|_2,
    \label{eq:approximation-l2-differences-low-rank}
\end{multline}
where $\eMat^0\in\PCmanifold$ is a reference conformation and $[\eMatB]_{\diamond \eMat^0}$ is the element $\eMatB\in [\eMatB]$ closest to $\eMat^0$ in a least-squares sense \cite{arun1987least}. The approximation holds up to first order since addition is a retraction (see \cref{sec:separation-protein-geometry}). Hence, since \cref{eq:optimal-approximation-log-protein} is now an optimal approximation of the right hand side, we expect that the left hand side error is small as well.

\begin{remark}
\label{rem:curvature-proteins-low-rank}
    The careful reader is cautious to conclude from the above discussion that the left hand side error in \cref{eq:approximation-l2-differences-low-rank} will always be small under the proposed low rank approximation scheme. That is, because of possible instabilities due to curvature effects. For the subclass of \emph{symmetric Riemannian manifolds} we know low rank approximation is seriously affected by negative curvature \cite[Theorem 3.4]{diepeveen2023curvature}. However, these results are at the moment only known with respect to the Riemannian distance -- even though we expect similar behaviour under the $\ell^2$-distance.
\end{remark}

\begin{remark}
\label{rem:curvature-proteins-geo}
    Similarly to the low rank case, when comparing geodesic interpolates to real data, we also want to use error metrics with respect to the $\ell^2$-distance. Here too, curvature can cause instabilities with respect to the end points in the case of positive curvature \cite[Lemma 1]{bergmann2019recent} in the Riemannian distance. Subsequently, these instabilities can also be noticeable 
    % we also expect similar behaviour 
    in the $\ell^2$-distance.
\end{remark}

In addition to \cref{rem:curvature-proteins-low-rank,rem:curvature-proteins-geo}, a complete characterization of all the errors made, i.e., curvature and all approximation errors due to inexactness of the separation-based mappings, and ways of correcting for them are still open, but beyond the scope of this work. So these discrepancies will need to be checked for numerically.

\section{Numerics}
\label{sec:numerics-protein-geometry}

In this section we test the suitability of proposed Riemannian protein geometry for basic data analysis tasks. Remember from \cref{sec:intro-protein-geometry} that ideal energy landscape-based data analysis tools should be able to interpolate and extrapolate over \emph{energy-minimising non-linear paths}, compute \emph{non-linear means on such energy-minimising paths} and compute \emph{low-rank approximations over curved subspaces spanned by energy-minimising paths}. So naturally, we test how well the separation-based Riemannian interpretation of these tools -- $\separation^\GyRaParam$-geodesic interpolation, the $\separation^\GyRaParam$-barycentre as a non-linear mean, and tangent space low-rank approximation from the $\separation^\GyRaParam$-barycentre using the $\separation^\GyRaParam$-logarithm -- capture the energy landscape. 

% \todo[inline]{Say that we expect to find out whether data is disordered and quantify this using low rank}

% \todo[inline]{Consider writing both in one paragraph so that we don't repeat ourselves. Maybe better to keep it separate}

\paragraph{Data.}
For the numerical evaluation, we include protein dynamics data sets
% The most elucidating way of testing this is through protein dynamics data sets 
with varying levels of (dis)order and varying deformation size. For that, we consider two molecular dynamics simulations: (i) the adenylate kinase protein under a (famously) ordered and medium-size closed-to-open transition \cite{Beckstein2018} and (ii) the SARS-CoV-2 helicase nsp 13 protein under a more disordered and large-deformation motion \cite{shaw2020molecular}. 

% If our proposed geometry were to capture the trajectory perfectly, $\separation^\GyRaParam$-geodesics between closed and open would overlap, the $\separation^\GyRaParam$-barycentre would coincide perfectly with the half-way point of the trajectory and the low-rank approximation would return a 1-dimensional subspace, because all data lie on 1 geodesic that goes through the $\separation^\GyRaParam$-barycentre. In the following we will see that \emph{our proposed geometry does not perfectly capture this behaviour, but is strikingly close to it}. 

% that is intrinsically 1-dimensional, i.e., a trajectory.

% \todo[inline]{To also get an idea how the geometry behaves for less structured motion, we will also consider ...}

\paragraph{Outline of experiments.}
Considering the trajectory data sets on a $\mathrm{C}_\alpha$-coarse-grained level, we will showcase in \cref{sec:numerics-protein-geometry-geodesics} to what extend the $\separation^\GyRaParam$-geodesics approximate the trajectories.
% that $\separation^\GyRaParam$-geodesics closely approximate the adenylate kinase trajectory, but unsurprisingly have a harder time approximating the disordered and large-deformation SARS-CoV-2 data set. 
Subsequently, in \cref{sec:numerics-protein-geometry-log-low-rank} we will compute a low rank approximation at the $\separation^\GyRaParam$-barycentre and infer dimensionality.
% see in the case of adenylate kinase, that it is very close to the trajectory mid-point and that we recover a 1-dimensional geodesic subspace that captures most of the data set, and in the case of SARS-CoV-2, that more dimensions -- but only 7 of them -- are needed to capture the full range of motions. 
In \cref{sec:intro-protein-geometry} we also argued that suitable data analysis tools need to be computationally feasible. Consequently, we will also show that our separation-based approximations are very fast due to linear convergence of Riemannian gradient descent predicted by theory in \cref{sec:separation-protein-geometry}. 

Besides these main experiments showcasing the practical use of Riemannian geometry for protein conformation analysis, we will provide several more technical sanity checks for the interested reader. In particular, we check the prediction of \cref{rem:nice-interpolation} that small distances -- i.e., adjacent $C_\alpha$ distances -- stay more or less invariant throughout all data analysis tasks (\cref{app:adjacent-distances}), and check that the curvature effects discussed in \cref{rem:curvature-proteins-low-rank,rem:curvature-proteins-geo} are negligible for both data sets (\cref{app:numerics-protein-geometry-curvature-effects}), indicating stability.

\begin{figure}[h!]
    \centering
    \begin{subfigure}{0.21\linewidth}
    \makebox[10pt]{\raisebox{40pt}{\rotatebox[origin=c]{90}{\small \shortstack{Molecular \\ dynamics} }}}
    \hfill
        \includegraphics[width=0.85714\linewidth]{"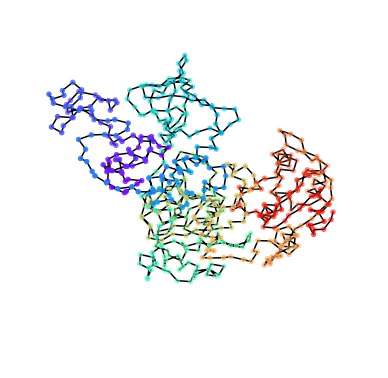"}
        \\
        \makebox[10pt]{\raisebox{40pt}{\rotatebox[origin=c]{90}{\small \shortstack{$\separation^\GyRaParam$-geodesic from \\ closed to open} }}}
        \hfill
        \includegraphics[width=0.85714\linewidth]{"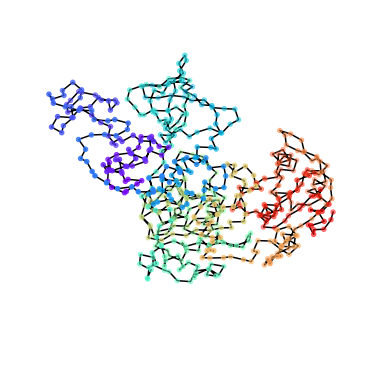"}
        \\
        \makebox[10pt]{\raisebox{40pt}{\rotatebox[origin=c]{90}{\small \shortstack{Rank 7 approx. at \\ the $\separation^\GyRaParam$-barycentre} }}}
        \hfill
        \includegraphics[width=0.85714\linewidth]{"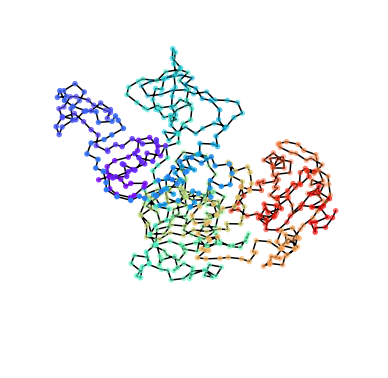"}
        \caption{$t=0$}
    \end{subfigure}
    \hfill
    \begin{subfigure}{0.18\linewidth}
        \includegraphics[width=\linewidth]{"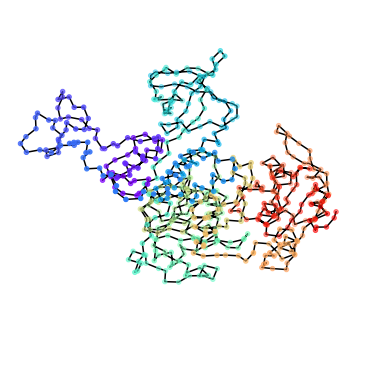"}
        \\
        \includegraphics[width=\linewidth]{"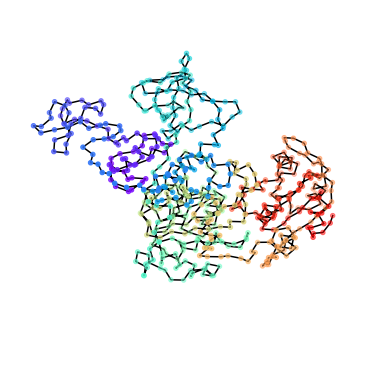"}
        \\
        \includegraphics[width=\linewidth]{"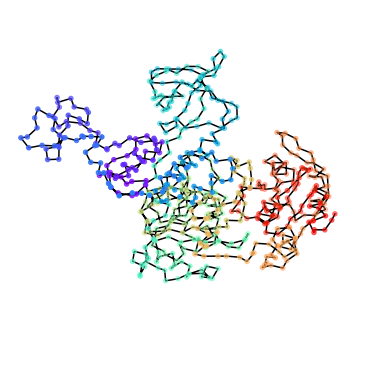"}
        \caption{$t=0.25$}
    \end{subfigure}
    \hfill
    \begin{subfigure}{0.18\linewidth}
        \includegraphics[width=\linewidth]{"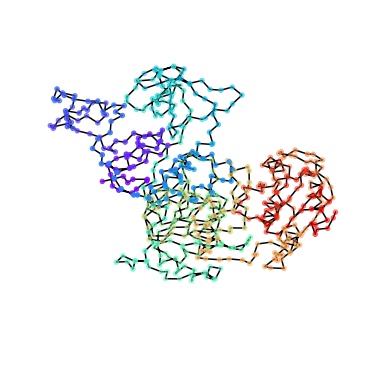"}
        \\
        \includegraphics[width=\linewidth]{"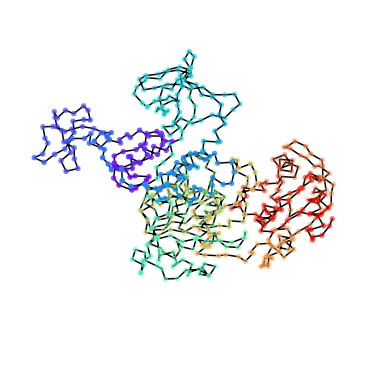"}
        \\
        \includegraphics[width=\linewidth]{"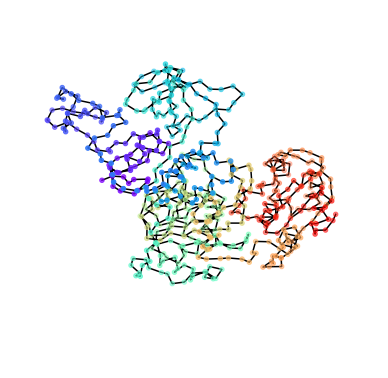"}
        \caption{$t=0.5$}
    \end{subfigure}
    \hfill
    \begin{subfigure}{0.18\linewidth}
        \includegraphics[width=\linewidth]{"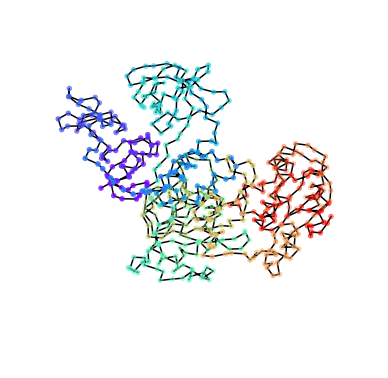"}
        \\
        \includegraphics[width=\linewidth]{"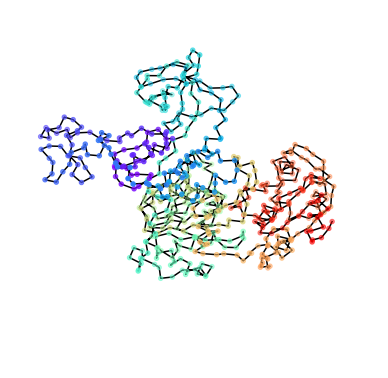"}
        \\
        \includegraphics[width=\linewidth]{"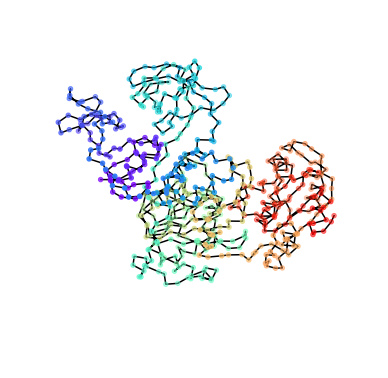"}
        \caption{$t=0.75$}
    \end{subfigure}
    \hfill
    \begin{subfigure}{0.18\linewidth}
        \includegraphics[width=\linewidth]{"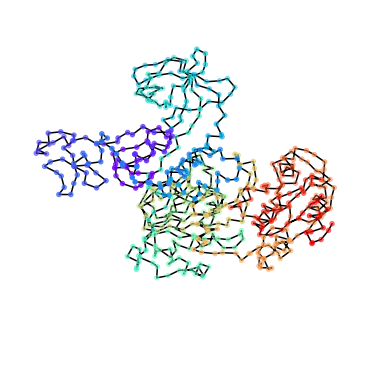"}
        \\
        \includegraphics[width=\linewidth]{"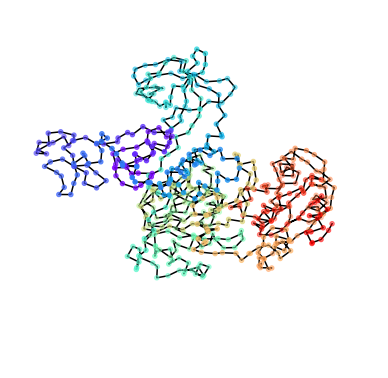"}
        \\
        \includegraphics[width=\linewidth]{"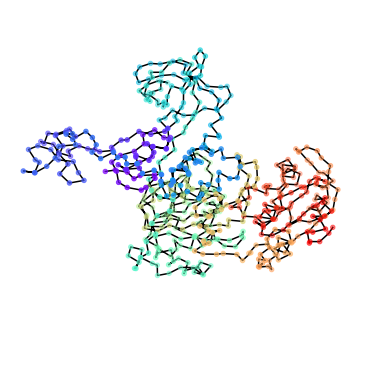"}
        \caption{$t=1$}
    \end{subfigure}
    \caption{Several snapshots of the $\mathrm{C}_\alpha$ atoms of the SARS-CoV-2 helicase nsp 13 protein under disordered motion. The top row is generated from a molecular dynamics simulation, the middle row is a $\separation^\GyRaParam$-geodesic between the end points and the bottom row a rank 7 approximation of the data at the $\separation^\GyRaParam$-barycentre. The begin-to-end $\separation^\GyRaParam$-geodesic does not capture the large-scale motion well due to significant errors at the top left dark blue part of the protein -- indicating that the motion is not 1-dimensional --, but rank 7 approximation captures the data almost perfectly.}
    \label{fig:md-geo-low-rank-covid}
\end{figure}

% \todo[inline]{Think about whether it is nice to call it time if we have a rank 7 approximation}

% \todo[inline]{Also do experiments for covid spike now that we know we have more space available.}
\paragraph{General experimental settings}
As mentioned above, we will consider a coarse-grained representation of the adenylate kinase protein and the SARS-CoV-2 helicase nsp 13 protein. In particular, we construct point clouds where the $\sumIndA$th point corresponds to the $\mathrm{C}_\alpha$ position of the $\sumIndA$th peptide. As adenylate kinase has 214 peptides and the data set has 102 frames, we obtain $\{[\eMat_\sumIndC^{AK}]\}_{\sumIndC=1}^{102} \subset \BlankPCmanifold(3, 214) /\mathbb{E}(3)$. As SARS-CoV-2 helicase nsp 13 has 590 peptides and the data set has 200 frames, we obtain $\{[\eMat_\sumIndC^{SC}]\}_{\sumIndC=1}^{200} \subset \BlankPCmanifold(3, 590) /\mathbb{E}(3)$. Several snapshots are shown in the top row of \cref{fig:md-geo-low-rank-4ake,fig:md-geo-low-rank-covid}, in which we have normalized time so that frame $1$ corresponds to $t=0$ and the last frame to $t=1$. Subsequently, we construct the Riemannian manifolds $(\BlankPCmanifold(3, 214) /\mathbb{E}(3), (\cdot, \cdot)^\GyRaParam)$ and $(\BlankPCmanifold(3, 590) /\mathbb{E}(3), (\cdot, \cdot)^\GyRaParam)$ with $\GyRaParam = 1$ and note that the dimensions of these spaces are 636 and 1764. 

% for each of the 102 snapshots in the closed-to-open transition data set \cite{Beckstein2018}

For computing $\separation^\GyRaParam$-geodesics and $\separation^\GyRaParam$-exponential mappings, we solve the corresponding optimisation problems \cref{eq:separation-geodesics} -- for their respective begin and end points $[\eMatB]$ and $[\eMatC]$ -- using Riemannian gradient descent with unit step size under addition as a retraction (see \cref{sec:separation-protein-geometry}) and initialized at the end point $[\eMatC]$. The parameter $M$ for the $\separation^\GyRaParam$-exponential map is chosen as the smallest integer larger than a quarter of the tangent vector norm. Riemannian gradients are linear combinations of the $\separation^\GyRaParam$-logarithm, which is provided in \cref{app:closed-form-separation-log}. As both the exponential mapping and geodesics solve the same problem, we stop the solver at iteration $\iterIndA$ if
\begin{equation}
    \frac{\|t\log^{\separation^\GyRaParam}_{[\eMat^{\iterIndA}]}([\eMatB]) + (1-t)\log^{\separation^\GyRaParam}_{[\eMat^{\iterIndA}]}([\eMatC])\|^\GyRaParam_{[\eMat^{\iterIndA}]}}{\separation^\GyRaParam([\eMatB],[\eMatC])} < 10^{-4},
\end{equation}
where $[\eMat^{\iterIndA}]$ is the $\iterIndA$th iterate.

The barycentre problem \cref{eq:4ake-bary} is also solved using Riemannian gradient descent with unit step size under addition as a retraction and initialized the trajectory midpoints $[\eMat_{51}^{AK}]$ and $[\eMat_{100}^{SC}]$ respectively. Here too, Riemannian gradients are linear combinations of the $\separation^\GyRaParam$-logarithm. The optimisation scheme is terminated at iteration $\iterIndA$ if 
\begin{equation}
    \frac{\|\Grad  B_{\separation^\GyRaParam} (\cdot)\|^\GyRaParam_{[\eMat^{\iterIndA}]}}{ B_{\separation^\GyRaParam}([\eMat^{0}])} < 10^{-4},
\end{equation}
where $[\eMat^{\iterIndA}]$ is the $\iterIndA$th iterate.

For visualisation of the resulting interpolates and exterpolates and of the data set itself, the point clouds are registered to the re-centered first frame, i.e., $\eMat^{AK,0} := \eMat_1^{AK} -  \eMat_1^{AK} \mathbf{1}_{214} \mathbf{1}_{214}^\top \in \BlankPCmanifold(3, 214)$ and $\eMat^{SC,0} := \eMat_1^{SC} -  \eMat_1^{SC} \mathbf{1}_{590}\mathbf{1}_{590}^\top \in \BlankPCmanifold(3, 590)$, in a least-squares sense \cite{arun1987least}.

Finally, all of the experiments are implemented using \texttt{PyTorch} in Python 3.8 and run on a 2 GHz Quad-Core Intel Core i5 with 16GB RAM. All reported time measurements have been made using the \texttt{\%timeit} magic command.

\begin{figure}[h!]
    \centering
    \begin{subfigure}{0.28\linewidth}
        \makebox[10pt]{\raisebox{40pt}{\rotatebox[origin=c]{90}{\small \shortstack{$\separation^\GyRaParam$-geodesic from \\ closed to open} }}}
        \hfill 
        \includegraphics[width=0.8214\linewidth]{"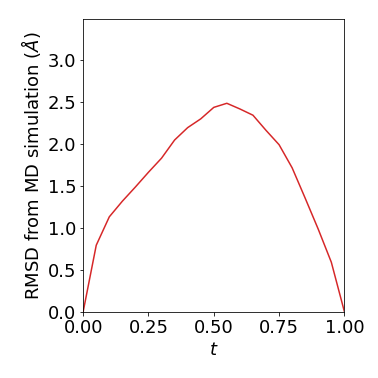"}
        \\
        \makebox[10pt]{\raisebox{40pt}{\rotatebox[origin=c]{90}{\small \shortstack{Rank 1 approx. at \\ the $\separation^\GyRaParam$-barycentre} }}}
        \hfill
        \includegraphics[width=0.8214\linewidth]{"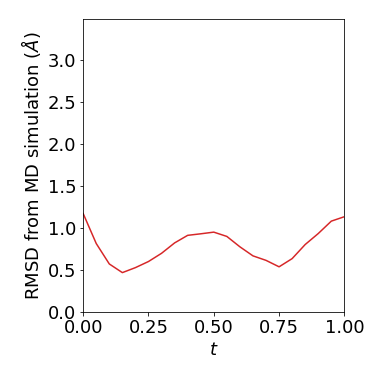"}
        \caption{Progression}
    \end{subfigure}
    \hfill
    \begin{subfigure}{0.23\linewidth}
        \includegraphics[width=\linewidth]{"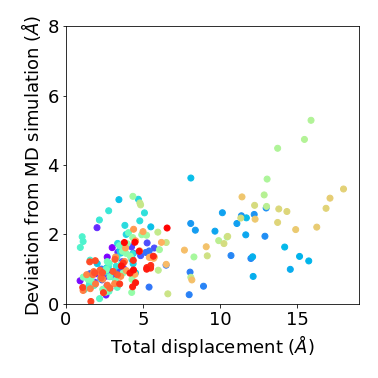"}
        \\
        \includegraphics[width=\linewidth]{"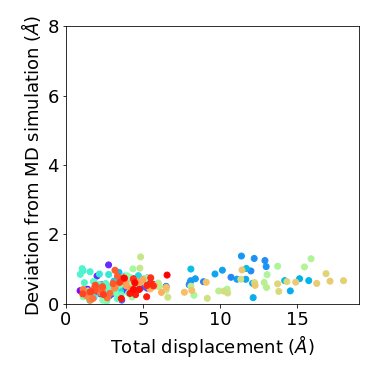"}
        \caption{$t=0.25$}
    \end{subfigure}
    \hfill
    \begin{subfigure}{0.23\linewidth}
        \includegraphics[width=\linewidth]{"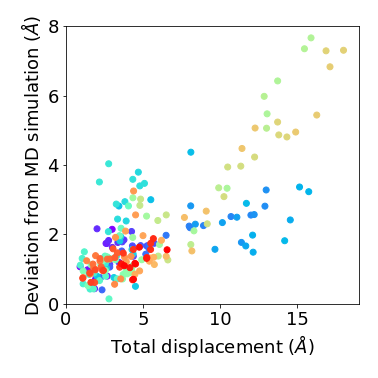"}
        \\
        \includegraphics[width=\linewidth]{"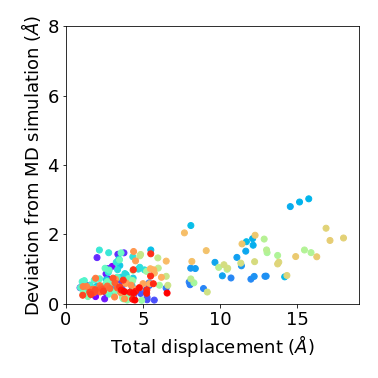"}
        \caption{$t=0.5$}
    \end{subfigure}
    \hfill
    \begin{subfigure}{0.23\linewidth}
        \includegraphics[width=\linewidth]{"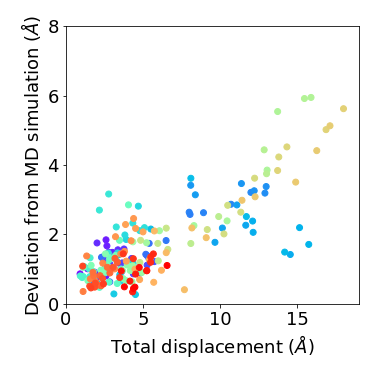"}
        \\
        \includegraphics[width=\linewidth]{"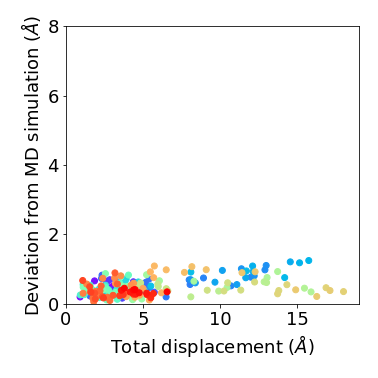"}
        \caption{$t=0.75$}
    \end{subfigure}
    \caption{Progessions of the RMSD from the adenylate kinase molecular dynamics simulation for the close to open $\separation^\GyRaParam$-geodesic and the rank 1 approximation, and several snapshots displaying the atom-specific displacement as a function of the total displacement. Both schemes show only minimal deviation from the ground truth as only the lower -- green and yellow -- part of the protein are not predicted within a deviation smaller than the 3.85 \AA\, distance between adjacent $\mathrm{C}_\alpha$ atoms.}
    \label{fig:error-geo-low-rank}
\end{figure}

\begin{figure}[h!]
    \centering
    \begin{subfigure}{0.28\linewidth}
        \makebox[10pt]{\raisebox{40pt}{\rotatebox[origin=c]{90}{\small \shortstack{$\separation^\GyRaParam$-geodesic from \\ closed to open} }}}
        \hfill 
        \includegraphics[width=0.8214\linewidth]{"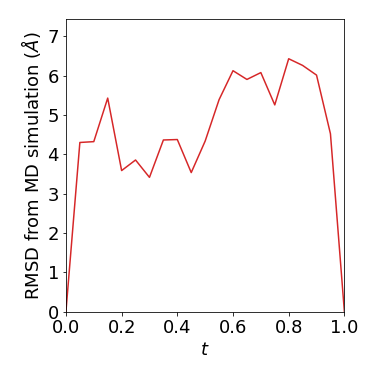"}
        \\
        \makebox[10pt]{\raisebox{40pt}{\rotatebox[origin=c]{90}{\small \shortstack{Rank 7 approx. at \\ the $\separation^\GyRaParam$-barycentre} }}}
        \hfill
        \includegraphics[width=0.8214\linewidth]{"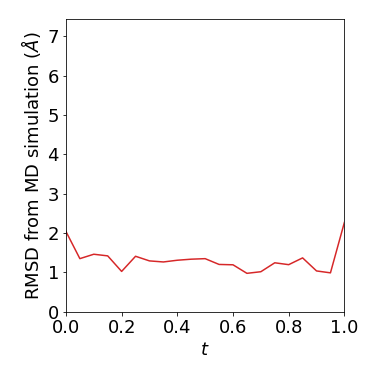"}
        \caption{Progression}
    \end{subfigure}
    \hfill
    \begin{subfigure}{0.23\linewidth}
        \includegraphics[width=\linewidth]{"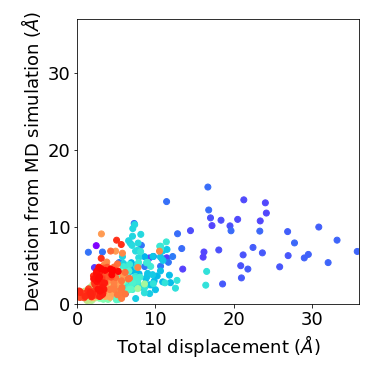"}
        \\
        \includegraphics[width=\linewidth]{"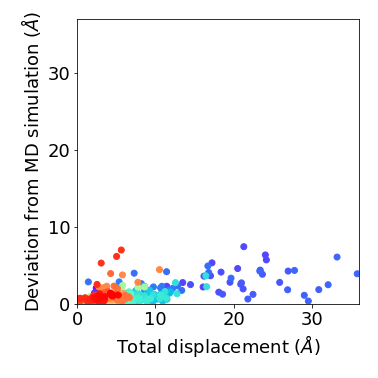"}
        \caption{$t=0.25$}
    \end{subfigure}
    \hfill
    \begin{subfigure}{0.23\linewidth}
        \includegraphics[width=\linewidth]{"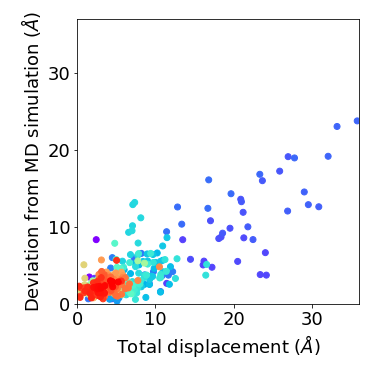"}
        \\
        \includegraphics[width=\linewidth]{"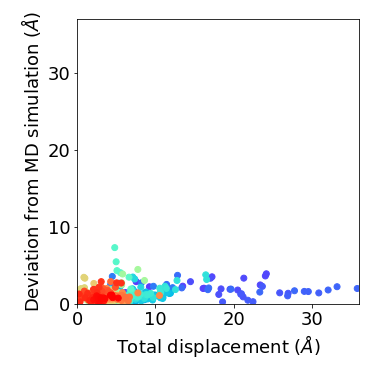"}
        \caption{$t=0.5$}
    \end{subfigure}
    \hfill
    \begin{subfigure}{0.23\linewidth}
        \includegraphics[width=\linewidth]{"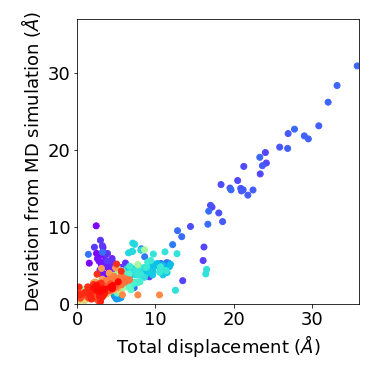"}
        \\
        \includegraphics[width=\linewidth]{"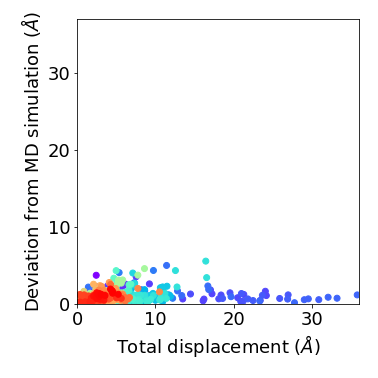"}
        \caption{$t=0.75$}
    \end{subfigure}
    \caption{Progessions of the RMSD from the SARS-CoV-2 helicase nsp 13 molecular dynamics simulation for the begin to end frame $\separation^\GyRaParam$-geodesic and the rank 7 approximation, and several snapshots displaying the atom-specific displacement as a function of the total displacement. One $\separation^\GyRaParam$-geodesic is clearly unable to capture the full range of motions, but a rank 7 approximation shows only minimal deviation from the ground truth as here most deviation is smaller than the 3.85 \AA\, distance between adjacent $\mathrm{C}_\alpha$ atoms.}
    \label{fig:error-geo-low-rank-covid}
\end{figure}

\subsection{$\separation^\GyRaParam$-geodesic interpolation}
\label{sec:numerics-protein-geometry-geodesics}

\paragraph{Adenylate kinase.}
Several snapshots of $\separation^\GyRaParam$-geodesic interpolation between the closed and open state -- $[\eMat_1^{AK}]$ and $[\eMat_{102}^{AK}]$ respectively -- are shown in the middle row of \cref{fig:md-geo-low-rank-4ake}. Computing the interpolates is very cheap because of the linear convergence of Riemannian gradient descent due to local strong geodesic convexity (\cref{thm:well-posed-ness-separation-geodesics}). For example, computing the mid-point at $t=0.5$ converges after 4 iterations and takes 245 ms $\pm$ 11.4 ms, which is much faster than it would have taken to compute the trajectory using molecular dynamics simulation. 
On that note, 
% even though the distance between adjacent $C_\alpha$ atoms are preserved (\cref{app:adjacent-distances-geo-AK}) throughout the $\separation^\GyRaParam$-geodesic as predicted in \cref{rem:nice-interpolation}, 
there are visible differences -- although strikingly minor -- between the top and middle row of \cref{fig:md-geo-low-rank-4ake}. These differences are more conveniently quantified by considering the deviation between corresponding atoms and their \emph{root-mean-square deviation} (RMSD), which are shown in the top row of \cref{fig:error-geo-low-rank}. From the left-most plot in the figure it is clear that the error increases the further away we are from either end point, but is much smaller than the distance between $\mathrm{C}_\alpha$ atoms -- which is approximately 3.85 \AA \, in this data set. Unsurprisingly, from the three scatter plots on the right side we see that the atoms undergoing larger deformation are hardest to predict, and that largest errors come from lower part of the protein -- indicated by the yellow and green atoms. 
The latter observation suggests that there is a discrepancy between the actual energy landscape and the Riemannian structure. This is to be expected as potentials like Lennard Jones are stronger at short-range than for example $\separation^\GyRaParam([\eMat^{AK,0}], \cdot)$, but weaker at long-range. 

Overall, this experiment suggests that the proposed Riemannian geometry is well-suited for interpolating in terms of efficiency and ability to approximate energy-minimising paths, when the protein is undergoing a reasonably ordered transition with medium-size deformations.

\paragraph{SARS-CoV-2 helicase nsp 13.}
Once again, computing the $\separation^\GyRaParam$-geodesic is cheap, e.g., the mid-point converges after 14 iterations and takes 10.1 s ± 43.3 ms. Contrary to the adenylate kinase results, the middle row of \cref{fig:md-geo-low-rank-covid} suggests that we are unable to approximate this protein's trajectory with just one $\separation^\GyRaParam$-geodesic between the first and last frame, i.e., $[\eMat_1^{SC}]$ and $[\eMat_{200}^{SC}]$. Realistically, the higher level of disorder is the most likely cause. However, on top of that it is very likely that the just mentioned modelling discrepancy between the underlying energy landscape from the simulation and the imposed Riemannian structure -- becoming inevitably more noticeable the larger the deformation is -- is an additional factor. This hypothesis is backed up by the experiment. That is, upon closer inspection and considering the top row of \cref{fig:error-geo-low-rank-covid}, the major source of the error is the dark blue lobe on the left side of the protein -- once again being the part of the protein with the largest deformation. The remaining $\mathrm{C}_\alpha$ atoms have a deviation that is around the acceptable inter $\mathrm{C}_\alpha$-distance of 3.85 \AA. 

Overall, this experiment showcases some limitations of the proposed Riemannian geometry for more disordered proteins and proteins undergoing very large-scale deformations. That is, our method is capable of quickly simulating realistic states, but we just cannot only use geodesics in such cases.

\subsection{Low rank approximation at the $\separation^\GyRaParam$-barycentre}
\label{sec:numerics-protein-geometry-log-low-rank}

\paragraph{Adenylate kinase.}
Next, we will attempt to find a better geodesic that captures the large scale motion through rank 1 approximation of the data set at the $\separation^\GyRaParam$-barycentre tangent space. Solving the barycentre problem \cref{eq:4ake-bary} is once again cheap due to linear convergence of Riemannian gradient descent under local strong geodesic convexity. In particular, the solver converges after 3 iterations and takes 425 ms $\pm$ 13.8 ms. 
% Also, adjacent distances are once again preserved (\cref{app:adjacent-distances-bary-AK}). 
Then, for the actual low rank approximation, we follow the procedure in \cref{sec:data-analysis-protein-geometry}. We only consider a rank 1 approximation\footnote{The tangent vectors corresponding to the largest three eigenvalues are shown in \cref{app:tangent-svd-vectors}.} and use the exponential mapping to retrieve the corresponding protein conformations, which can be done cheaply (\cref{thm:well-posed-ness-separation-exp}), e.g., for the approximation of $[\eMat_{102}^{AK}]$ this takes 657 ms $\pm$ 19.6 ms. The approximations are shown in the bottom row of \cref{fig:md-geo-low-rank-4ake}, in which we see that this approximation is almost a perfect reconstruction of the original data set. 
% As expected, adjacent distances are preserved over the geodesic (\cref{app:adjacent-distances-bary-low-rank-exp-AK}). 
This is backed up by the RMSD and the per-residue-deviation shown in the bottom row of \cref{fig:error-geo-low-rank}. Here we see that the error is about 1 \AA. We remind the reader again that the distance between adjacent $\mathrm{C}_\alpha$ atoms is approximately 3.85 \AA. From the plot we also observe that the barycentre does not go through the data -- otherwise there would at least be one point with zero error --, but is close to it. 

Overall, this experiment confirms that the proposed Riemannian geometry enables us to find non-linear means close to energy-minimising paths for medium-sized and relatively ordered transitions and enables us to compute low-rank approximations over curved subspaces spanned by energy-minimising paths that actually capture the intrinsic data dimensionality in a numerically highly efficient manner.
%-- all within mere seconds.

\paragraph{SARS-CoV-2 helicase nsp 13.} 
For the disordered case we will try to find a geodesic subspace that will capture the full motion range. In the above case a rank 1 approximation covers about 85 \% of the variance. Using this threshold for SARS-CoV-2 tells us that we need at least a rank 7 approximation. Similarly to above, we compute the $\separation^\GyRaParam$-barycentre, which converges quickly after 4 iterations and takes 7.08 s $\pm$ 289 ms. 
% Adjacent $C_\alpha$ distances once again deteriorate moderately, but remain in the same order of magnitude (\cref{app:adjacent-distances-bary-SC}). 
After computing a rank 7 approximation\footnote{The tangent vectors corresponding to the largest nine eigenvalues are shown in \cref{app:tangent-svd-vectors}.} in the tangent space, the corresponding protein conformation approximations are computed using the $\separation^\GyRaParam$-exponential mapping, which can be done reasonably cheaply, e.g., in 25.2 s $\pm$ 164 ms for the approximation of $[\eMat_{200}^{SC}]$. 
% Adjacent $C_\alpha$ distances are slightly deteriorated once again \cref{app:adjacent-distances-bary-low-rank-exp-SC}, 
% Now, we do get a better
We obtain a significantly better approximation than with a single $\separation^\GyRaParam$-geodesic:
% . That is, 
the bottom row of \cref{fig:md-geo-low-rank-covid} gives us a near-perfect reconstruction and the averaged deviations and atom-specific individual deviations in the bottom row of \cref{fig:error-geo-low-rank-covid} confirm this.

Overall, this experiment showcases that the proposed Riemannian geometry still enables us to find low rank behaviour in seemingly disordered large-size deformations within a reasonable time frame.

\section{Conclusions}
\label{sec:conclusions-protein-geometry}

With this investigation, we hope to contribute to the development of an energy landscape-based Riemannian geometry for efficient analysis of protein dynamics. The primary challenges were that
% We set out to use an energy landscape-based Riemannian geometry for efficient analysis of protein dynamics data and ran into several challenges. That is, 
there was -- without extra structure -- a trade-off between accurate modelling and computational feasibility, but even with this sorted there was no readily available manifold of protein conformations, and existing Riemannian structures did not cater to protein energy landscapes.

% \todo[inline]{Say something more about what we exactly found in the experiments}

\paragraph{Computationally feasible Riemannian geometry.}
To overcome the computational feasibility issue, we have proposed to consider a relaxation of the Riemannian distance -- called a separation -- from which approximations of all essential manifold mappings for data analysis can be obtained in closed form (\cref{cor:log-approx-separation}) or constructed in a provably efficient way (\cref{thm:well-posed-ness-separation-geodesics,thm:well-posed-ness-separation-exp}).

\paragraph{Reverse-engineering Riemannian geometry for point cloud conformations.}
Subsequently, for an energy landscape-based Riemannian structure we have proposed a smooth manifold of point cloud conformations modulo rigid body motion to model (coarse-grained) protein data (\cref{thm:quotient-manifold-Rdnd-mod-Ed}), and have proposed guidelines for constructing Riemannian metric tensor fields with an accompanying separation (\cref{thm:class-of-separations-on-pointcloud-manifold}).

\paragraph{Riemannian geometry for efficient analysis of protein dynamics data.}
Within this framework we have used several best practices from normal mode analysis and proposed an energy landscape-based Riemannian protein geometry (\cref{thm:protein-geometry-tensor-separation}) that arguably resolves basic locality and linearity problems in the normal modes framework. In numerical experiments with molecular dynamics simulations undergoing transitions, we observed that the proposed geometry is exceedingly useful for data analysis, but could be improved on -- within the proposed framework -- to cater for even larger-scale deformations. In particular, approximate geodesics with respect to the separation approximate the molecular dynamics simulation data very well for low-disorder transitions with medium-sized deformations, separation-based barycentres give a meaningful representation of the data for both ordered and disordered transitions under any deformation size, and low rank approximation can recover the underlying dimensionality of such data sets. In particular, our approach broke down 636-dimensional conformations to just a non-linear 1-dimensional space and broke down 1764-dimensional conformations to a non-linear 7-dimensional space without a significant loss in approximation accuracy.
In addition,
%on top of that 
the approximations are very fast and can be computed in seconds on a laptop.

\section*{Acknowledgments}
CBS acknowledges support from the Philip Leverhulme Prize, the Royal Society Wolfson Fellowship, the EPSRC advanced career fellowship EP/V029428/1, EPSRC grants EP/S026045/1 and EP/T003553/1, EP/N014588/1, EP/T017961/1, the Wellcome Innovator Awards 215733/Z/19/Z and 221633/Z/20/Z, the European Union Horizon 2020 research and innovation programme under the Marie Skodowska-Curie grant agreement No. 777826 NoMADS, the Cantab Capital Institute for the Mathematics of Information and the Alan Turing Institute.

\bibliographystyle{plain}
\bibliography{references} 

\clearpage
\appendix
\section{Proofs for the results from \cref{sec:separation-protein-geometry}}
\label{app:proofs-separation-protein-geometry}
% \todo[inline]{We need more of a story here as we really skip all the details}

\subsection{Proof of \cref{thm:separation-and-distance}}
\paragraph{Auxiliary lemma}

\begin{lemma}
\label{lem:cov-log-along-geodesic}
    Let $(\manifold, (\cdot, \cdot))$ be a Riemannian manifold, let $\mPoint, \mPointB\in \manifold$ be points, and assume that the geodesic $\geodesic_{\mPoint, \mPointB}:[0,1] \to \manifold$ exists. We have
    \begin{equation}
        \nabla_{\dot{\geodesic}_{\mPoint, \mPointB}(t)}\log_{(\cdot)} (\mPoint) = - \dot{\geodesic}_{\mPoint, \mPointB}(t).
        \label{eq:lem-cov-log-along-geodesic}
    \end{equation}
\end{lemma}

\begin{proof}
    We will show the validity of \cref{eq:lem-cov-log-along-geodesic} using Jacobi fields.

    We fix $t\in (0,1]$, define $\mPointC:= \geodesic_{\mPoint, \mPointB}(t)$ and construct a new geodesic $\geodesicB_{\mPointC, \mPoint}:[0,1]\to\manifold$ from $\mPointC$ to $ \mPoint$. Note that
    \begin{equation}
        \dot{\geodesic}_{\mPoint, \mPointB}(t) = - \frac{\distance_{\manifold}(\mPoint, \mPointB)}{\distance_{\manifold}(\mPoint, \mPointC)} \dot{\geodesicB}_{\mPointC, \mPoint}(0)
    \end{equation}
    With this definition, \cref{eq:lem-cov-log-along-geodesic} is equivalent
    %In other words, for showing \cref{eq:lem-cov-log-along-geodesic} it is sufficient 
    to show that
    \begin{equation}
        \nabla_{\dot{\geodesicB}_{\mPointC, \mPoint}(0)}\log_{(\cdot)} (\mPoint) = - \dot{\geodesicB}_{\mPointC, \mPoint}(0).
        \label{eq:lem-cov-log-along-geodesic-reparametrized}
    \end{equation}

    For showing \cref{eq:lem-cov-log-along-geodesic-reparametrized} we will look at a particular geodesic variation along $\geodesicB_{\mPointC, \mPoint}$. That is, for some small $\epsilon>0$ we define the function $\Gamma:[0,1]\times (-\epsilon, \epsilon) \to \manifold$ given by 
    \begin{equation}
        \Gamma(\tau, \sigma):= \exp_{\exp_{\mPointC}(\sigma \dot{\geodesicB}_{\mPointC, \mPoint}(0))}(\tau \log_{\exp_{\mPointC}(\sigma \dot{\geodesicB}_{\mPointC, \mPoint}(0))} (\mPoint)).
    \end{equation}
    Then, $\Gamma(\tau, 0) = \geodesicB_{\mPointC, \mPoint}(\tau)$ and $\frac{\partial}{\partial \tau} \Gamma(\tau, \sigma) \mid_{\tau = 0} = \log_{\exp_{\mPointC}(\sigma \dot{\geodesicB}_{\mPointC, \mPoint}(0))} (\mPoint)$. Also, the vector field $J(\tau):= \frac{\partial}{\partial \sigma}\Gamma(\tau, \sigma)\mid_{\sigma = 0}$ solves the Jacobi field equation with boundary conditions $J(0) = \dot{\geodesicB}_{\mPointC, \mPoint}(0)$ and $J(1) = 0$.

    Now, note that
    \begin{equation}
        \nabla_{\dot{\geodesicB}_{\mPointC, \mPoint}(0)}\log_{(\cdot)} (\mPoint) = \frac{D}{\mathrm{d} \sigma} \log_{\exp_{\mPointC}(\sigma \dot{\geodesicB}_{\mPointC, \mPoint}(0))} (\mPoint) \mid_{\sigma = 0} = \frac{D}{\mathrm{d} \sigma} \frac{\partial}{\partial \tau} \Gamma(\tau, \sigma) \mid_{\tau = 0, \sigma = 0} \overset{\text{symmetry}}{=} \frac{D}{\mathrm{d} \tau} \frac{\partial}{\partial \sigma} \Gamma(\tau, \sigma) \mid_{\sigma = 0, \tau = 0} = \frac{D}{\mathrm{d} \tau} J(0)
        \label{eq:lem-jacobi-diff-log-relation}
    \end{equation}
    So it remains to solve the Jacobi field equation with boundary conditions $J(0) = \dot{\geodesicB}_{\mPointC, \mPoint}(0)$ and $J(1) = 0$ and evaluate $\frac{D}{\mathrm{d} \tau} J(0)$.

    It is easily verified that $J(\tau) = (1 - \tau)  \dot{\geodesicB}_{\mPointC, \mPoint} (\tau)$ solves the equation. Indeed, the boundary conditions are satisfied and both terms in the Jacobi field equation vanish. That is, 
    \begin{equation}
        \frac{D^2}{\mathrm{d} \tau^2} J(\tau)  = \frac{D}{\mathrm{d} \tau} \bigl( - \dot{\geodesicB}_{\mPointC, \mPoint} (\tau) + (1 - \tau) \frac{D}{\mathrm{d} \tau} \dot{\geodesicB}_{\mPointC, \mPoint} (\tau) \bigr) \overset{\frac{D}{\mathrm{d} \tau} \dot{\geodesicB}_{\mPointC, \mPoint} (\tau) = 0}{=} - \frac{D}{\mathrm{d} \tau} \dot{\geodesicB}_{\mPointC, \mPoint} (\tau) \overset{\frac{D}{\mathrm{d} \tau} \dot{\geodesicB}_{\mPointC, \mPoint} (\tau) = 0}{=} 0,
    \end{equation}
    and 
    \begin{equation}
        \curvature(J(\tau), \dot{\geodesicB}_{\mPointC, \mPoint}(\tau))\dot{\geodesicB}_{\mPointC, \mPoint}(\tau) = (1 - \tau) \curvature(\dot{\geodesicB}_{\mPointC, \mPoint} (\tau), \dot{\geodesicB}_{\mPointC, \mPoint}(\tau))\dot{\geodesicB}_{\mPointC, \mPoint}(\tau) = 0.
    \end{equation}

    Finally, evaluating
    \begin{equation}
        \frac{D}{\mathrm{d} \tau} J(0) = - \dot{\geodesicB}_{\mPointC, \mPoint} (0)
    \end{equation}
    gives \cref{eq:lem-cov-log-along-geodesic-reparametrized} when combined with \cref{eq:lem-jacobi-diff-log-relation}, from which we conclude that \cref{eq:lem-cov-log-along-geodesic} holds.
\end{proof}

\paragraph{Proof of the theorem}

\begin{proof}[Proof of \cref{thm:separation-and-distance}]
    We will show (i) \cref{eq:thm-relation-sepa-dist-i} and (ii) \cref{eq:thm-relation-sepa-dist-ii} using a Taylor expansion along the geodesic $\geodesic_{\mPoint,\mPointB}: [0,1]\to \manifold$
    \begin{multline}
        \separation(\mPoint, \geodesic_{\mPoint,\mPointB}(t))^2 = \separation(\mPoint, \geodesic_{\mPoint,\mPointB}(0))^2 + t\frac{\mathrm{d}}{\mathrm{d}\tau} \separation(\mPoint, \geodesic_{\mPoint,\mPointB}(\tau))^2 \mid_{\tau=0} \\
        + \frac{t^2}{2} \frac{\mathrm{d}^2}{\mathrm{d}\tau^2} \separation(\mPoint, \geodesic_{\mPoint,\mPointB}(\tau))^2 \mid_{\tau=0} + \frac{1}{2}\int_0^t \frac{\mathrm{d}^3}{\mathrm{d}\tau^3} \separation(\mPoint, \geodesic_{\mPoint,\mPointB}(\tau))^2 (t - \tau)^2 \mathrm{d}\tau,
        \label{eq:thm-sepa-dist-taylor}
    \end{multline}
    which will be evaluated at $t=1$. Note that these derivatives exist by the second property in \cref{def:separation}.

    Before we prove the statements, we will evaluate the different terms. Because the separation mapping is a metric on $\manifold$, the constant
    %zeroth order 
    term vanishes:
    \begin{equation}
        \separation(\mPoint, \geodesic_{\mPoint,\mPointB}(0))^2 = \separation(\mPoint, \mPoint)^2 = 0.
        \label{eq:thm-relation-sep-dist-zeroth-order}
    \end{equation}
    The first-order term can be rewritten as
    \begin{equation}
        \frac{\mathrm{d}}{\mathrm{d}\tau} \separation(\mPoint, \geodesic_{\mPoint,\mPointB}(\tau))^2 = (\Grad \separation(\mPoint, \cdot)^2, \dot{\geodesic}_{\mPoint,\mPointB})_{\geodesic_{\mPoint,\mPointB}(\tau)},
        \label{eq:thm-relation-sep-dist-first-order}
    \end{equation}
    and the second-order term as
    \begin{multline}
        \frac{\mathrm{d}^2}{\mathrm{d}\tau^2} \separation(\mPoint, \geodesic_{\mPoint,\mPointB}(\tau))^2 \overset{\text{\cref{eq:thm-relation-sep-dist-first-order}}}{=} \frac{\mathrm{d}}{\mathrm{d}\tau} (\Grad \separation(\mPoint, \cdot)^2, \dot{\geodesic}_{\mPoint,\mPointB} )_{\geodesic_{\mPoint,\mPointB}(\tau)} 
        = \dot{\geodesic}_{\mPoint,\mPointB}(\tau) (\Grad \separation(\mPoint, \cdot)^2, \dot{\geodesic}_{\mPoint,\mPointB})_{(\cdot)} \\
        \overset{\text{metric compatibility}}{=} (\nabla_{\dot{\geodesic}_{\mPoint,\mPointB}}\Grad \separation(\mPoint, \cdot)^2 , \dot{\geodesic}_{\mPoint,\mPointB})_{\geodesic_{\mPoint,\mPointB}(\tau)} + (\Grad \separation(\mPoint, \cdot)^2 , \nabla_{\dot{\geodesic}_{\mPoint,\mPointB}} \dot{\geodesic}_{\mPoint,\mPointB})_{\geodesic_{\mPoint,\mPointB}(\tau)}\\
        \overset{\nabla_{\dot{\geodesic}_{\mPoint,\mPointB}} \dot{\geodesic}_{\mPoint,\mPointB}=0}{=} (\nabla_{\dot{\geodesic}_{\mPoint,\mPointB}}\Grad \separation(\mPoint, \cdot)^2 , \dot{\geodesic}_{\mPoint,\mPointB})_{\geodesic_{\mPoint,\mPointB}(\tau)}.
        \label{eq:thm-relation-sep-dist-second-order}
    \end{multline}
    Finally, the third-order term can also be rewritten in a more convenient form
    \begin{multline}
        \frac{\mathrm{d}^3}{\mathrm{d}\tau^3} \separation(\mPoint, \geodesic_{\mPoint,\mPointB}(\tau))^2 = \frac{\mathrm{d}}{\mathrm{d}\tau} (\nabla_{\dot{\geodesic}_{\mPoint,\mPointB}}\Grad \separation(\mPoint, \cdot)^2 , \dot{\geodesic}_{\mPoint,\mPointB})_{\geodesic_{\mPoint,\mPointB}(\tau)}
        = \dot{\geodesic}_{\mPoint,\mPointB}(\tau) (\nabla_{\dot{\geodesic}_{\mPoint,\mPointB}}\Grad \separation(\mPoint, \cdot)^2 , \dot{\geodesic}_{\mPoint,\mPointB})_{(\cdot)}\\
        \overset{\text{metric compatibility}}{=}  (\nabla_{\dot{\geodesic}_{\mPoint,\mPointB}} \nabla_{\dot{\geodesic}_{\mPoint,\mPointB}}\Grad \separation(\mPoint, \cdot)^2 , \dot{\geodesic}_{\mPoint,\mPointB})_{\geodesic_{\mPoint,\mPointB}(\tau)} + (\nabla_{\dot{\geodesic}_{\mPoint,\mPointB}}\Grad \separation(\mPoint, \cdot)^2 , \nabla_{\dot{\geodesic}_{\mPoint,\mPointB}} \dot{\geodesic}_{\mPoint,\mPointB})_{\geodesic_{\mPoint,\mPointB}(\tau)}\\
        \overset{\nabla_{\dot{\geodesic}_{\mPoint,\mPointB}} \dot{\geodesic}_{\mPoint,\mPointB}=0}{=} (\nabla_{\dot{\geodesic}_{\mPoint,\mPointB}} \nabla_{\dot{\geodesic}_{\mPoint,\mPointB}}\Grad \separation(\mPoint, \cdot)^2 , \dot{\geodesic}_{\mPoint,\mPointB})_{\geodesic_{\mPoint,\mPointB}(\tau)}.
        \label{eq:thm-relation-sep-dist-third-order}
    \end{multline}
    We can now move on to proving statement (i) and (ii).
    
    (i) The first order term \cref{eq:thm-relation-sep-dist-first-order} vanishes at $\tau=0$, since $\tau \mapsto \separation(\mPoint,  \geodesic_{\mPoint,\mPointB}(\tau))^2$ attains a global minimum by the first property in \cref{def:separation}, from which we conclude that $\Grad \separation(\mPoint, \cdot)^2\mid_{\mPoint} = 0$ by first-order-optimality conditions. By the third property in \cref{def:separation} the second order term \cref{eq:thm-relation-sep-dist-second-order} reduces to
    \begin{multline}
        \frac{\mathrm{d}^2}{\mathrm{d}\tau^2} \separation(\mPoint, \geodesic_{\mPoint,\mPointB}(\tau))^2 \mid_{\tau=0} = (\nabla_{\dot{\geodesic}_{\mPoint,\mPointB}}\Grad \separation(\mPoint, \cdot)^2 , \dot{\geodesic}_{\mPoint,\mPointB})_{\geodesic_{\mPoint,\mPointB}(0)} \\
        \overset{\dot{\geodesic}_{\mPoint,\mPointB}(0) = \log_\mPoint (\mPointB)}{=} (\nabla_{\log_\mPoint (\mPointB)}\Grad \separation(\mPoint, \cdot)^2 , \log_\mPoint (\mPointB))_{\mPoint} \overset{\text{\cref{def:separation}}}{=} 2 (\log_\mPoint (\mPointB), \log_\mPoint (\mPointB))_{\mPoint} = 2 \distance_\manifold(\mPoint, \mPointB)^2.
    \end{multline}
    It is easy to verify that the third order term scales as $\|\log_{\mPoint}(\mPointB)\|_\mPoint^3 = \distance_\manifold (\mPoint, \mPointB)^3$. Substituting the above results in the Taylor expansion \cref{eq:thm-sepa-dist-taylor} at $t=1$ gives us \cref{eq:thm-relation-sepa-dist-i} as claimed.

    (ii) \enquote{$\Rightarrow$} We start with assuming that $\separation(\mPoint, \mPointB)^2 = \distance_{\manifold}(\mPoint, \mPointB)^2$. Then, all conditions are automatically satisfied.
    % as $\distance_{\manifold}$
    % \begin{equation}
    %     \nabla_{\mTVector_{\mPoint}} \Grad \distance_{\manifold}(\mPoint, \cdot)^2 = \nabla_{\mTVector_{\mPoint}} \Grad \distance_{\manifold}(\mPoint, \cdot)^2 = - 2 \nabla_{\mTVector_{\mPoint}} \log_{(\cdot)}\mPoint = 2 \mTVector_{\mPoint}.
    % \end{equation}

    \enquote{$\Leftarrow$} Finally, we assume that $\separation$ is a separation and assume that $\Grad \separation(\mPoint, \cdot)^2\mid_{\mPointB} = - 2 \log_{\mPointB}(\mPoint)$ for any $\mPointB\in \manifold$. To prove the claim, we need to show that the third order term vanishes. This follows as a result of \cref{lem:cov-log-along-geodesic}. Indeed,
    \begin{multline}
        \frac{\mathrm{d}^3}{\mathrm{d}\tau^3} \separation(\mPoint, \geodesic_{\mPoint,\mPointB}(\tau))^2 = (\nabla_{\dot{\geodesic}_{\mPoint,\mPointB}} \nabla_{\dot{\geodesic}_{\mPoint,\mPointB}}\Grad \separation(\mPoint, \cdot)^2 , \dot{\geodesic}_{\mPoint,\mPointB})_{\geodesic_{\mPoint,\mPointB}(\tau)} \\
        = -2 (\nabla_{\dot{\geodesic}_{\mPoint,\mPointB}} \nabla_{\dot{\geodesic}_{\mPoint,\mPointB}}\log_{\mPointB}(\mPoint), \dot{\geodesic}_{\mPoint,\mPointB})_{\geodesic_{\mPoint,\mPointB}(\tau)} \overset{\cref{eq:lem-cov-log-along-geodesic}}{=} 2 (\nabla_{\dot{\geodesic}_{\mPoint,\mPointB}} \dot{\geodesic}_{\mPoint,\mPointB}, \dot{\geodesic}_{\mPoint,\mPointB})_{\geodesic_{\mPoint,\mPointB}(\tau)} \overset{\nabla_{\dot{\geodesic}_{\mPoint,\mPointB}} \dot{\geodesic}_{\mPoint,\mPointB}= 0}{= } 0
    \end{multline}
\end{proof}

\subsection{Proof of \cref{thm:well-posed-ness-separation-geodesics}}

\begin{proof}
    For notational convenience we define the mapping $F: \manifold \to \Real$ as
    \begin{equation}
        F(\mPointC) := \frac{1 - t}{2}\separation (\mPoint, \mPointC)^2 + \frac{t}{2} \separation (\mPointC, \mPointB)^2, \quad t\in [0,1].
    \end{equation}
    (i) We first note that $F$ is lower bounded by $0$. Also for $t=0$ this value is attained at $\mPointC = \mPoint$, and fort $t=1$ at $\mPointC = \mPointB$. So it remains to show existence of minimisers for $t\in (0,1)$.
    
    For that we next note that it is easy to see that $F$ is 1-coercive at  $\mPoint$, i.e., 
    \begin{equation}
        \lim_{\separation(\mPoint,\mPointC) \to \infty} \frac{F(\mPointC)}{\separation(\mPoint,\mPointC)} \geq \lim_{\separation(\mPoint,\mPointC) \to \infty} \frac{\frac{1 - t}{2}\separation (\mPoint, \mPointC)^2}{\separation(\mPoint,\mPointC)} = \lim_{\separation(\mPoint,\mPointC) \to \infty} \frac{1 - t}{2}\separation (\mPoint, \mPointC) = \infty.
    \end{equation}
    % In particular, we conclude that $\lim_{\separation([\eMat],[\eMatC]) \to \infty} F([\eMatC]) = \infty$. 
    So, there exists a radius $r > 0$ such that for all $\mPointC$ with $\separation(\mPoint, \mPointC) > r$ the inequality $F(\mPoint) \leq F(\mPointC)$ holds. Now consider the metric ball $\ball^{\separation}_{\mPoint}(r) := \{\mPointC\in  \manifold \; \mid \; \separation(\mPoint, \mPointC) \leq r\}$, which consequently has to contain the minimiser if it exists. Since $\separation$ is complete, $\ball^{\separation}_{\mPoint}(r)$ is compact. Finally, since $F$ is continuous there exists a $\mPointC^*\in \ball^{\separation}_{\mPoint}(r)$ such that $F(\mPointC^*) \leq F(\mPointC)$ for all $\mPointC\in \manifold$ and we conclude that $\mPointC^*$ is a global minimiser on $\manifold$.

    (ii) It suffices to show that in a neighbourhood of $\mPoint$ and $\mPointB$ the Riemannian Hessian of $F$ is positive definite. To see why this is true for $\mPoint$ and $\mPointB$ close enough, note that
    \begin{equation}
        \Hess \separation (\mPoint, \cdot)^2 (\mTVector_{\mPoint}, \mTVector_{\mPoint}) = (\nabla_{\mTVector_{\mPoint}} \Grad \separation(\mPoint, \cdot)^2, \mTVector_{\mPoint})_{\mPoint}  \overset{\text{\cref{def:separation}}}{=} 2 \|\mTVector_{\mPoint}\|_{\mPoint}^2, 
    \end{equation}
    and
    \begin{equation}
        \Hess \separation (\mPointB, \cdot)^2 (\mTVector_{\mPointB}, \mTVector_{\mPointB}) = (\nabla_{\mTVector_{\mPointB}} \Grad \separation(\mPointB, \cdot)^2, \mTVector_{\mPointB})_{\mPointB}  \overset{\text{\cref{def:separation}}}{=} 2 \|\mTVector_{\mPointB}\|_{\mPointB}^2. 
    \end{equation}
    In other words, the Riemannian Hessian of the first term in $F$ will be positive definite in a neighbourhood of $\mPoint$ and vice versa, the Riemannian Hessian of the second term in a neighbourhood of $\mPointB$. Since the separation is locally smooth, higher order derivatives are locally bounded. So we conclude that for $\mPoint$ and $\mPointB$ close enough there exists a neighbourhood where both terms are positive definite.
\end{proof}

\subsection{Proof of \cref{thm:well-posed-ness-separation-exp}}

\begin{proof}
    The proof is analogous to the proof of \cref{thm:well-posed-ness-separation-geodesics}. So again, for notational convenience we define the mapping $F: \manifold \to \Real$ as
    \begin{equation}
        F(\mPointC) := - \frac{1}{2}\separation (\mPoint, \mPointC)^2 + \separation (\mPointC, \mPointB)^2
    \end{equation}
    (i) We note that it is easy to see that $F$ is 1-coercive at  $\mPointB$, i.e., 
    \begin{multline}
        \lim_{\separation(\mPointB,\mPointC) \to \infty} \frac{F(\mPointC)}{\separation(\mPointB,\mPointC)} = \lim_{\separation(\mPointB,\mPointC) \to \infty} \frac{1}{2} \frac{-\separation (\mPoint, \mPointC)^2 + 2 \separation (\mPointC, \mPointB)^2}{\separation(\mPointB,\mPointC)} = \lim_{\separation(\mPointB,\mPointC) \to \infty} \frac{1}{2} \frac{-\separation (\mPoint, \mPointC)^2 + 2 \separation (\mPointB, \mPointC)^2}{\separation(\mPointB,\mPointC)} \\
        \geq \lim_{\separation(\mPointB,\mPointC) \to \infty} \frac{1}{2} \frac{-(\separation (\mPoint, \mPointB) + \separation (\mPointB, \mPointC))^2 + 2 \separation (\mPointB, \mPointC)^2}{\separation(\mPointB,\mPointC)} = \lim_{\separation(\mPointB,\mPointC) \to \infty} \frac{1}{2} \frac{-\separation (\mPoint, \mPointB)^2 - 2\separation (\mPoint, \mPointB)\separation (\mPointB, \mPointC) +  \separation (\mPointB, \mPointC)^2}{\separation(\mPointB,\mPointC)}\\
        = \lim_{\separation(\mPointB,\mPointC) \to \infty} -   \separation (\mPoint, \mPointB) +  \frac{1}{2} \separation (\mPointB, \mPointC) = \infty.
    \end{multline}
    % In particular, we conclude that $\lim_{\separation([\eMat],[\eMatC]) \to \infty} F([\eMatC]) = \infty$. 
    Following the same line of argument as in the proof of \cref{thm:well-posed-ness-separation-geodesics}, there exists a radius $r > 0$ such that for all $\mPointC$ with $\separation(\mPointB, \mPointC) > r$ the inequality $F(\mPointB) \leq F(\mPointC)$ holds. Now consider the metric ball $\ball^{\separation}_{\mPointB}(r) := \{\mPointC\in  \manifold \; \mid \; \separation(\mPointB, \mPointC) \leq r\}$. Since $\separation$ is complete, $\ball^{\separation}_{\mPointB}(r)$ is compact. Finally, since $F$ is continuous there exists a $\mPointC^*\in \ball^{\separation}_{\mPointB}(r)$ such that $F(\mPointC^*) \leq F(\mPointC)$ for all $\mPointC\in \manifold$ and we conclude that $\mPointC^*$ is a global minimiser on $\manifold$.

    (ii) It suffices to show that in a neighbourhood of $\mPoint$ and $\mPointB$ the Riemannian Hessian of $F$ is positive definite. To see why this is true for $\mPoint$ and $\mPointB$ close enough, note again  that
    \begin{equation}
        \Hess \separation (\mPoint, \cdot)^2 (\mTVector_{\mPoint}, \mTVector_{\mPoint}) \overset{\text{\cref{def:separation}}}{=} 2 \|\mTVector_{\mPoint}\|_{\mPoint}^2, \quad \text{and} \quad \Hess \separation (\mPointB, \cdot)^2 (\mTVector_{\mPointB}, \mTVector_{\mPointB}) \overset{\text{\cref{def:separation}}}{=} 2 \|\mTVector_{\mPointB}\|_{\mPointB}^2. 
    \end{equation}
    In particular, if $\mPoint = \mPointB$ we have $\Hess F (\mTVector_{\mPointB}, \mTVector_{\mPointB}) = - \|\mTVector_{\mPointB}\|_{\mPointB}^2 + 2 \|\mTVector_{\mPointB}\|_{\mPointB}^2 = \|\mTVector_{\mPointB}\|_{\mPointB}^2$. So there is a neighbourhood where $\Hess F$ will be positive definite. Since the separation is locally smooth, higher order derivatives are locally bounded. Thus, for $\mPoint$ close enough to $\mPointB$ such a neighbourhood still exists, which proves the claim.
\end{proof}
\clearpage
\section{Proofs for the results from \cref{sec:riemannian-protein-geometry}}
\label{app:proofs-quotient-manifold}

% \todo[inline]{Go over proofs once more}
\subsection{Proof of \cref{thm:point-cloud-manifold}}
\begin{proof}
It is sufficient to show that $\PCmanifold$ is a non-empty open subset of $\Real^{\dimInd \times \proteinLen}$ if $ \proteinLen \geq \dimInd +1$. 

First, assuming $ \proteinLen \geq \dimInd +1$, we will show that $\PCmanifold$ is non-empty, for which it is sufficient to show the case $\proteinLen = \dimInd+1$. For that consider the standard orthonormal basis $\{\mathbf{e}^{\sumIndA}\}_{\sumIndA=1}^\dimInd$ in $\Real^\dimInd$ and note that we have $\eMat := (\mathbf{e}^{1}, \ldots, \mathbf{e}^{\dimInd}, - \frac{1}{\dimInd}\sum_{\sumIndA}\mathbf{e}^{\sumIndA})\in \BlankPCmanifold^{\dimInd\times (\dimInd+1)}$, which proves the non-emptiness.

In order to show openness, we assume a general
%Next, assuming 
$\proteinLen \geq \dimInd + 1$ and consider the definition of $\PCmanifold$ as the intersection of two sets once more:
\begin{equation}
    \PCmanifold =  \PCmanifoldB \cap \PCmanifoldA.
\end{equation}
We have $\PCmanifoldB = \{\eMat = (\ePoint_1, \ldots, \ePoint_\proteinLen) \in \Real^{\dimInd\times \proteinLen} \mid \eMat - \frac{1}{\proteinLen}\eMat\mathbf{1}_{\proteinLen} \mathbf{1}_{\proteinLen}^{\top} \in \Real^{\dimInd\times \proteinLen}_\dimInd \} \neq \emptyset$ and the set $\PCmanifoldA = \{\eMat = (\ePoint_1, \ldots, \ePoint_\proteinLen) \in \Real^{\dimInd\times \proteinLen} \mid \ePoint_\sumIndA\neq \ePoint_\sumIndB \} \neq \emptyset$.
Since the intersection of two open sets is an open set, it remains and suffices to show that (i) $\PCmanifoldB$ is open in $\Real^{\dimInd\times\proteinLen}$ and that (ii) $\PCmanifoldA$ is open in $\Real^{\dimInd\times\proteinLen}$.

(i) We can see that $\PCmanifoldB$ is open through rewriting the set in a more convenient form
\begin{multline}
    \PCmanifoldB = \{\eMat \in  \Real^{\dimInd\times \proteinLen} \;\mid \; \det ((\eMat - \frac{1}{\proteinLen}\eMat\mathbf{1}_{\proteinLen} \mathbf{1}_{\proteinLen}^{\top}) (\eMat - \frac{1}{\proteinLen}\eMat\mathbf{1}_{\proteinLen} \mathbf{1}_{\proteinLen}^{\top})^\top) \neq 0\} \\
    = \Bigl(\eMat \mapsto \det \bigl((\eMat - \frac{1}{\proteinLen}\eMat\mathbf{1}_{\proteinLen} \mathbf{1}_{\proteinLen}^{\top}) (\eMat - \frac{1}{\proteinLen}\eMat\mathbf{1}_{\proteinLen} \mathbf{1}_{\proteinLen}^{\top})^\top\bigr)\Bigr)^{-1}\bigl((0, \infty)\bigr).
    \label{eq:thm-smooth-manifold-open-set-1}
\end{multline}
Note that we indeed only have to consider the inverse image of the positive real line in \cref{eq:thm-smooth-manifold-open-set-1} as the symmetric matrix $(\eMat - \frac{1}{\proteinLen}\eMat\mathbf{1}_{\proteinLen} \mathbf{1}_{\proteinLen}^{\top}) (\eMat - \frac{1}{\proteinLen}\eMat\mathbf{1}_{\proteinLen} \mathbf{1}_{\proteinLen}^{\top})^\top \in \Real^{\dimInd\times\dimInd}$ is semi-definite by construction. In other words, the eigenvalues are positive or zero, and hence the determinant mapping will be positive or zero. However, due to the full rank constraint the determinant cannot be zero, which leaves us with the positive real line.

Then, we have proven our claim as the right hand side of \cref{eq:thm-smooth-manifold-open-set-1} is the inverse image of an open set under a continuous function, i.e., it is an open set. We conclude that $\PCmanifoldB$ is open in $\Real^{\dimInd\times \proteinLen}$.

(ii) Next, we choose any matrix $\eMat = (\ePoint_1, \ldots, \ePoint_\proteinLen) \in \PCmanifoldA$. It suffices to show that there exists an open neighbourhood $\ball(\eMat) \subseteq \Real^{\dimInd\times \proteinLen}$ of $\eMat$ such that $\ball(\eMat) \subseteq \PCmanifoldA$. 

First, pick any $0 < \delta < \frac{1}{2}\min_{\sumIndA,\sumIndB} \|\ePoint_\sumIndA - \ePoint_\sumIndB\|_2$ and define the open set
\begin{equation}
    \ball_\delta (\eMat):= \{\eMatB = (\ePointB_1, \ldots, \ePointB_\proteinLen) \in \Real^{\dimInd\times\proteinLen} \; \mid \;  \|\ePointB_\sumIndA - \ePoint_\sumIndA\|_2 < \delta \text{ for } \sumIndA= 1, \ldots, n\}.
    \label{eq:prop-Rdnd-smooth manifold-open-ball}
\end{equation}
To prove the claim (ii) we will pick $\ball(\eMat) := \ball_\delta(\eMat)$ and will show that for any $\eMatB\in \ball_\delta (\eMat)$ we have $\eMatB \in \PCmanifoldA$. Then, since $\eMatB$ is arbitrary we must have that $\ball_\delta(\eMat) \subseteq \PCmanifoldA$, which proves the claim. 

So it remains to choose $\eMatB = (\ePointB_1, \ldots, \ePointB_\proteinLen) \in \ball_\delta (\eMat)$ and consider any $\ePointB_\sumIndA$ and $ \ePointB_\sumIndB$. We have
\begin{multline}
    \|\ePointB_\sumIndA - \ePointB_\sumIndB\|_2 = \|\ePointB_\sumIndA - \ePoint_\sumIndA + \ePoint_\sumIndA - \ePoint_\sumIndB + \ePoint_\sumIndB - \ePointB_\sumIndB\|_2 \\
    \overset{\text{reverse triangle ineq}}\geq \bigl| \|\ePoint_\sumIndA - \ePoint_\sumIndB\|_2 - \|\ePointB_\sumIndA - \ePoint_\sumIndA - (\ePointB_\sumIndB - \ePoint_\sumIndB)\|_2 \bigr| 
    \geq \|\ePoint_\sumIndA - \ePoint_\sumIndB\|_2 - \|\ePointB_\sumIndA - \ePoint_\sumIndA - (\ePointB_\sumIndB - \ePoint_\sumIndB)\|_2\\
    \overset{\text{definition $\delta$}}{>} 2\delta - \|\ePointB_\sumIndA - \ePoint_\sumIndA - (\ePointB_\sumIndB - \ePoint_\sumIndB)\|_2\\
    \overset{\text{triangle ineq}}{\geq} 2 \delta - (\|\ePointB_\sumIndA - \ePoint_\sumIndA\| + \|\ePointB_\sumIndB - \ePoint_\sumIndB\|_2) \overset{\cref{eq:prop-Rdnd-smooth manifold-open-ball}}{>} 2 \delta - (\delta + \delta) = 0.
\end{multline}
So $\|\ePointB_\sumIndA - \ePointB_\sumIndB\|_2 >0$ and $\ePointB_\sumIndA \neq\ePointB_\sumIndB$ for all $\sumIndA, \sumIndB = 1, \ldots, n$ and hence $\eMatB \in \PCmanifoldA$, which proves the claim.
\end{proof}

\subsection{Proof of \cref{thm:quotient-manifold-Rdnd-mod-Ed}}

\paragraph{Auxiliary lemmas}
If $\group$ is a Lie group and $\manifold$ is a smooth manifold, a  group action $\leftGroupAction: \group \times \manifold \rightarrow \manifold$ is \emph{smooth} if $\theta$ is smooth as a mapping. 

\begin{lemma}
\label{lem:smooth-left-group-action-Rdnd}
    The mapping $\leftGroupAction: \mathbb{E}(\dimInd) \times \PCmanifold \to \PCmanifold$ in \cref{prop:left-group-action-Rdnd} defines a smooth left group action if $\proteinLen \geq \dimInd +1$.
\end{lemma} 

\begin{proof}
    First, note that because $\proteinLen \geq \dimInd +1$ the set $\PCmanifold \neq \emptyset$ by \cref{thm:point-cloud-manifold}, which allows us to talk about left group actions. Next, it is easy to check that $\leftGroupAction$ satisfies the identity and the compatibility properties \cref{eq:identity-compatibility-lie-group}, from which we conclude that $\leftGroupAction$ is a left group action. Furthermore, because $\PCmanifold$ is locally diffeomorphic to a subset of $\Real^{\dimInd\times \proteinLen}$, smoothness of $\leftGroupAction$ follows from the fact that the mapping $\tilde{\leftGroupAction}: \mathbb{E}(\dimInd) \times \Real^{\dimInd\times \proteinLen} \to \Real^{\dimInd\times \proteinLen}$ defined analogously to \cref{prop:left-group-action-Rdnd} is smooth.
\end{proof}

In order for this space to be a quotient manifold as well, we need the notion of \emph{free and proper group actions} \cite[Def. 9.15 \& 9.16]{boumal2023introduction}.

A group action $\leftGroupAction$ is \emph{free} if only the identity fixes any $\mPoint\in \manifold$. In particular, a left action is free if $\leftGroupAction(\gPoint, \mPoint)=\mPoint \Rightarrow \gPoint=\idPoint$ for all $\mPoint \in \manifold$. 

\begin{lemma}
\label{lem:free-left-group-action-Rdnd}
    The group action $\leftGroupAction: \mathbb{E}(\dimInd) \times \PCmanifold \to \PCmanifold$ in \cref{prop:left-group-action-Rdnd} is free if $\proteinLen \geq \dimInd +1$.
\end{lemma}

\begin{proof}
    The assumption $\proteinLen \geq \dimInd +1$ tells us that $\PCmanifold \neq \emptyset$ by \cref{thm:point-cloud-manifold}, which is important because we do not get a free group action otherwise. So having eliminated problematic cases, we will show that the action is free for the remaining cases in two steps. First, we will show that for any point $\eMat \in \PCmanifold$ that is fixed under $(\EdOrthoref, \EdTrans)\in \mathbb{E}(\dimInd)$, we have the implication
\begin{equation}
    \leftGroupAction((\EdOrthoref, \EdTrans), \eMat) = \eMat \quad \Rightarrow \quad  \EdTrans = \mathbf{0}_\dimInd \in \Real^\dimInd.
    \label{eq:thm-quotient-manifold-Rdnd-mod-Ed-free-t}
\end{equation}
Subsequently, we will show that
\begin{equation}
    \leftGroupAction((\EdOrthoref, \mathbf{0}_\dimInd), \eMat) = \eMat \quad \Rightarrow \quad  \EdOrthoref = \mathbf{I}_{\dimInd} \in \Real^{\dimInd\times \dimInd}.
    \label{eq:thm-quotient-manifold-Rdnd-mod-Ed-free-O}
\end{equation}
Then, combining \cref{eq:thm-quotient-manifold-Rdnd-mod-Ed-free-t} and \cref{eq:thm-quotient-manifold-Rdnd-mod-Ed-free-O} immediately gives us the desired result
\begin{equation}
    \leftGroupAction((\EdOrthoref, \EdTrans), \eMat) = \eMat \quad \Rightarrow \quad  (\EdOrthoref, \EdTrans) = (\mathbf{I}_{\dimInd}, \mathbf{0}_\dimInd)=: \idPoint \in \mathbb{E}(\dimInd).
\end{equation}

(i) For the first step, i.e., proving \cref{eq:thm-quotient-manifold-Rdnd-mod-Ed-free-t}, we note that
\begin{equation}
    \EdOrthoref \eMat + \EdTrans \mathbf{1}_{\proteinLen}^{\top} = \eMat \quad \Leftrightarrow \quad \EdOrthoref \ePoint_\sumIndA + \EdTrans = \ePoint_\sumIndA \quad  \Rightarrow \quad \|\EdOrthoref \ePoint_\sumIndA + \EdTrans\|_2^2 = \|\ePoint_\sumIndA\|_2^2, \quad \text{for $\sumIndA = 1, \ldots, \proteinLen$}.
\end{equation}
Rewriting the last equality
%right hand side 
gives
\begin{equation}
    \|\EdOrthoref \ePoint_\sumIndA + \EdTrans\|_2^2 = \|\ePoint_\sumIndA\|_2^2
    \overset{\EdOrthoref\in \mathbb{O}(\dimInd)}{\Leftrightarrow} \| \ePoint_\sumIndA + \EdOrthoref^\top \EdTrans\|_2^2 = \|\ePoint_\sumIndA\|_2^2 \Leftrightarrow 2 (\ePoint_\sumIndA, \EdOrthoref^\top t)_2 + \|\EdOrthoref^\top \EdTrans\|_2^2 = 0 
    % \\
    % \overset{\EdOrthoref\in \mathbb{O}(\dimInd)}{\Leftrightarrow} 2 (\EdOrthoref \ePoint_\sumIndA, \EdTrans)_2 + \|\EdTrans\|_2^2 = 0, \quad \text{for $\sumIndA = 1, \ldots, n$}.
    \label{eq:thm-quotient-manifold-Rdnd-mod-Ed-free-t-step1}
\end{equation}
It is clear that the last equality
%right hand side 
of \cref{eq:thm-quotient-manifold-Rdnd-mod-Ed-free-t-step1} is satisfied if $\EdTrans = 0$. However, for  $t\neq \mathbf{0}_\dimInd \in \Real^\dimInd$ the above tells us that all $\ePoint_\sumIndA$ lie in an affine subspace. It follows that $\eMat \notin \PCmanifold$ due to the rank constraint, which is a contradiction. In other words, $\EdTrans = \mathbf{0}_\dimInd$ is the only valid solution to \cref{eq:thm-quotient-manifold-Rdnd-mod-Ed-free-t-step1}, which proves the claim \cref{eq:thm-quotient-manifold-Rdnd-mod-Ed-free-t}.

(ii) For the second step, i.e., proving \cref{eq:thm-quotient-manifold-Rdnd-mod-Ed-free-O}, we will solve for $\EdOrthoref$. That is,
\begin{multline}
    \EdOrthoref \eMat = \eMat \quad \Leftrightarrow \quad \eMat^\top \EdOrthoref^\top = \eMat^\top \quad \Rightarrow \quad \eMat\eMat^\top \EdOrthoref^\top = \eMat\eMat^\top \quad \\
    \overset{\text{$\eMat\eMat^\top$ full rank}}{\Leftrightarrow} \quad \EdOrthoref^\top = (\eMat\eMat^\top)^{-1}\eMat\eMat^\top = I \quad  \Rightarrow \quad \EdOrthoref = I,
\end{multline}
which proves the claim \cref{eq:thm-quotient-manifold-Rdnd-mod-Ed-free-O}.

\end{proof}

A left group action $\leftGroupAction$ is \emph{proper} if the function $\leftGroupActionMap: \group \times \manifold \rightarrow \manifold \times \manifold$ given by 
\begin{equation}
    \leftGroupActionMap(\gPoint, \mPoint): =(\leftGroupAction(\gPoint, \mPoint), \mPoint),
\end{equation}
is a proper map, i.e., if inverse images of compact subsets of $\manifold \times \manifold$ under $\leftGroupActionMap$ are compact in $\group \times \manifold$.

\begin{lemma}
\label{lem:proper-left-group-action-Rdnd}
    The group action $\leftGroupAction: \mathbb{E}(\dimInd) \times \PCmanifold \to \PCmanifold$ in \cref{prop:left-group-action-Rdnd} is proper if $\proteinLen \geq \dimInd +1$.
\end{lemma}

\begin{proof}
    Consider any compact set $L\subseteq \PCmanifold \times \PCmanifold $ and let $K:=$ $\pi_1(L) \cup \pi_2(L) \subseteq \PCmanifold$, where $\pi_1, \pi_2: \PCmanifold \times \PCmanifold \rightarrow \PCmanifold$ are the projections onto the first and second factors. Since $\leftGroupActionMap^{-1} (L) \subseteq \leftGroupActionMap^{-1} (K \times K)$, it suffices to show that there is some compact set $G_K \subseteq \mathbb{E}(\dimInd)$ such that
    \begin{equation}
        \leftGroupActionMap^{-1} (K \times K) \subseteq G_K \times K.
        \label{lem:compactness-property-GK-proper}
    \end{equation}
    Indeed, since $\leftGroupActionMap^{-1} (L)$ is closed by continuity (even smoothness) of $\leftGroupActionMap$, it is a closed subset of the compact set $G_K \times K$ and therefore compact. So it remains to show that there is such a $G_K$ satisfying \cref{lem:compactness-property-GK-proper}.

    The main step will be to show that the $\ell^2$-norm of admissible translations $\EdTrans\in \Real^\dimInd$ is uniformly bounded, i.e., uniformly over all $\EdOrthoref\in \mathbb{O}(\dimInd)$ and $\eMat\in K$. For that, define $F: \PCmanifold\times \PCmanifold \to \Real$ given by 
    \begin{equation}
        F(\eMat, \eMatB) := \max_{\sumIndA = 1, \ldots, \proteinLen} \|\ePoint_\sumIndA - \ePointB_\sumIndA\|_2.
    \end{equation}
    Next, we define $r^1_K\geq 0$ and $r^2_K\geq 0$ as
    \begin{equation}
        r^1_K := \max_{(\eMat,\eMatB) \in K \times K} \; F(\eMat, \eMatB), \quad \text{and} \quad r^2_K := \max_{(\EdOrthoref, \eMat) \in \mathbb{O}(\dimInd) \times K} \; F(\EdOrthoref \eMat, \eMat),
    \end{equation}
    which are well-defined and finite, because the function $F$ is continuous and both $\mathbb{O}(\dimInd)$ and $K$ are compact. Then, choose any $(\EdOrthoref, \EdTrans)\in \mathbb{E}(\dimInd)$ and $\eMat\in K$ such that $\EdOrthoref\eMat + \EdTrans\mathbf{1}_{\proteinLen}^\top\in K$. We find that
    \begin{equation}
        \|\EdTrans\|_2 = \|\EdTrans + \EdOrthoref\ePoint_1 - \ePoint_1 - (\EdOrthoref\ePoint_1 - \ePoint_1)\|_2 \leq \|\EdTrans + \EdOrthoref\ePoint_1 - \ePoint_1\|_2 + \|\EdOrthoref\ePoint_1 - \ePoint_1\|_2\leq r^1_K + r^2_K.
    \end{equation}
    Finally, for any $((\EdOrthoref, \EdTrans), \eMat) \in \leftGroupActionMap^{-1} (K \times K)$ we must have that $((\EdOrthoref, \EdTrans), \eMat) \in (\mathbb{O}(\dimInd) \times \bar{\ball}_{r_K}^\dimInd(0)) \times K$, where $\bar{\ball}_{r_K}^\dimInd(0)$ is the closed ball centered at $0\in \Real^\dimInd$ with radius $r_K := r^1_K + r^2_K$. 
    Due to the compactness of the orthogonal group $\mathbb{O}(\dimInd)$,
    %Also, note that 
    the set $\mathbb{O}(\dimInd) \times \bar{\ball}_{r_K}^\dimInd(0)$ is compact. Consequently, $G_K := \mathbb{O}(\dimInd) \times \bar{\ball}_{r_K}^\dimInd(0)$ satisfies \cref{lem:compactness-property-GK-proper} and proves the claim.
\end{proof}

\paragraph{Proof of the theorem}
\begin{proof}[Proof of \cref{thm:quotient-manifold-Rdnd-mod-Ed}]
The left action $\leftGroupAction: \mathbb{E}(\dimInd) \times \PCmanifold \to \PCmanifold$ in \cref{prop:left-group-action-Rdnd} is smooth by \cref{lem:smooth-left-group-action-Rdnd}, free by \cref{lem:free-left-group-action-Rdnd} and proper by \cref{lem:proper-left-group-action-Rdnd}. The claim follows from the quotient manifold theorem \cite[Theorem~9.18]{boumal2023introduction}.
\end{proof}

\subsection{Proof of \cref{thm:class-of-separations-on-pointcloud-manifold}}
% \todo[inline]{Go over proofs once more.}
% \todo[inline]{We don't have to say what C is, just that $\MetricTensor_{\diamond \eMat}: \Real^{\dimInd\times \proteinLen} \to \Real^{\dimInd\times \proteinLen}$, but has range $\horizontal_{\eMat} \Real^{\dimInd\times \proteinLen}_{\dimInd, \star, *}$}

%For the following we define the $\ell^2$-inner product on $\Real^{\dimInd\times \proteinLen}$ in the canonical way as
%\begin{equation}
%    (\mathbf{U}, \mathbf{V})_2 := \trace (\mathbf{U}^\top \mathbf{V}), \quad \mathbf{U}, \mathbf{V} \in \Real^{\dimInd\times \proteinLen}.
%\end{equation}

\paragraph{Auxiliary lemmas}

The quotient manifold $\PCmanifold /\mathbb{E}(\dimInd)$ has a vertical space
\begin{equation}
    \vertical_{\eMat} \PCmanifold = \{ \EdOrthorefTV \eMat + \EdTransTV \mathbf{1}_{\proteinLen}^\top \in \Real^{\dimInd\times\proteinLen}\mid \EdOrthorefTV^\top = - \EdOrthorefTV \in \Real^{\dimInd\times \dimInd} ,\; \EdTransTV\in \Real^\dimInd\}
    \label{eq:vertical-space}
\end{equation}
and horizontal space
\begin{align}
    \horizontal_{\eMat} \PCmanifold = \{\mTVector_{\diamond\eMat}:= (\mTVectorCompon_1,\ldots,\mTVectorCompon_\proteinLen) \in \Real^{\dimInd\times \proteinLen} \mid \sum_{\sumIndA}\mTVectorCompon_\sumIndA = 0, \: \sum_{\sumIndA}(\EdOrthorefTV \ePoint_\sumIndA  \mTVectorCompon_\sumIndA)_2=0 \; \forall \; \EdOrthorefTV\in \Real^{\dimInd\times \dimInd} \; s.t. \; \EdOrthorefTV^\top = - \EdOrthorefTV \} .
    \label{eq:horizontal-space}
\end{align}
For the horizontal space, we have the natural identification 
\begin{equation}
    \mTVector_{\diamond \leftGroupAction((\EdOrthoref, \EdTrans), \eMat)} =  \leftGroupAction((\EdOrthoref, \mathbf{0}_\dimInd), \mTVector_{\diamond\eMat}).
    \label{eq:natural-vertical-parallel}
\end{equation}

In the following, we will often consider metric tensor fields $(\cdot, \cdot): \vectorfield (\PCmanifold /\mathbb{E}(\dimInd)) \times \vectorfield (\PCmanifold /\mathbb{E}(\dimInd)) \to C^\infty(\PCmanifold /\mathbb{E}(\dimInd))$ of the form
\begin{equation}
    (\mTVector, \mTVectorB)_{[\eMat]} := (\MetricTensor_{\diamond \eMat} (\mTVector_{\diamond \eMat}),  \mTVectorB_{\diamond \eMat})_2, \quad \eMat \in [\eMat],
    \label{eq:standard-metric-tensor-field-pointclouds}
\end{equation}
where $\MetricTensor_{\diamond \eMat}: \Real^{\dimInd\times \proteinLen} \to \Real^{\dimInd\times \proteinLen}$ is a symmetric linear mapping parametrized by $\eMat \in \PCmanifold$ that satisfies 
\begin{equation}
    \MetricTensor_{\diamond \leftGroupAction((\EdOrthoref, \EdTrans), \eMat)} \circ \leftGroupAction((\EdOrthoref, \mathbf{0}_\dimInd), \cdot) = \leftGroupAction((\EdOrthoref, \mathbf{0}_\dimInd), \cdot) \circ \MetricTensor_{\diamond \eMat} \quad \text{for all } (\EdOrthoref, \EdTrans)\in \mathbb{E}(\dimInd)
    \label{eq:equivariance-like-metric-tensor}
\end{equation}
as to ensure well-posedness of the inner product regarding invariance of the choice of $\eMat \in [\eMat]$.

Similarly, cotangent vectors $\mCTVector_{[\eMat]} \in \tangent_{[\eMat]}^* \PCmanifold /\mathbb{E}(\dimInd)$ can naturally be represented as
\begin{equation}
    \mCTVector_{[\eMat]}(\mTVector_{[\eMat]}):= \mCTVector_{\diamond \eMat}(\mTVector_{\diamond \eMat}), \quad \eMat \in [\eMat],
    \label{eq:standard-covector-pointclouds}
\end{equation}
where $\mCTVector_{\diamond \eMat}: \Real^{\dimInd \times \proteinLen} \to \Real$ is a linear mapping parametrized by $\eMat \in \PCmanifold$ that satisfies
    \begin{equation}
        \mCTVector_{\diamond \leftGroupAction((\EdOrthoref, \EdTrans), \eMat)} \circ \leftGroupAction((\EdOrthoref, \mathbf{0}_\dimInd), \cdot) = \mCTVector_{\diamond \eMat}, \quad \text{for all } (\EdOrthoref, \EdTrans)\in \mathbb{E}(\dimInd)
        \label{eq:equivariance-like-one-form}
    \end{equation}
    for ensureing well-posedness regarding invariance of the choice of $\eMat\in [\eMat]$.

% \todo[inline]{Define horizontal and vertical spaces here already and discuss the natural identification Xi gX = G Xi X + discuss how we will often use an inner product of the form C. Naturally, we will need a certain symmetry for well-posedness of such inner products.}

% \todo[inline]{Notation : $\MetricTensor_{\diamond \eMat}(\mTVector_{\diamond \eMat}) := \sum_{\sumIndB} ((\MetricTensor_{\diamond \eMat})_{1,\sumIndB} (\mTVector_{\diamond \eMat})_\sumIndB, \ldots, (\MetricTensor_{\diamond \eMat})_{\sumIndA,\sumIndB} (\mTVector_{\diamond \eMat})_\sumIndB, \ldots, (\MetricTensor_{\diamond \eMat})_{\proteinLen,\sumIndB} (\mTVector_{\diamond \eMat})_\sumIndB) \in \horizontal_{\eMat} \Real^{\dimInd\times \proteinLen}_{\dimInd, \star, *}$ and [the double product like the inner product]}

\begin{lemma}
\label{lem:rewrite-linear-mapping-as-tangent-vector}
    Consider the smooth manifold $\PCmanifold /\mathbb{E}(\dimInd)$ for $\proteinLen \geq \dimInd+1$ under the metric tensor field $(\cdot, \cdot): \vectorfield (\PCmanifold /\mathbb{E}(\dimInd)) \times \vectorfield (\PCmanifold /\mathbb{E}(\dimInd)) \to C^\infty(\PCmanifold /\mathbb{E}(\dimInd))$ of the form \cref{eq:standard-metric-tensor-field-pointclouds}, i.e., given by 
    \begin{equation}
        (\mTVector, \mTVectorB)_{[\eMat]} := (\MetricTensor_{\diamond \eMat} (\mTVector_{\diamond \eMat}),  \mTVectorB_{\diamond \eMat})_2, \quad \eMat \in [\eMat],
    \end{equation}
    for some symmetric linear mapping $\MetricTensor_{\diamond \eMat}: \Real^{\dimInd\times \proteinLen} \to \Real^{\dimInd\times \proteinLen}$ parametrized by $\eMat \in \PCmanifold$ that satisfies \cref{eq:equivariance-like-metric-tensor} and is positive definite when restricted to the horizontal space $\horizontal_{\eMat} \PCmanifold$ for all $\eMat \in \PCmanifold$. Additionally, consider a cotangent vector $\mCTVector_{[\eMat]} \in \tangent_{[\eMat]}^* \PCmanifold /\mathbb{E}(\dimInd)$ of the form \cref{eq:standard-covector-pointclouds}, i.e., given by
    \begin{equation}
        \mCTVector_{[\eMat]}(\mTVector_{[\eMat]}):= \mCTVector_{\diamond \eMat}(\mTVector_{\diamond \eMat}), \quad \eMat \in [\eMat],
    \end{equation}
    for some linear mapping  $\mCTVector_{\diamond \eMat}: \Real^{\dimInd \times \proteinLen} \to \Real$ parametrized by $\eMat \in \PCmanifold$ that satisfies \cref{eq:equivariance-like-one-form} for all $\eMat \in \PCmanifold$.
    
    Then, 
    \begin{equation}
        \mCTVector_{\diamond \eMat}(\mTVector_{\diamond \eMat}) = (\MetricTensor_{\diamond \eMat}(\MetricTensor_{\diamond \eMat}^\dagger (\bigl[ \mCTVector_{\diamond \eMat} (\mathrm{e}^\sumIndA \otimes \mathrm{f}^\sumIndB ) \bigr]_{\sumIndA,\sumIndB=1}^{\dimInd,\proteinLen})), \mTVector_{\diamond \eMat})_2, \quad \text{for all } \mTVector_{\diamond \eMat} \in \horizontal_{\eMat} \PCmanifold,
        \label{eq:tvector-representation-linear-functional}
    \end{equation}
where $\MetricTensor_{\diamond \eMat}^\dagger$ is the pseudo-inverse of $\MetricTensor_{\diamond \eMat}$, $\mathrm{e}^\sumIndA \in \Real^\dimInd$ for $\sumIndA = 1, \ldots, \dimInd$ such that $(\mathrm{e}^\sumIndA)_\sumIndC =\delta_{\sumIndA,\sumIndC}$ and $\mathrm{f}^\sumIndB \in \Real^\proteinLen$ for $\sumIndB = 1, \ldots, \proteinLen$ such that $(\mathrm{f}^\sumIndB)_\sumIndD =\delta_{\sumIndB,\sumIndD}$.
\end{lemma}

\begin{proof}
    First, we invoke linearity of $\mCTVector_{\diamond \eMat}$ and rewrite the left hand term in \cref{eq:tvector-representation-linear-functional}
    \begin{multline}
        \mCTVector_{\diamond \eMat}(\mTVector_{\diamond \eMat}) = \mCTVector_{\diamond \eMat}\Bigl(\sum_{\sumIndA, \sumIndB=1}^{\dimInd, \proteinLen} (\mathrm{e}^\sumIndA \otimes \mathrm{f}^\sumIndB, \mTVector_{\diamond \eMat})_2 \; \mathrm{e}^\sumIndA \otimes \mathrm{f}^\sumIndB \Bigr) = \sum_{\sumIndA, \sumIndB=1}^{\dimInd, \proteinLen} (\mathrm{e}^\sumIndA \otimes \mathrm{f}^\sumIndB, \mTVector_{\diamond \eMat})_2 \;\mCTVector_{\diamond \eMat}( \mathrm{e}^\sumIndA \otimes \mathrm{f}^\sumIndB ) \\
        = \sum_{\sumIndA, \sumIndB=1}^{\dimInd, \proteinLen} (\mCTVector_{\diamond \eMat}( \mathrm{e}^\sumIndA \otimes \mathrm{f}^\sumIndB ) \; \mathrm{e}^\sumIndA \otimes \mathrm{f}^\sumIndB, \mTVector_{\diamond \eMat})_2  = (\bigl[ \mCTVector_{\diamond \eMat}( \mathrm{e}^\sumIndA \otimes \mathrm{f}^\sumIndB ) \bigr]_{\sumIndA,\sumIndB=1}^{\dimInd,\proteinLen}, \mTVector_{\diamond \eMat})_2.
        \label{eq:lemma-sort-of-sharp-1}
    \end{multline}
    Next, since $\MetricTensor_{\diamond \eMat}$ satisfies \cref{eq:equivariance-like-metric-tensor} and is positive definite when restricted to the horizontal space $\horizontal_{\eMat} \PCmanifold$ for all $\eMat \in \PCmanifold$ we can write
    \begin{equation}
        (\bigl[ \mCTVector_{\diamond \eMat}( \mathrm{e}^\sumIndA \otimes \mathrm{f}^\sumIndB ) \bigr]_{\sumIndA,\sumIndB=1}^{\dimInd,\proteinLen}, \mTVector_{\diamond \eMat})_2 = 
        % (\bigl[ \mCTVector_{\diamond \eMat}( \mathrm{e}^\sumIndA \otimes \mathrm{f}^\sumIndB ) \bigr]_{\sumIndA,\sumIndB=1}^{\dimInd,\proteinLen}, P_{\horizontal_{\eMat} \Real^{\dimInd\times \proteinLen}_{\dimInd, \star, *}} (\mTVector_{\diamond \eMat}))_2 \\
        % = (P_{\horizontal_{\eMat} \Real^{\dimInd\times \proteinLen}_{\dimInd, \star, *}} (\bigl[ \mCTVector_{\diamond \eMat}( \mathrm{e}^\sumIndA \otimes \mathrm{f}^\sumIndB ) \bigr]_{\sumIndA,\sumIndB=1}^{\dimInd,\proteinLen}), \mTVector_{\diamond \eMat})_2  
        % = \MetricTensor_{\diamond \eMat}(\MetricTensor_{\diamond \eMat}^\dagger (P_{\horizontal_{\eMat} \Real^{\dimInd\times \proteinLen}_{\dimInd, \star, *}} (\bigl[ \mCTVector_{\diamond \eMat} (\mathrm{e}^\sumIndA \otimes \mathrm{f}^\sumIndB ) \bigr]_{\sumIndA,\sumIndB=1}^{\dimInd,\proteinLen})), \mTVector_{\diamond \eMat}) \\
        % = 
        (\MetricTensor_{\diamond \eMat}(\MetricTensor_{\diamond \eMat}^\dagger (\bigl[ \mCTVector_{\diamond \eMat} (\mathrm{e}^\sumIndA \otimes \mathrm{f}^\sumIndB ) \bigr]_{\sumIndA,\sumIndB=1}^{\dimInd,\proteinLen})), \mTVector_{\diamond \eMat})_2.
        \label{eq:lemma-sort-of-sharp-2}
    \end{equation}
    The claim follows from combining \cref{eq:lemma-sort-of-sharp-1,eq:lemma-sort-of-sharp-2}.
\end{proof}

\begin{lemma}[Levi-Civita connection]
\label{lem:cov-derivative}
Consider the smooth manifold $\PCmanifold /\mathbb{E}(\dimInd)$ for $\proteinLen \geq \dimInd+1$ under the metric tensor field $(\cdot, \cdot): \vectorfield (\PCmanifold /\mathbb{E}(\dimInd)) \times \vectorfield (\PCmanifold /\mathbb{E}(\dimInd)) \to C^\infty(\PCmanifold /\mathbb{E}(\dimInd))$ of the form \cref{eq:standard-metric-tensor-field-pointclouds}, i.e., given by  
    \begin{equation}
        (\mTVector, \mTVectorB)_{[\eMat]} := (\MetricTensor_{\diamond \eMat} (\mTVector_{\diamond \eMat}),  \mTVectorB_{\diamond \eMat})_2, \quad \eMat \in [\eMat],
        \label{eq:template-metric-tensor-field}
    \end{equation}
    for some symmetric linear mapping $\MetricTensor_{\diamond \eMat}: \Real^{\dimInd\times \proteinLen} \to \Real^{\dimInd\times \proteinLen}$ parametrized by $\eMat \in \PCmanifold$ that satisfies \cref{eq:equivariance-like-metric-tensor}, is positive definite when restricted to the horizontal space $\horizontal_{\eMat}\PCmanifold$ and has kernel $\ker(\MetricTensor_{\diamond \eMat} ) = \vertical_{\eMat} \PCmanifold$ for all $\eMat \in \PCmanifold$.

Then, the Levi-Civita connection on $(\PCmanifold /\mathbb{E}(\dimInd), (\cdot, \cdot))$ is
\begin{multline}
    (\nabla_{\mTVector} \mTVectorB)_{\diamond \eMat} = D_{\eMat}(\mTVectorB_{\diamond (\cdot)}) [\mTVector_{\diamond \eMat}] + \frac{1}{2} \MetricTensor_{\diamond \eMat}^\dagger \Bigl( \bigl[(D_{\eMat}\MetricTensor_{\diamond (\cdot)}   (\mTVectorB_{\diamond \eMat}) [\mTVector_{\diamond \eMat}], \mathrm{e}^\sumIndA \otimes \mathrm{f}^\sumIndB)_2]_{\sumIndA,\sumIndB=1}^{\dimInd,\proteinLen} \\
        + \bigl[(D_{\eMat}\MetricTensor_{\diamond (\cdot)}   (\mTVector_{\diamond \eMat}) [\mTVectorB_{\diamond \eMat}], \mathrm{e}^\sumIndA \otimes \mathrm{f}^\sumIndB)_2]_{\sumIndA,\sumIndB=1}^{\dimInd,\proteinLen} 
        - \bigl[(D_{\eMat} \MetricTensor_{\diamond (\cdot)} (\mTVector_{\diamond \eMat})[\mathrm{e}^\sumIndA \otimes \mathrm{f}^\sumIndB]) ,\mTVectorB_{\diamond \eMat})_2 \bigr]_{\sumIndA,\sumIndB=1}^{\dimInd,\proteinLen}\Bigr),
     \label{eq:lem-covariant-derivative}
\end{multline}
% \begin{multline}
%     (\nabla_{\mTVector}^\alpha \mTVectorB, \mTVectorC)_{[\eMat]}^\alpha = (D_{\eMat} \mTVectorB_{\diamond (\cdot)} [\mTVector_{\diamond \eMat}], \mTVectorC_{\diamond \eMat})_{\diamond \eMat}^\alpha + \frac{1}{2}\Bigl(((\MetricTensor^{\alpha}_{\diamond \eMat})^\dagger (D_{\eMat} \MetricTensor^{\alpha}_{\diamond (\cdot)} [\mTVector_{\diamond \eMat}]) \mTVectorB_{\diamond \eMat}, \mTVectorC_{\diamond \eMat})_{\diamond \eMat}^\alpha \\
%     + ((\MetricTensor^{\alpha}_{\diamond \eMat})^\dagger (D_{\eMat} \MetricTensor^{\alpha}_{\diamond (\cdot)} [\mTVectorB_{\diamond \eMat}]) \mTVector_{\diamond \eMat}, \mTVectorC_{\diamond \eMat})_{\diamond \eMat}^\alpha - ((\MetricTensor^{\alpha}_{\diamond \eMat})^\dagger (D_{\eMat} \MetricTensor^{\alpha}_{\diamond (\cdot)} [\star]) (\mTVector_{\diamond \eMat},\mTVectorB_{\diamond \eMat}), \mTVectorC_{\diamond \eMat})_{\diamond \eMat}^\alpha \Bigr)
% \end{multline}
where $(\MetricTensor_{\diamond \eMat})^\dagger$ is the pseudo-inverse of $\MetricTensor_{\diamond \eMat}$, $\mathrm{e}^\sumIndA \in \Real^\dimInd$ for $\sumIndA = 1, \ldots, \dimInd$ such that $(\mathrm{e}^\sumIndA)_\sumIndC =\delta_{\sumIndA,\sumIndC}$ and $\mathrm{f}^\sumIndB \in \Real^\proteinLen$ for $\sumIndB = 1, \ldots, \proteinLen$ such that $(\mathrm{f}^\sumIndB)_\sumIndD =\delta_{\sumIndB,\sumIndD}$.
\end{lemma}

\begin{proof}
    We will show the identity \cref{eq:lem-covariant-derivative} using Koszul's formula \cite[(5.11)]{boumal2023introduction}. That is, the Levi-Civita connection uniquely satisfies
    \begin{equation}
        2 (\nabla_{\mTVector} \mTVectorB, \mTVectorC)_{[\eMat]}= \mTVector_{[\eMat]}(\mTVectorB, \mTVectorC)_{(\cdot)}+\mTVectorB_{[\eMat]}(\mTVector, \mTVectorC)_{(\cdot)}-\mTVectorC_{[\eMat]}(\mTVector, \mTVectorB)_{(\cdot)}-([\mTVectorB, \mTVector], \mTVectorC)_{[\eMat]}-([\mTVector, \mTVectorC], \mTVectorB)_{[\eMat]}-([\mTVectorB, \mTVectorC], \mTVector)_{[\eMat]},
        \label{eq:lem-cov-deriv-koszul}
    \end{equation}
    where $\mTVector,\mTVectorB,\mTVectorC \in \vectorfield (\PCmanifold /\mathbb{E}(\dimInd))$ and $[\eMat]\in \PCmanifold /\mathbb{E}(\dimInd)$. In the following we will evaluate each of the terms.

    By symmetry in $\mTVector,\mTVectorB$, and $\mTVectorC$ it suffices to expand
    \begin{multline}
        \mTVector_{[\eMat]}(\mTVectorB, \mTVectorC)_{(\cdot)} = D_{\eMat}(\MetricTensor_{\diamond (\cdot)} (\mTVectorB_{\diamond (\cdot)}), \mTVectorC_{\diamond (\cdot)} )_2 [\mTVector_{\diamond \eMat}] \\
        \overset{\text{product rule}}{=}  (D_{\eMat}\MetricTensor_{\diamond (\cdot)}   (\mTVectorB_{\diamond \eMat}) [\mTVector_{\diamond \eMat}], \mTVectorC_{\diamond \eMat})_2  
        +  (\MetricTensor_{\diamond \eMat} (D_{\eMat}\mTVectorB_{\diamond (\cdot)} [\mTVector_{\diamond \eMat}]), \mTVectorC_{\diamond \eMat})_2  
        + (\MetricTensor_{\diamond \eMat} (\mTVectorB_{\diamond \eMat}),  D_{\eMat}\mTVectorC_{\diamond (\cdot)} [\mTVector_{\diamond \eMat}])_2,
        \label{eq:lem-cov-deriv-term-1}
    \end{multline}
    and
    \begin{equation}
        ([\mTVectorB, \mTVector], \mTVectorC)_{[\eMat]} = (\MetricTensor_{\diamond \eMat} (D_{\eMat}\mTVector_{\diamond (\cdot)} [\mTVectorB_{\diamond \eMat}]), \mTVectorC_{\diamond \eMat})_2 - (\MetricTensor_{\diamond \eMat} (D_{\eMat}\mTVectorB_{\diamond (\cdot)} [\mTVector_{\diamond \eMat}]), \mTVectorC_{\diamond \eMat})_2.
        \label{eq:lem-cov-deriv-term-2}
    \end{equation}

    Substituting \cref{eq:lem-cov-deriv-term-1,eq:lem-cov-deriv-term-2} into \cref{eq:lem-cov-deriv-koszul} will result in most terms cancelling due to the symmetry of $\MetricTensor_{\diamond \eMat}$, i.e., in the sense that $\MetricTensor_{\diamond \eMat}$ is a symmetric linear mapping. Indeed,
    \begin{multline}
        2 (\nabla_{\mTVector} \mTVectorB, \mTVectorC)_{[\eMat]} = 
        \Bigl( (D_{\eMat}\MetricTensor_{\diamond (\cdot)}   (\mTVectorB_{\diamond \eMat}) [\mTVector_{\diamond \eMat}], \mTVectorC_{\diamond \eMat})_2  +  (\MetricTensor_{\diamond \eMat} (D_{\eMat}\mTVectorB_{\diamond (\cdot)} [\mTVector_{\diamond \eMat}]), \mTVectorC_{\diamond \eMat})_2  + \cancel{(\MetricTensor_{\diamond \eMat} (\mTVectorB_{\diamond \eMat}),  D_{\eMat}\mTVectorC_{\diamond (\cdot)} [\mTVector_{\diamond \eMat}])_2}\Bigr) \\
        + \Bigl( (D_{\eMat}\MetricTensor_{\diamond (\cdot)}   (\mTVector_{\diamond \eMat}) [\mTVectorB_{\diamond \eMat}], \mTVectorC_{\diamond \eMat})_2  +  (\MetricTensor_{\diamond \eMat} (D_{\eMat}\mTVector_{\diamond (\cdot)} [\mTVectorB_{\diamond \eMat}]), \mTVectorC_{\diamond \eMat})_2  + (\MetricTensor_{\diamond \eMat} (\mTVector_{\diamond \eMat}),  D_{\eMat}\mTVectorC_{\diamond (\cdot)} [\mTVectorB_{\diamond \eMat}])_2\Bigr) \\
        - \Bigl((D_{\eMat}\MetricTensor_{\diamond (\cdot)}   (\mTVector_{\diamond \eMat}) [\mTVectorC_{\diamond \eMat}], \mTVectorB_{\diamond \eMat})_2  +  (\MetricTensor_{\diamond \eMat} (D_{\eMat}\mTVector_{\diamond (\cdot)} [\mTVectorC_{\diamond \eMat}]), \mTVectorB_{\diamond \eMat})_2  + \bcancel{(\MetricTensor_{\diamond \eMat} (\mTVector_{\diamond \eMat}),  D_{\eMat}\mTVectorB_{\diamond (\cdot)} [\mTVectorC_{\diamond \eMat}])_2} \Bigr) \\
        - \Bigl( (\MetricTensor_{\diamond \eMat} (D_{\eMat}\mTVector_{\diamond (\cdot)} [\mTVectorB_{\diamond \eMat}]), \mTVectorC_{\diamond \eMat})_2 - (\MetricTensor_{\diamond \eMat} (D_{\eMat}\mTVectorB_{\diamond (\cdot)} [\mTVector_{\diamond \eMat}]), \mTVectorC_{\diamond \eMat})_2 \Bigr) \\
        - \Bigl( \cancel{(\MetricTensor_{\diamond \eMat} (D_{\eMat}\mTVectorC_{\diamond (\cdot)} [\mTVector_{\diamond \eMat}]), \mTVectorB_{\diamond \eMat})_2} - (\MetricTensor_{\diamond \eMat} (D_{\eMat}\mTVector_{\diamond (\cdot)} [\mTVectorC_{\diamond \eMat}]), \mTVectorB_{\diamond \eMat})_2  \Bigr) \\
        - \Bigl( (\MetricTensor_{\diamond \eMat} (D_{\eMat}\mTVectorC_{\diamond (\cdot)} [\mTVectorB_{\diamond \eMat}]), \mTVector_{\diamond \eMat})_2 - \bcancel{(\MetricTensor_{\diamond \eMat} (D_{\eMat}\mTVectorB_{\diamond (\cdot)} [\mTVectorC_{\diamond \eMat}]), \mTVector_{\diamond \eMat})_2 } \Bigr)\\
         =  (D_{\eMat}\MetricTensor_{\diamond (\cdot)}   (\mTVectorB_{\diamond \eMat}) [\mTVector_{\diamond \eMat}], \mTVectorC_{\diamond \eMat})_2  +  (\MetricTensor_{\diamond \eMat} (D_{\eMat}\mTVectorB_{\diamond (\cdot)} [\mTVector_{\diamond \eMat}]), \mTVectorC_{\diamond \eMat})_2 +  (D_{\eMat}\MetricTensor_{\diamond (\cdot)}   (\mTVector_{\diamond \eMat}) [\mTVectorB_{\diamond \eMat}], \mTVectorC_{\diamond \eMat})_2 \\
        +  \cancel{(\MetricTensor_{\diamond \eMat} (D_{\eMat}\mTVector_{\diamond (\cdot)} [\mTVectorB_{\diamond \eMat}]), \mTVectorC_{\diamond \eMat})_2}  + (\MetricTensor_{\diamond \eMat} (\mTVector_{\diamond \eMat}),  D_{\eMat}\mTVectorC_{\diamond (\cdot)} [\mTVectorB_{\diamond \eMat}])_2
        - (D_{\eMat}\MetricTensor_{\diamond (\cdot)}   (\mTVector_{\diamond \eMat}) [\mTVectorC_{\diamond \eMat}], \mTVectorB_{\diamond \eMat})_2
        \\ -   \bcancel{(\MetricTensor_{\diamond \eMat} (D_{\eMat}\mTVector_{\diamond (\cdot)} [\mTVectorC_{\diamond \eMat}]), \mTVectorB_{\diamond \eMat})_2} 
        - \cancel{(\MetricTensor_{\diamond \eMat} (D_{\eMat}\mTVector_{\diamond (\cdot)} [\mTVectorB_{\diamond \eMat}]), \mTVectorC_{\diamond \eMat})_2 } + (\MetricTensor_{\diamond \eMat} (D_{\eMat}\mTVectorB_{\diamond (\cdot)} [\mTVector_{\diamond \eMat}]), \mTVectorC_{\diamond \eMat})_2 \\
        +  \bcancel{(\MetricTensor_{\diamond \eMat} (D_{\eMat}\mTVector_{\diamond (\cdot)} [\mTVectorC_{\diamond \eMat}]), \mTVectorB_{\diamond \eMat})_2} - (\MetricTensor_{\diamond \eMat} (D_{\eMat}\mTVectorC_{\diamond (\cdot)} [\mTVectorB_{\diamond \eMat}]), \mTVector_{\diamond \eMat})_2 \\
        = (D_{\eMat}\MetricTensor_{\diamond (\cdot)}   (\mTVectorB_{\diamond \eMat}) [\mTVector_{\diamond \eMat}], \mTVectorC_{\diamond \eMat})_2  +  (\MetricTensor_{\diamond \eMat} (D_{\eMat}\mTVectorB_{\diamond (\cdot)} [\mTVector_{\diamond \eMat}]), \mTVectorC_{\diamond \eMat})_2 +  (D_{\eMat}\MetricTensor_{\diamond (\cdot)}   (\mTVector_{\diamond \eMat}) [\mTVectorB_{\diamond \eMat}], \mTVectorC_{\diamond \eMat})_2 \\
        + \cancel{(\MetricTensor_{\diamond \eMat} (\mTVector_{\diamond \eMat}),  D_{\eMat}\mTVectorC_{\diamond (\cdot)} [\mTVectorB_{\diamond \eMat}])_2}
        - (D_{\eMat}\MetricTensor_{\diamond (\cdot)}   (\mTVector_{\diamond \eMat}) [\mTVectorC_{\diamond \eMat}], \mTVectorB_{\diamond \eMat})_2 + (\MetricTensor_{\diamond \eMat} (D_{\eMat}\mTVectorB_{\diamond (\cdot)} [\mTVector_{\diamond \eMat}]), \mTVectorC_{\diamond \eMat})_2\\
         - \cancel{(\MetricTensor_{\diamond \eMat} (D_{\eMat}\mTVectorC_{\diamond (\cdot)} [\mTVectorB_{\diamond \eMat}]), \mTVector_{\diamond \eMat})_2}\\
         = 2  (\MetricTensor_{\diamond \eMat} (D_{\eMat}\mTVectorB_{\diamond (\cdot)} [\mTVector_{\diamond \eMat}]), \mTVectorC_{\diamond \eMat})_2 + (D_{\eMat}\MetricTensor_{\diamond (\cdot)}   (\mTVectorB_{\diamond \eMat}) [\mTVector_{\diamond \eMat}], \mTVectorC_{\diamond \eMat})_2 + (D_{\eMat}\MetricTensor_{\diamond (\cdot)}   (\mTVector_{\diamond \eMat}) [\mTVectorB_{\diamond \eMat}], \mTVectorC_{\diamond \eMat})_2 \\
         - (D_{\eMat}\MetricTensor_{\diamond (\cdot)}   (\mTVector_{\diamond \eMat}) [\mTVectorC_{\diamond \eMat}], \mTVectorB_{\diamond \eMat})_2.
         \label{eq:lema-eval-cov-derivative-terms}
    \end{multline}

    In order to retrieve \cref{eq:lem-covariant-derivative}, we need to write the remaining terms in inner product form $(\MetricTensor_{\diamond \eMat}(\mTVectorD_{\diamond \eMat}), \mTVectorC_{\diamond \eMat})_2$ for some $\mTVectorD_{\diamond \eMat}\in  \horizontal_{\eMat} \PCmanifold$. 
    For that, we first note that
    \begin{multline}
        (D_{\leftGroupAction((\EdOrthoref, \EdTrans), \eMat)}\MetricTensor_{\diamond (\cdot)}   (\mTVectorB_{\leftGroupAction((\EdOrthoref, \EdTrans), \eMat)}) [\mTVector_{\diamond \leftGroupAction((\EdOrthoref, \EdTrans), \eMat)}], \mTVectorC_{\diamond \leftGroupAction((\EdOrthoref, \EdTrans), \eMat)})_2 \\
        \overset{\text{\cref{eq:natural-vertical-parallel}}}{=} (D_{\eMat}\MetricTensor_{\diamond \leftGroupAction((\EdOrthoref, \EdTrans), \cdot)}   (\leftGroupAction((\EdOrthoref, \mathbf{0}_\dimInd), \mTVectorB_{\eMat})) [ \mTVector_{\diamond \eMat}], \leftGroupAction((\EdOrthoref, \mathbf{0}_\dimInd),\mTVectorC_{\eMat}))_2\\
        \overset{\text{\cref{eq:equivariance-like-metric-tensor}}}{=} (\leftGroupAction((\EdOrthoref, \mathbf{0}_\dimInd),D_{\eMat}\MetricTensor_{\diamond (\cdot)}   (\mTVectorB_{\diamond \eMat}) [\mTVector_{\diamond \eMat}]),  \leftGroupAction((\EdOrthoref, \mathbf{0}_\dimInd),\mTVectorC_{\eMat}))_2\\
        \overset{\EdOrthoref\in \mathbb{O}(\dimInd)}{=} (D_{\eMat}\MetricTensor_{\diamond (\cdot)}   (\mTVectorB_{\diamond \eMat}) [\mTVector_{\diamond \eMat}], \mTVectorC_{\diamond \eMat})_2.
    \end{multline}
    In other words, the cotangent vector corresponding to the mapping $\mTVector_{\diamond \eMat}\mapsto (D_{\eMat}\MetricTensor_{\diamond (\cdot)}   (\mTVectorB_{\diamond \eMat}) [\mTVector_{\diamond \eMat}], \mTVectorC_{\diamond \eMat})_2$ satisfies \cref{eq:equivariance-like-one-form}. So we can use \cref{lem:rewrite-linear-mapping-as-tangent-vector} and rewrite the second term in the remainder of \cref{eq:lema-eval-cov-derivative-terms} as
    \begin{equation}
        (D_{\eMat}\MetricTensor_{\diamond (\cdot)}   (\mTVectorB_{\diamond \eMat}) [\mTVector_{\diamond \eMat}], \mTVectorC_{\diamond \eMat})_2 
        \overset{\cref{eq:tvector-representation-linear-functional}}{=} (\MetricTensor_{\diamond \eMat} (\MetricTensor_{\diamond \eMat}^\dagger (\bigl[(D_{\eMat}\MetricTensor_{\diamond (\cdot)}   (\mTVectorB_{\diamond \eMat}) [\mTVector_{\diamond \eMat}], \mathrm{e}^\sumIndA \otimes \mathrm{f}^\sumIndB)_2]_{\sumIndA,\sumIndB=1}^{\dimInd,\proteinLen})), \mTVectorC_{\diamond \eMat})_2.
        \label{eq:lema-eval-cov-derivative-terms-1}
    \end{equation}
    Similarly,
    \begin{equation}
        (D_{\eMat}\MetricTensor_{\diamond (\cdot)}   (\mTVector_{\diamond \eMat}) [\mTVectorB_{\diamond \eMat}], \mTVectorC_{\diamond \eMat})_2 \overset{\cref{eq:tvector-representation-linear-functional}}{=} (\MetricTensor_{\diamond \eMat} (\MetricTensor_{\diamond \eMat}^\dagger (\bigl[(D_{\eMat}\MetricTensor_{\diamond (\cdot)}   (\mTVector_{\diamond \eMat}) [\mTVectorB_{\diamond \eMat}], \mathrm{e}^\sumIndA \otimes \mathrm{f}^\sumIndB)_2]_{\sumIndA,\sumIndB=1}^{\dimInd,\proteinLen})), \mTVectorC_{\diamond \eMat})_2,
        \label{eq:lema-eval-cov-derivative-terms-2}
    \end{equation}
    and
    \begin{equation}
        (D_{\eMat}\MetricTensor_{\diamond (\cdot)}   (\mTVector_{\diamond \eMat}) [\mTVectorC_{\diamond \eMat}], \mTVectorB_{\diamond \eMat})_2 \overset{\cref{eq:tvector-representation-linear-functional}}{=}  (\MetricTensor_{\diamond \eMat} (\MetricTensor_{\diamond \eMat}^\dagger (\bigl[(D_{\eMat} \MetricTensor_{\diamond (\cdot)} (\mTVector_{\diamond \eMat})[\mathrm{e}^\sumIndA \otimes \mathrm{f}^\sumIndB]) ,\mTVectorB_{\diamond \eMat})_2 \bigr]_{\sumIndA,\sumIndB=1}^{\dimInd,\proteinLen})), \mTVectorC_{\diamond \eMat})_2.
        \label{eq:lema-eval-cov-derivative-terms-3}
    \end{equation}

    So
    \begin{multline}
        2 \MetricTensor_{\diamond \eMat} ((\nabla_{\mTVector} \mTVectorB)_{\diamond \eMat}, \mTVectorC_{\diamond \eMat}) = 2 (\nabla_{\mTVector} \mTVectorB, \mTVectorC)_{[\eMat]} 
        \overset{\cref{eq:lema-eval-cov-derivative-terms}}{=} 2  (\MetricTensor_{\diamond \eMat} (D_{\eMat}\mTVectorB_{\diamond (\cdot)} [\mTVector_{\diamond \eMat}]), \mTVectorC_{\diamond \eMat})_2 \\
        + (D_{\eMat}\MetricTensor_{\diamond (\cdot)}   (\mTVectorB_{\diamond \eMat}) [\mTVector_{\diamond \eMat}], \mTVectorC_{\diamond \eMat})_2 + (D_{\eMat}\MetricTensor_{\diamond (\cdot)}   (\mTVector_{\diamond \eMat}) [\mTVectorB_{\diamond \eMat}], \mTVectorC_{\diamond \eMat})_2 
         - (D_{\eMat}\MetricTensor_{\diamond (\cdot)}   (\mTVector_{\diamond \eMat}) [\mTVectorC_{\diamond \eMat}], \mTVectorB_{\diamond \eMat})_2\\
         \overset{\cref{eq:lema-eval-cov-derivative-terms-1,eq:lema-eval-cov-derivative-terms-2,eq:lema-eval-cov-derivative-terms-3}}{=}  2  (\MetricTensor_{\diamond \eMat} (D_{\eMat}\mTVectorB_{\diamond (\cdot)} [\mTVector_{\diamond \eMat}]), \mTVectorC_{\diamond \eMat})_2
         + (\MetricTensor_{\diamond \eMat} (\MetricTensor_{\diamond \eMat}^\dagger (\bigl[(D_{\eMat}\MetricTensor_{\diamond (\cdot)}   (\mTVectorB_{\diamond \eMat}) [\mTVector_{\diamond \eMat}], \mathrm{e}^\sumIndA \otimes \mathrm{f}^\sumIndB)_2]_{\sumIndA,\sumIndB=1}^{\dimInd,\proteinLen})), \mTVectorC_{\diamond \eMat})_2 \\
         + (\MetricTensor_{\diamond \eMat} (\MetricTensor_{\diamond \eMat}^\dagger (\bigl[(D_{\eMat}\MetricTensor_{\diamond (\cdot)}   (\mTVector_{\diamond \eMat}) [\mTVectorB_{\diamond \eMat}], \mathrm{e}^\sumIndA \otimes \mathrm{f}^\sumIndB)_2]_{\sumIndA,\sumIndB=1}^{\dimInd,\proteinLen})), \mTVectorC_{\diamond \eMat})_2
         \\
         - (\MetricTensor_{\diamond \eMat} (\MetricTensor_{\diamond \eMat}^\dagger (\bigl[(D_{\eMat} \MetricTensor_{\diamond (\cdot)} (\mTVector_{\diamond \eMat})[\mathrm{e}^\sumIndA \otimes \mathrm{f}^\sumIndB]) ,\mTVectorB_{\diamond \eMat})_2 \bigr]_{\sumIndA,\sumIndB=1}^{\dimInd,\proteinLen})), \mTVectorC_{\diamond \eMat})_2.
    \end{multline}
    Since $\mTVectorC_{\diamond \eMat}$ is arbitrary and  $\ker(\MetricTensor_{\diamond \eMat} ) = \vertical_{\eMat} \PCmanifold$, we conclude that
    \begin{multline}
        (\nabla_{\mTVector} \mTVectorB)_{\diamond \eMat} = D_{\eMat}(\mTVectorB_{\diamond (\cdot)}) [\mTVector_{\diamond \eMat}] + \frac{1}{2} \MetricTensor_{\diamond \eMat}^\dagger \Bigl( \bigl[(D_{\eMat}\MetricTensor_{\diamond (\cdot)}   (\mTVectorB_{\diamond \eMat}) [\mTVector_{\diamond \eMat}], \mathrm{e}^\sumIndA \otimes \mathrm{f}^\sumIndB)_2]_{\sumIndA,\sumIndB=1}^{\dimInd,\proteinLen} \\
        + \bigl[(D_{\eMat}\MetricTensor_{\diamond (\cdot)}   (\mTVector_{\diamond \eMat}) [\mTVectorB_{\diamond \eMat}], \mathrm{e}^\sumIndA \otimes \mathrm{f}^\sumIndB)_2]_{\sumIndA,\sumIndB=1}^{\dimInd,\proteinLen} 
        - \bigl[(D_{\eMat} \MetricTensor_{\diamond (\cdot)} (\mTVector_{\diamond \eMat})[\mathrm{e}^\sumIndA \otimes \mathrm{f}^\sumIndB]) ,\mTVectorB_{\diamond \eMat})_2 \bigr]_{\sumIndA,\sumIndB=1}^{\dimInd,\proteinLen}\Bigr),
    \end{multline}
    which proves the claim.
    
\end{proof}

\begin{lemma}
\label{lem:equivariance-separation-class}
    Consider a a mapping $\tilde{\separation}: \PCmanifold \times  \PCmanifold\to \Real$ for $\proteinLen \geq \dimInd+1$ that is invariant under $\mathbb{E}(\dimInd)$ action in both arguments. Additionally, assume that $\tilde{\separation}(\eMat,\cdot)^2$ is smooth in a neighbourhood of $\eMat$ for any $\eMat\in \PCmanifold$.
    
    Then, the Euclidean Hessian $D_{\eMat} \nabla \tilde{\separation}(\eMat, \cdot)^2 : \Real^{\dimInd\times \proteinLen} \to \Real^{\dimInd\times \proteinLen}$ satisfies \cref{eq:equivariance-like-metric-tensor}.
\end{lemma}

\begin{proof}
First note that
\begin{multline}
    (\nabla \tilde{\separation}(\eMat, \cdot)^2 (\eMat), \mTVector_{\diamond \eMat})_2 
    =  D_{\eMat} \tilde{\separation}(\eMat,  \cdot)^2 [\mTVector_{\diamond \eMat}]
    \overset{\mathbb{E}(\dimInd)-\text{invariance}}{=}  D_{\eMat} \tilde{\separation}(\eMat, \leftGroupAction((\EdOrthoref, \EdTrans), \cdot))^2 [\mTVector_{\diamond \eMat}]\\
    \overset{\text{chain rule}}{=} D_{\leftGroupAction((\EdOrthoref, \EdTrans),\eMat)} \tilde{\separation}(\eMat,  \cdot)^2 [\mTVector_{\diamond \leftGroupAction((\EdOrthoref, \EdTrans),\eMat)}] 
    = (\nabla \tilde{\separation}(\eMat, \cdot)^2 (\leftGroupAction((\EdOrthoref, \EdTrans),\eMat) ), \mTVector_{\diamond \leftGroupAction((\EdOrthoref, \EdTrans),\eMat)})_2.
\end{multline}
So we have
\begin{equation}
    \nabla \tilde{\separation}(\eMat, \cdot)^2 (\leftGroupAction((\EdOrthoref, \EdTrans),\eMat)) = \leftGroupAction((\EdOrthoref, \mathbf{0}_\dimInd), \nabla \tilde{\separation}(\eMat, \cdot)^2 (\eMat) ).
    \label{eq:lem-gradient-equivariance}
\end{equation}

For proving that $D_{\eMat} \nabla \tilde{\separation}(\eMat, \cdot)^2$ satisfies \cref{eq:equivariance-like-metric-tensor} we must to show that
\begin{equation}
    D_{\leftGroupAction((\EdOrthoref, \EdTrans),\eMat)} \nabla \tilde{\separation}(\eMat, \cdot)^2 [\mTVector_{\diamond \leftGroupAction((\EdOrthoref, \EdTrans),\eMat)}] = \leftGroupAction((\EdOrthoref, \mathbf{0}_\dimInd), D_{\eMat} \nabla \tilde{\separation}(\eMat, \cdot)^2 [\mTVector_{\diamond \eMat}]),
\end{equation}
which follows directly from \cref{eq:lem-gradient-equivariance}:
\begin{multline}
    D_{\leftGroupAction((\EdOrthoref, \EdTrans),\eMat)} \nabla \tilde{\separation}(\eMat, \cdot)^2 [\mTVector_{\diamond \leftGroupAction((\EdOrthoref, \EdTrans),\eMat)}] \overset{\text{\cref{eq:lem-gradient-equivariance}}}{=} D_{\eMat} \leftGroupAction((\EdOrthoref, \mathbf{0}_\dimInd), \nabla \tilde{\separation}(\eMat, \cdot)^2) [\mTVector_{\diamond \eMat}] \\
    \overset{\text{linearity}}{=}  \leftGroupAction((\EdOrthoref, \mathbf{0}_\dimInd), D_{\eMat}\nabla \tilde{\separation}(\eMat, \cdot)^2 [\mTVector_{\diamond \eMat}]).
\end{multline}
    
\end{proof}

\paragraph{Proof of the theorem}

\begin{proof}[Proof of \cref{thm:class-of-separations-on-pointcloud-manifold}]
    We need to show the defining properties of a separation. Since (i) and (ii) in \cref{def:separation} are satisfied by assumption, it remains to verify (iii) $(\nabla_{\mTVector} \Grad \separation([\eMat], \cdot)^2)_{\diamond \eMat} = 2 \mTVector_{\diamond \eMat}$. For notational convenience we write $\MetricTensor_{\diamond \eMat} := \frac{1}{2} D_{\eMat} \nabla \tilde{\separation}(\eMat, \cdot)^2$. 

    By definition of the Riemannian gradient
    \begin{equation}
        (\Grad \separation([\eMat], \cdot)^2, \mTVector)_{[\eMatB]} = D_{[\eMatB]} \separation([\eMat], \cdot)^2 [\mTVector_{[\eMatB]}] = D_{\eMatB} \tilde{\separation}(\eMat, \cdot)^2 [\mTVector_{\diamond \eMatB}], \quad \mTVector_{[\eMatB]} \in \tangent_{[\eMatB]} \PCmanifold /\mathbb{E}(\dimInd).
        \label{eq:riemann-gradient-def-quotient}
    \end{equation}
    We know that $\MetricTensor_{\diamond \eMat}$ is symmetric and that it satisfies \cref{eq:equivariance-like-metric-tensor} by \cref{lem:equivariance-separation-class}. Then, by \cref{lem:rewrite-linear-mapping-as-tangent-vector} we have that
    \begin{equation}
        D_{\eMatB} \tilde{\separation}(\eMat, \cdot)^2 [\mTVector_{\diamond \eMatB}] \overset{\cref{eq:tvector-representation-linear-functional}}{=} (\MetricTensor_{\diamond \eMatB}(\MetricTensor_{\diamond \eMatB}^\dagger (\bigl[ D_{\eMatB} \tilde{\separation}(\eMat, \cdot)^2 [\mathrm{e}^\sumIndA \otimes \mathrm{f}^\sumIndB ] \bigr]_{\sumIndA,\sumIndB=1}^{\dimInd,\proteinLen})), \mTVector_{\diamond \eMatB})_2 = (\MetricTensor_{\diamond \eMatB}(\MetricTensor_{\diamond \eMatB}^\dagger (\nabla \tilde{\separation}(\eMat, \cdot)^2 (\eMatB))), \mTVector_{\diamond \eMatB})_2,
        \label{eq:riemann-gradient-def-quotient-eval}
    \end{equation}
    and subsequently find that
    \begin{multline}
        (\MetricTensor_{\diamond \eMatB}((\Grad \separation([\eMat], \cdot)^2)_{\diamond \eMatB}), \mTVector_{\diamond \eMatB})_2 = (\Grad \separation([\eMat], \cdot)^2, \mTVector)_{[\eMatB]} \overset{\cref{eq:riemann-gradient-def-quotient}}{=} D_{\eMatB} \tilde{\separation}(\eMat, \cdot)^2 [\mTVector_{\diamond \eMatB}] \\
        \overset{\cref{eq:riemann-gradient-def-quotient-eval}}{=} (\MetricTensor_{\diamond \eMatB}(\MetricTensor_{\diamond \eMatB}^\dagger (\nabla \tilde{\separation}(\eMat, \cdot)^2 (\eMatB))), \mTVector_{\diamond \eMatB})_2.
    \end{multline}
    Since $\mTVector_{\diamond \eMatB}$ is arbitrary we conclude that
    \begin{equation}
        (\Grad \separation([\eMat], \cdot)^2)_{\diamond \eMatB} = \MetricTensor_{\diamond \eMatB}^\dagger (\nabla \tilde{\separation}(\eMat, \cdot)^2 (\eMatB)), \quad \eMat \in [\eMat].
        \label{eq:Riemannian-grad-separation-sq}
    \end{equation}
    Note that
    \begin{equation}
        (\Grad \separation([\eMat], \cdot)^2)_{\diamond \eMat} \overset{\cref{eq:Riemannian-grad-separation-sq}}{=} \MetricTensor_{\diamond \eMat}^\dagger (\nabla \tilde{\separation}(\eMat, \cdot)^2 (\eMat)) = 0,
        \label{eq:Riemannian-grad-separation-sq-at-X}
    \end{equation}
    as for the Euclidean gradient it holds that $\nabla \tilde{\separation}(\eMat, \cdot)^2 (\eMat) = 0$ due to first-order optimality conditions, since $\eMatB\mapsto\tilde{\separation}(\eMat, \eMatB)^2$ attains a global minimum at $\eMatB = \eMat$.

    The covariant derivative identity (iii) follows immediately using \cref{lem:cov-derivative}, which we can use because of the fact that $\MetricTensor_{\diamond \eMat}$ is symmetric and that it satisfies \cref{eq:equivariance-like-metric-tensor} by \cref{lem:equivariance-separation-class}:
    \begin{multline}
        (\nabla_{\mTVector} \Grad \separation([\eMat], \cdot)^2)_{\diamond \eMat} \\
        \overset{\cref{eq:lem-covariant-derivative}}{=} D_{\eMat}((\Grad \separation([\eMat], \cdot)^2)_{\diamond (\star)}) [\mTVector_{\diamond \eMat}] + \frac{1}{2} \MetricTensor_{\diamond \eMat}^\dagger \Bigl( \bigl[(D_{\eMat}\MetricTensor_{\diamond (\star)} ((\Grad \separation([\eMat], \cdot)^2)_{\diamond \eMat}) [\mTVector_{\diamond \eMat}], \mathrm{e}^\sumIndA \otimes \mathrm{f}^\sumIndB)_2]_{\sumIndA,\sumIndB=1}^{\dimInd,\proteinLen} \\
        + \bigl[(D_{\eMat}\MetricTensor_{\diamond (\star)}(\mTVector_{\diamond \eMat})  [(\Grad \separation([\eMat], \cdot)^2)_{\diamond \eMat}], \mathrm{e}^\sumIndA \otimes \mathrm{f}^\sumIndB)_2]_{\sumIndA,\sumIndB=1}^{\dimInd,\proteinLen}
        - \bigl[(D_{\eMat} \MetricTensor_{\diamond (\star)} (\mTVector_{\diamond \eMat})[\mathrm{e}^\sumIndA \otimes \mathrm{f}^\sumIndB] ,(\Grad \separation([\eMat], \cdot)^2)_{\diamond \eMat})_2 \bigr]_{\sumIndA,\sumIndB=1}^{\dimInd,\proteinLen}\Bigr)\\
        \overset{\text{\cref{eq:Riemannian-grad-separation-sq,eq:Riemannian-grad-separation-sq-at-X}}}{=}  D_{\eMat}(\MetricTensor_{\diamond (\star)}^\dagger (\nabla \tilde{\separation}(\eMat, \cdot)^2 (\star))) [\mTVector_{\diamond \eMat}] \\
        =  (D_{\eMat}\MetricTensor_{\diamond (\star)}^\dagger (\nabla \tilde{\separation}(\eMat, \cdot)^2 (\eMat))[\mTVector_{\diamond \eMat}])  + \MetricTensor_{\diamond \eMat}^\dagger (D_{\eMat}\nabla \tilde{\separation}(\eMat, \cdot)^2 (\mTVector_{\diamond \eMat}))\\
        \overset{\cref{eq:Riemannian-grad-separation-sq-at-X}}{=} \MetricTensor_{\diamond \eMat}^\dagger (2\MetricTensor_{\diamond \eMat} (\mTVector_{\diamond \eMat})) = 2 \mTVector_{\diamond \eMat}.
    \end{multline}
\end{proof}
\clearpage
\section{Proofs for the results from \cref{sec:riemannian-protein-geometry-separation}}
\label{app:proofs-chosen-Riemannian}

\subsection{Proof of \cref{thm:candidate-metric}}

\begin{proof}
    We prove the statement through checking the metric axioms.

    [positivity] By construction we have $\separation([\eMat], [\eMatB]) \geq 0$ for all $[\eMat], [\eMatB]\in \PCmanifold /\mathbb{E}(\dimInd)$ and we have equality if and only if 
    \begin{equation}
        \distance_{>0}(\|\ePoint_\sumIndA - \ePoint_\sumIndB\|_2, \|\ePointB_\sumIndA - \ePointB_\sumIndB\|_2) = 0, \quad \text{for all } \sumIndA= 1, \ldots, \proteinLen, \text{ and }\sumIndB = \sumIndA+1, \ldots, \proteinLen.
        \label{eq:thm-metric-candidate-positivity}
    \end{equation}
    Since $\distance_{>0}$ is a metric, the equality \cref{eq:thm-metric-candidate-positivity} can only hold if
    \begin{equation}
        \|\ePoint_\sumIndA - \ePoint_\sumIndB\|_2 =  \|\ePointB_\sumIndA - \ePointB_\sumIndB\|_2, \quad \text{for all } \sumIndA= 1, \ldots, \proteinLen, \text{ and } \sumIndB = \sumIndA+1, \ldots, \proteinLen,
        \label{eq:thm-metric-candidate-positivity-2}
    \end{equation}
    which is equivalent to $\eMat$ and $\eMatB$ being equal up to Euclidean group actions \cite{dokmanic2015euclidean}. In other words, we must have that $\separation([\eMat], [\eMatB]) = 0$ implies $[\eMat] = [\eMatB]$, which proves positivity.

    [symmetry] Symmetry of $\separation$ follows from symmetry of $\distance_{>0}$.

    [triangle inequality] Finally, triangle inequality also follows 
    \begin{multline}
        \separation([\eMat], [\eMatB])= \sqrt{\sum_{\sumIndA} \sum_{\sumIndB > \sumIndA} \distance_{>0}(\|\ePoint_\sumIndA - \ePoint_\sumIndB\|_2, \|\ePointB_\sumIndA - \ePointB_\sumIndB\|_2)^2} \\
        \overset{\substack{\sqrt{\cdot}\text{ increasing \&} \\ \text{triangle inequality}}}{\leq} \sqrt{\sum_{\sumIndA} \sum_{\sumIndB > \sumIndA} (\distance_{>0}(\|\ePoint_\sumIndA - \ePoint_\sumIndB\|_2, \|\ePointC_\sumIndA - \ePointC_\sumIndB\|_2) + \distance_{>0}(\|\ePointC_\sumIndA - \ePointC_\sumIndB\|_2, \|\ePointB_\sumIndA - \ePointB_\sumIndB\|_2))^2}\\
        \overset{\text{Minkowski's inequality}}{\leq} \sqrt{\sum_{\sumIndA} \sum_{\sumIndB > \sumIndA} \distance_{>0}(\|\ePoint_\sumIndA - \ePoint_\sumIndB\|_2, \|\ePointC_\sumIndA - \ePointC_\sumIndB\|_2)^2} + \sqrt{\sum_{\sumIndA} \sum_{\sumIndB > \sumIndA}  \distance_{>0}(\|\ePointC_\sumIndA - \ePointC_\sumIndB\|_2, \|\ePointB_\sumIndA - \ePointB_\sumIndB\|_2)^2}\\
        = \separation([\eMat], [\eMatC]) + \separation([\eMatC], [\eMatB]).
    \end{multline}
    
    % [uses Minkowski's Inequality for Sums]
\end{proof}

\subsection{Proof of \cref{thm:protein-geometry-tensor-separation}}

% It is easy to check that the quotient manifold $\PCmanifold /\mathbb{E}(\dimInd)$ has a vertical space
% \begin{equation}
%     \vertical_{\eMat} \PCmanifold = \{ \EdOrthorefTV \eMat + \EdTransTV \mathbf{1}_{\proteinLen}^\top \in \Real^{\dimInd\times\proteinLen}\mid \EdOrthorefTV^\top = - \EdOrthorefTV \in \Real^{\dimInd\times \dimInd} ,\; \EdTransTV\in \Real^\dimInd\}
%     \label{eq:vertical-space}
% \end{equation}
% and horizontal space
% \begin{align}
%     \horizontal_{\eMat} \PCmanifold = \{\mTVector_{\diamond\eMat}:= (\mTVectorCompon_1,\ldots,\mTVectorCompon_\proteinLen) \in \Real^{\dimInd\times \proteinLen} \mid \sum_{\sumIndA}\mTVectorCompon_\sumIndA = 0, \: \sum_{\sumIndA}(\EdOrthorefTV \ePoint_\sumIndA  \mTVectorCompon_\sumIndA)_2=0 \; \forall \; \EdOrthorefTV\in \Real^{\dimInd\times \dimInd} \; s.t. \; \EdOrthorefTV^\top = - \EdOrthorefTV \} 
%     \label{eq:horizontal-space}
% \end{align}

% \paragraph{Proof of the theorem}

\begin{proof}
The metric tensor field \cref{thm:candidate-metric-tensor-field} is the Euclidean Hessian of the mapping $\eMatB\mapsto \tilde{\separation}^\GyRaParam (\eMat, \eMatB)^2$ at $\eMat$, i.e.,
\begin{equation}
    \pwdMetricTensor_{\diamond\eMat} + \GyRaParam \MetricTensorCorr_{\diamond\eMat} = \frac{1}{2} D_{\eMat} \nabla \tilde{\separation}^\GyRaParam (\eMat, \cdot)^2,
    \label{eq:thm-candidate-metric-tensor-field-as-hessian}
\end{equation}
where $\tilde{\separation}^\GyRaParam :\PCmanifold\times \PCmanifold\to \Real$ is given by
\begin{equation}
    \tilde{\separation}^\GyRaParam (\eMat, \eMatB) := \sqrt{\sum_{\sumIndA} \sum_{\sumIndB > \sumIndA} \Bigl(\log \Bigl(\frac{\|\ePoint_\sumIndA - \ePoint_\sumIndB\|_2}{\|\ePointB_\sumIndA - \ePointB_\sumIndB\|_2} \Bigr)\Bigr)^2 + \GyRaParam \Bigl(\log \Bigl(\frac{\det(\sum_{\sumIndA} (\ePoint_\sumIndA - \frac{1}{\proteinLen} \eMat \mathbf{1}_\proteinLen) \otimes (\ePoint_\sumIndA - \frac{1}{\proteinLen} \eMat \mathbf{1}_\proteinLen))}{\det(\sum_{\sumIndA} (\ePointB_\sumIndA - \frac{1}{\proteinLen} \eMatB \mathbf{1}_\proteinLen) \otimes (\ePointB_\sumIndA - \frac{1}{\proteinLen} \eMatB \mathbf{1}_\proteinLen))} \Bigr)\Bigr)^2}.
\end{equation}
We will prove the claim through invoking \cref{thm:class-of-separations-on-pointcloud-manifold}. Since $\tilde{\separation}^\GyRaParam$ has the right form with the appropriate symmetry properties and satisfies the smoothness property, it remains to show positive definiteness of its Euclidean Hessian when restricting ourselves to the horizontal space $\horizontal_{\eMat} \PCmanifold$ and that its kernel is $\vertical_{\eMat} \PCmanifold$.

% In other words, the left hand side in \cref{eq:thm-candidate-metric-tensor-field-as-hessian} is a symmetric positive semidefinite operator on $\Real^{\dimInd\times \proteinLen}$, since $\eMatB\mapsto \tilde{\separation}^\GyRaParam (\eMat, \eMatB)^2$ attains a (global) minimum at $\eMat$. So it is sufficient to show positive definiteness when restricting ourselves to the horizontal space $\horizontal_{\eMat} \Real^{\dimInd\times \proteinLen}_{\dimInd, \star, *}$ and that its kernel is $\vertical_{\eMat} \Real^{\dimInd\times \proteinLen}_{\dimInd, \star, *}$.

[\emph{positive definiteness}] To simplify the proof, we can restrict ourselves to showing that only $\pwdMetricTensor_{\diamond\eMat}$ is positive definite on $\horizontal_{\eMat} \PCmanifold$. Indeed, both terms in $\eMatB\mapsto \tilde{\separation}^\GyRaParam (\eMat, \eMatB)^2$ attain their global minimum at $\eMat$, so we have positive semidefiniteness for both $\pwdMetricTensor_{\diamond\eMat}$ and $\MetricTensorCorr_{\diamond\eMat}$ by the second-order optimality condition.

For showing positive definiteness of $\pwdMetricTensor_{\diamond\eMat}$, we rewrite the bi-linear operator in a more convenient form, i.e., it is straightforward to check that
\begin{equation}
    (\pwdMetricTensor_{\diamond\eMat}(\mTVector_{\diamond \eMat}),\mTVector_{\diamond \eMat})_2 =  \sum_{\sumIndA} \sum_{\sumIndB > \sumIndA} \Bigl(\frac{(\ePoint_\sumIndA- \ePoint_\sumIndB, \mTVectorCompon_\sumIndA - \mTVectorCompon_\sumIndB)_2}{\|\ePoint_\sumIndA- \ePoint_\sumIndB\|_2^2} \Bigr)^2, \quad \mTVector_{\diamond\eMat}:= (\mTVectorCompon_1,\ldots,\mTVectorCompon_\proteinLen).
    \label{eq:thm-metric-tensor-field-convenient-A}
\end{equation}
From \cref{eq:thm-metric-tensor-field-convenient-A} we find that
\begin{equation}
    (\pwdMetricTensor_{\diamond\eMat}(\mTVector_{\diamond \eMat}),\mTVector_{\diamond \eMat})_2 = 0 \Leftrightarrow (\ePoint_\sumIndA- \ePoint_\sumIndB, \mTVectorCompon_\sumIndA - \mTVectorCompon_\sumIndB)_2 = 0, \text{ for all } \sumIndA= 1, \ldots, \proteinLen, \text{ and }\sumIndB = \sumIndA+1, \ldots, \proteinLen.
    \label{eq:thm-equivalence-all-zero-A}
\end{equation}
In other words, if we can show that
\begin{equation}
    (\ePoint_\sumIndA- \ePoint_\sumIndB, \mTVectorCompon_\sumIndA - \mTVectorCompon_\sumIndB)_2 = 0  \text{ for all } \sumIndA= 1, \ldots, \proteinLen, \text{ and }\sumIndB = \sumIndA+1, \ldots, \proteinLen \Rightarrow \mTVector_{\diamond \eMat} \in \vertical_{\eMat} \PCmanifold,
    \label{eq:thm-candidate-tensor-field-implication-main}
\end{equation}
we are done. The remainder of this proof will be showing the implication \cref{eq:thm-candidate-tensor-field-implication-main}.

For proving \cref{eq:thm-candidate-tensor-field-implication-main} we assume that the left-hand side in \cref{eq:thm-candidate-tensor-field-implication-main} holds and that $\mTVector_{\diamond \eMat} \in \horizontal_{\eMat} \PCmanifold$ is non-zero, and show that this results in a contradiction. 

Consider the family of functions $F^{\sumIndA,\sumIndB}:\PCmanifold \to \Real$ given by 
\begin{equation}
    F^{\sumIndA,\sumIndB}(\eMat):= (\ePoint_\sumIndA - \ePoint_1, \ePoint_\sumIndB - \ePoint_1)_2, \quad \text{for } \sumIndA= 1, \ldots, \proteinLen, \text{ and }\sumIndB = \sumIndA+1, \ldots, \proteinLen,
\end{equation}
and the action of $\mTVector_{\diamond \eMat}$ on each $F^{\sumIndA,\sumIndB}(\eMat)$. That is,
\begin{multline}
    \mTVector_{\diamond \eMat} F^{\sumIndA,\sumIndB} = D_{\eMat} F(\cdot) [\mTVector_{\diamond \eMat}] = \frac{\mathrm{d}}{\mathrm{d} \epsilon} F(\eMat + \epsilon \mTVector_{\diamond \eMat}) \mid_{\epsilon=0} = \frac{\mathrm{d}}{\mathrm{d} \epsilon} (\ePoint_\sumIndA - \ePoint_1 + \epsilon(\mTVectorCompon_\sumIndA - \mTVectorCompon_1), \ePoint_\sumIndB - \ePoint_1 + \epsilon(\mTVectorCompon_\sumIndB - \mTVectorCompon_1))_2 \mid_{\epsilon =0} \\
    = (\mTVectorCompon_\sumIndA - \mTVectorCompon_1, \ePoint_\sumIndB - \ePoint_1 )_2 + (\ePoint_\sumIndA - \ePoint_1, \mTVectorCompon_\sumIndB - \mTVectorCompon_1)_2 \overset{\text{\cref{eq:thm-candidate-tensor-field-implication-main}}}{=} (\mTVectorCompon_\sumIndA - \mTVectorCompon_1, \ePoint_\sumIndB - \ePoint_1 )_2 + (\ePoint_\sumIndA - \ePoint_1, \mTVectorCompon_\sumIndB - \mTVectorCompon_1)_2 + (\ePoint_\sumIndA- \ePoint_\sumIndB, \mTVectorCompon_\sumIndA - \mTVectorCompon_\sumIndB)_2 \\
    = (\mTVectorCompon_\sumIndA - \mTVectorCompon_1, \ePoint_\sumIndB - \ePoint_1 )_2 + (\ePoint_\sumIndA - \ePoint_1, \mTVectorCompon_\sumIndB - \mTVectorCompon_1)_2 + (\ePoint_\sumIndA- \ePoint_1 + \ePoint_1 - \ePoint_\sumIndB, \mTVectorCompon_\sumIndA - \mTVectorCompon_1 + \mTVectorCompon_1 - \mTVectorCompon_\sumIndB)_2\\
    = (\mTVectorCompon_\sumIndA - \mTVectorCompon_1, \ePoint_\sumIndB - \ePoint_1 )_2 + (\ePoint_\sumIndA - \ePoint_1, \mTVectorCompon_\sumIndB - \mTVectorCompon_1)_2 + (\ePoint_\sumIndA- \ePoint_1, \mTVectorCompon_\sumIndA - \mTVectorCompon_1)_2 + ( \ePoint_1 - \ePoint_\sumIndB, \mTVectorCompon_\sumIndA - \mTVectorCompon_1 )_2 + (\ePoint_\sumIndA- \ePoint_1 , \mTVectorCompon_1 - \mTVectorCompon_\sumIndB)_2 + (\ePoint_1 - \ePoint_\sumIndB, \mTVectorCompon_1 - \mTVectorCompon_\sumIndB)_2\\
    \overset{\text{\cref{eq:thm-candidate-tensor-field-implication-main}}}{=}
    (\mTVectorCompon_\sumIndA - \mTVectorCompon_1, \ePoint_\sumIndB - \ePoint_1 )_2 + (\ePoint_\sumIndA - \ePoint_1, \mTVectorCompon_\sumIndB - \mTVectorCompon_1)_2 + ( \ePoint_1 - \ePoint_\sumIndB, \mTVectorCompon_\sumIndA - \mTVectorCompon_1 )_2 + (\ePoint_\sumIndA- \ePoint_1 , \mTVectorCompon_1 - \mTVectorCompon_\sumIndB)_2 \\
    = (\ePoint_\sumIndB - \ePoint_1 , \mTVectorCompon_\sumIndA - \mTVectorCompon_1)_2 + (\ePoint_\sumIndA - \ePoint_1, \mTVectorCompon_\sumIndB - \mTVectorCompon_1)_2 - ( \ePoint_\sumIndB - \ePoint_1, \mTVectorCompon_\sumIndA - \mTVectorCompon_1 )_2 - (\ePoint_\sumIndA- \ePoint_1 , \mTVectorCompon_\sumIndB - \mTVectorCompon_1)_2 = 0.
\end{multline}
In other words, the tangent vector $\mTVector_{\diamond \eMat} - \mTVectorCompon_1 \mathbf{1}_\proteinLen^\top$ is infinitesimally angle-preserving. Next, due to the rank constraint on $\eMat$ we must have that the dimension of $\operatorname{span}\{\ePoint_\sumIndA - \ePoint_1\}_{\sumIndA=2}^\proteinLen$ is $\dimInd$, from which we conclude that $\mTVector_{\diamond \eMat} - \mTVectorCompon_1 \mathbf{1}_\proteinLen^\top$ must be an infinitesimal orthogonal transformation, i.e., $\mTVector_{\diamond \eMat} - \mTVectorCompon_1 \mathbf{1}_\proteinLen^\top \in \vertical_{\eMat - \ePoint_1 \mathbf{1}_\proteinLen^\top} \PCmanifold$. In other words, $\mTVector_{\diamond \eMat} - \mTVectorCompon_1 \mathbf{1}_\proteinLen^\top$ has the form
\begin{equation}
    \mTVector_{\diamond \eMat} - \mTVectorCompon_1 \mathbf{1}_\proteinLen^\top =  \EdOrthorefTV (\eMat - \ePoint_1 \mathbf{1}_\proteinLen^\top) + \EdTransTV \mathbf{1}_{\proteinLen}^\top, \quad \text{for some } \EdOrthorefTV\in \Real^{\dimInd\times \dimInd} \text{ such that } \EdOrthorefTV^\top = - \EdOrthorefTV, \text{ and } \EdTransTV\in \Real^\dimInd.
    \label{eq:thm-form-xi-xi1}
\end{equation}
Considering the first column on both sides of \cref{eq:thm-form-xi-xi1}, we see that
\begin{equation}
    \mathbf{0}_{\dimInd} = \mTVectorCompon_1 - \mTVectorCompon_1 = \EdOrthorefTV(\ePoint_1 - \ePoint_1) + \EdTransTV = \EdTransTV.
\end{equation}
From hereon it is easily verified that
\begin{equation}
    \mTVectorCompon_\sumIndA - \mTVectorCompon_\sumIndB = \EdOrthorefTV(\ePoint_\sumIndA- \ePoint_\sumIndB), \quad  \text{ for all } \sumIndA, \sumIndB= 1, \ldots, \proteinLen.
    \label{eq:thm-candidatate-tensor-all-same-S}
\end{equation}
% [different cases for d=1 and all other cases]

Finally, combining the equalities \cref{eq:thm-candidatate-tensor-all-same-S} will give us our contradiction. Indeed, for every $\sumIndA = 1, \ldots, \proteinLen$
\begin{multline}
    \sum_{\sumIndB \neq \sumIndA}\mTVectorCompon_\sumIndA - \mTVectorCompon_\sumIndB = \sum_{\sumIndB \neq \sumIndA}\EdOrthorefTV(\ePoint_\sumIndA- \ePoint_\sumIndB) \Leftrightarrow (\proteinLen - 1)\mTVectorCompon_\sumIndA - \sum_{\sumIndB \neq \sumIndA}\mTVectorCompon_\sumIndB = (\proteinLen-1)\EdOrthorefTV\ePoint_\sumIndA - \EdOrthorefTV\sum_{\sumIndB \neq \sumIndA} \ePoint_\sumIndB\\
    \overset{\mTVector_{\diamond \eMat} \in \horizontal_{\eMat} \PCmanifold}{\Leftrightarrow} (\proteinLen - 1)\mTVectorCompon_\sumIndA - (- \mTVectorCompon_\sumIndA) = (\proteinLen-1)\EdOrthorefTV\ePoint_\sumIndA - \EdOrthorefTV\Bigl(\sum_\sumIndB \ePoint_\sumIndB - \ePoint_\sumIndA \Bigr)\\
    \Leftrightarrow \proteinLen \mTVectorCompon_\sumIndA  = \proteinLen \EdOrthorefTV\ePoint_\sumIndA - \proteinLen\EdOrthorefTV\Bigl(\frac{1}{\proteinLen}\sum_\sumIndC \ePoint_\sumIndC\Bigr) 
    \Leftrightarrow \mTVectorCompon_\sumIndA  = \EdOrthorefTV\ePoint_\sumIndA - \EdOrthorefTV\Bigl(\frac{1}{\proteinLen}\sum_\sumIndB \ePoint_\sumIndB\Bigr) \quad \Rightarrow \quad \mTVector_{\diamond \eMat} \in \vertical_{\eMat} \PCmanifold,
\end{multline}
which contradicts our assumption that $\mTVector_{\diamond \eMat} \in \horizontal_{\eMat} \PCmanifold$ and is non-zero. We conclude that $\mTVector_{\diamond \eMat} \in \vertical_{\eMat} \PCmanifold$.

Summarized, we have found that \cref{eq:thm-candidate-tensor-field-implication-main} holds, which proves positive definiteness of $\pwdMetricTensor_{\diamond\eMat}$ and gives well-definedness of the metric tensor field \cref{thm:candidate-metric-tensor-field}.

[\emph{vertical space kernel}] 
We choose any $\mTVector_{\diamond \eMat} = \EdOrthorefTV \eMat + \EdTransTV \mathbf{1}_{\proteinLen}^\top \in \vertical_{\eMat} \PCmanifold$ for some $\EdOrthorefTV \in \Real^{\dimInd\times \dimInd}$ and $\EdTransTV \in \Real^{\dimInd}$. Then, \cref{eq:thm-equivalence-all-zero-A} gives us directly that $\mTVector_{\diamond \eMat} \in \ker(\pwdMetricTensor_{\diamond\eMat})$. Next, for showing that $\mTVector_{\diamond \eMat} \in \ker(\MetricTensorCorr_{\diamond\eMat})$ also holds, we rewrite the vector in a more convenient form $\mTVector_{\diamond \eMat} = \EdOrthorefTV (\eMat - \frac{1}{\proteinLen} \eMat \mathbf{1}_\proteinLen \mathbf{1}_\proteinLen^\top) + \tilde{\EdTransTV} \mathbf{1}_{\proteinLen}^\top$, where $\tilde{\EdTransTV} := \EdTransTV +\frac{1}{\proteinLen} \EdOrthorefTV \eMat \mathbf{1}_\proteinLen$. The claim now follows from directly evaluating the bi-linear form

\begin{equation}
    (\MetricTensorCorr_{\diamond\eMat}(\mTVector_{\diamond \eMat}),\mTVector_{\diamond \eMat})_2 =  (\bigl((\eMat - \frac{1}{\proteinLen} \eMat \mathbf{1}_\proteinLen\mathbf{1}_\proteinLen^\top) (\eMat - \frac{1}{\proteinLen} \eMat \mathbf{1}_\proteinLen\mathbf{1}_\proteinLen^\top)^\top \bigr)^{-1} (\eMat - \frac{1}{\proteinLen} \eMat \mathbf{1}_\proteinLen \mathbf{1}_\proteinLen^\top), \mTVector_{\diamond \eMat})_2^2,
\end{equation}
and realizing that
\begin{multline}
     (\bigl((\eMat - \frac{1}{\proteinLen} \eMat \mathbf{1}_\proteinLen\mathbf{1}_\proteinLen^\top) (\eMat - \frac{1}{\proteinLen} \eMat \mathbf{1}_\proteinLen\mathbf{1}_\proteinLen^\top)^\top \bigr)^{-1} (\eMat - \frac{1}{\proteinLen} \eMat \mathbf{1}_\proteinLen \mathbf{1}_\proteinLen^\top), \mTVector_{\diamond \eMat})_2 \\
     =  (\bigl((\eMat - \frac{1}{\proteinLen} \eMat \mathbf{1}_\proteinLen\mathbf{1}_\proteinLen^\top) (\eMat - \frac{1}{\proteinLen} \eMat \mathbf{1}_\proteinLen\mathbf{1}_\proteinLen^\top)^\top \bigr)^{-1} (\eMat - \frac{1}{\proteinLen} \eMat \mathbf{1}_\proteinLen \mathbf{1}_\proteinLen^\top), \EdOrthorefTV (\eMat - \frac{1}{\proteinLen} \eMat \mathbf{1}_\proteinLen \mathbf{1}_\proteinLen^\top) + \tilde{\EdTransTV} \mathbf{1}_{\proteinLen}^\top)_2\\
     = \trace \Bigl( \bigl((\eMat - \frac{1}{\proteinLen} \eMat \mathbf{1}_\proteinLen\mathbf{1}_\proteinLen^\top) (\eMat - \frac{1}{\proteinLen} \eMat \mathbf{1}_\proteinLen\mathbf{1}_\proteinLen^\top)^\top \bigr)^{-1} (\eMat - \frac{1}{\proteinLen} \eMat \mathbf{1}_\proteinLen \mathbf{1}_\proteinLen^\top) ((\eMat - \frac{1}{\proteinLen} \eMat \mathbf{1}_\proteinLen \mathbf{1}_\proteinLen^\top)^\top \EdOrthorefTV^\top + \mathbf{1}_{\proteinLen}\tilde{\EdTransTV}^\top )\Bigr) \\
     = \trace ( \mathbf{I}_{\dimInd} \EdOrthorefTV^\top) = \trace (\EdOrthorefTV^\top) = 0,
\end{multline}
since $\EdOrthorefTV$ has an all-zero diagonal. So we find that $\mTVector_{\diamond \eMat} \in \ker(\MetricTensorCorr_{\diamond\eMat})$ and conclude that $\mTVector_{\diamond \eMat} \in \ker(\pwdMetricTensor_{\diamond\eMat} + \GyRaParam \MetricTensorCorr_{\diamond\eMat})$ as claimed.

% [At some point the proof will rely on for any two points there being a third point that is not co-linear with the difference of the two points. We get this from the fixed rank assumption ]

\end{proof}

\clearpage
\section{Supplementary material to \cref{sec:numerics-protein-geometry}}

\subsection{A note on approximately preserving adjacent $\mathrm{C}_\alpha$ distances}
\label{app:adjacent-distances}

\Cref{rem:nice-interpolation} argues that small invariant distances in the data set should stay -- reasonably -- invariant when interpolating through $\separation^\GyRaParam$-geodesics. One can argue similarly for summary statistics such as the $\separation^\GyRaParam$-barycentre. To test this, we will consider the adjacent distances at the $\separation^\GyRaParam$-geodesic midpoints from \cref{sec:numerics-protein-geometry-geodesics} and the $\separation^\GyRaParam$-barycentres from \cref{sec:numerics-protein-geometry-log-low-rank}, i.e., for both the adenylate kinase data set and the SARS-CoV-2 helicase nsp 13 data set.

% \subsubsection{Adjacent pairwise distances of begin-to-end $\separation^\GyRaParam$-geodesics}

\paragraph{Adenylate kinase.}

\begin{figure}[h!]
    \centering
    \begin{subfigure}{\linewidth}
        \includegraphics[width=\linewidth]{"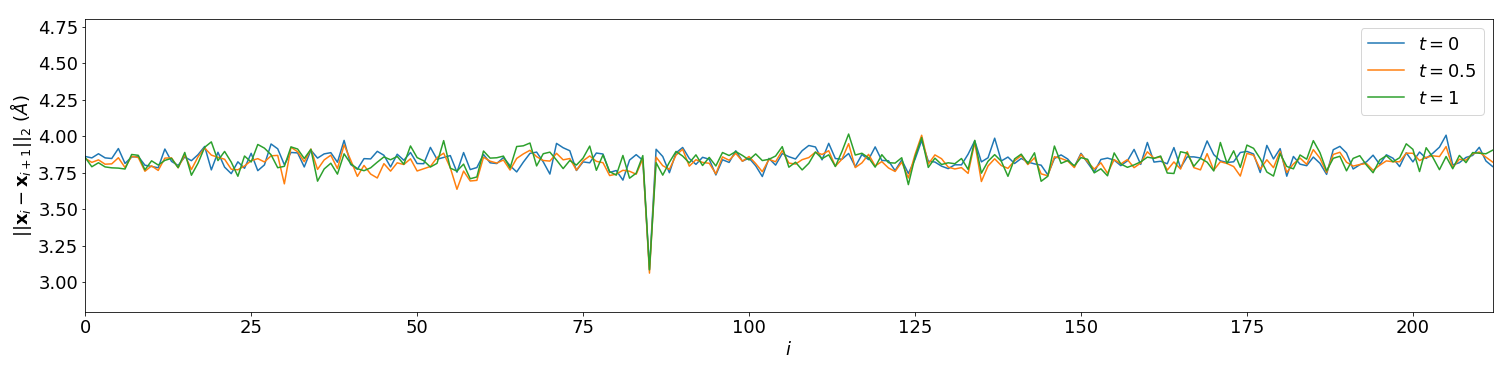"}
        \includegraphics[width=\linewidth]{"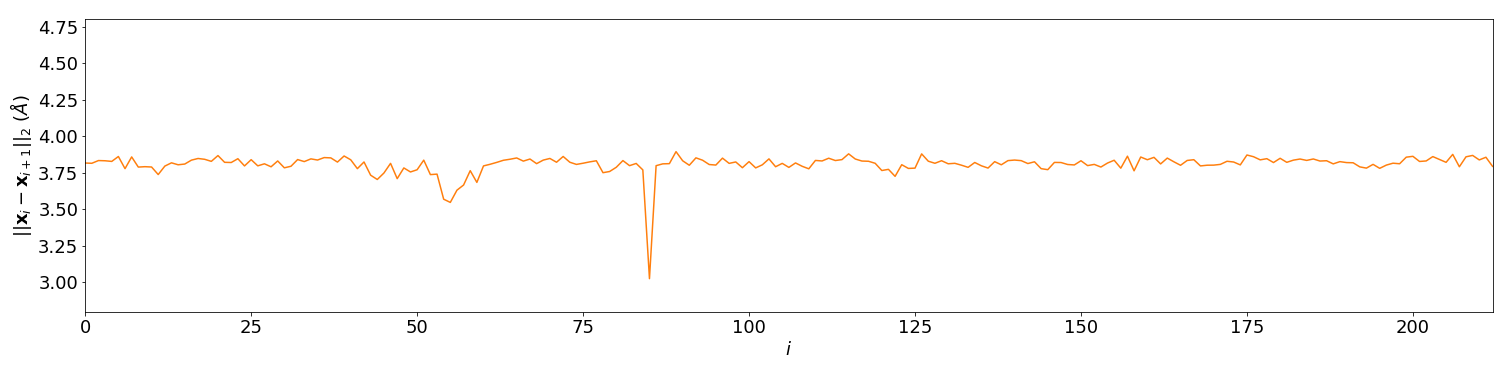"}
    \end{subfigure}
    \caption{The adjacent $\mathrm{C}_\alpha$ distances for the adenylate kinase $\separation^\GyRaParam$-geodesic midpoint (top) and $\separation^\GyRaParam$-barycentre (bottom) are more or less preserved as predicted by \cref{rem:nice-interpolation}. The small dip after atom 50 in the $\separation^\GyRaParam$-barycentre plot indicates a discrepancy between the energy landscape-based Riemannian structure and the actual physical potential, which is stronger in the short-range.}
    \label{fig:pwd-4ake}
\end{figure}

The adjacent $\mathrm{C}_\alpha$ distances for the adenylate kinase $\separation^\GyRaParam$-geodesic midpoint and $\separation^\GyRaParam$-barycentre are shown in \cref{fig:pwd-4ake}. 

Overall, the results are in line with \cref{rem:nice-interpolation}, i.e., small invariant distances stay more or less invariant. The only -- even though small -- discrepancy is the dip in the $\separation^\GyRaParam$-barycentre plot just after atom 50, which indicates that a larger deformation somewhere else in the protein outweighed keeping this distance invariant. In turn, this indicates -- as also pointed out in \cref{sec:numerics-protein-geometry-geodesics} -- that the discrepancy between Lennard-Jones potential and $\separation^\GyRaParam$ is also noticeable for medium-range deformations, meaning once again that the interactions modeled by $\separation^\GyRaParam([\eMat^{AK,0}], \cdot)$ for the short-range is not strong enough compared to long-range interactions.

% [does as expected, but just after atom 50 there is this dip. This indicates that a larger deformation somewhere outweighed [having to keep this invariant]. Which in turn indicates -- as also pointed out in \cref{sec:numerics-protein-geometry-geodesics} -- that [the discrepancy between Lennard-Jones and metric -- the metric for low distances is not strong enough compared to long-range.]

% \begin{figure}[h!]
%     \centering
%     \begin{subfigure}{\linewidth}
%         \includegraphics[width=\linewidth]{"experiments/4ake/4ake_adjacent_residue_5f20.png"}
%     % \caption{$t=0.25$}
    
%         \includegraphics[width=\linewidth]{"experiments/4ake/4ake_adjacent_residue_10f20.png"}
%     % \caption{$t=0.5$}
    
%         \includegraphics[width=\linewidth]{"experiments/4ake/4ake_adjacent_residue_15f20.png"}
%     % \caption{$t=0.75$}
%     \end{subfigure}
%     \caption{}
%     \label{fig:enter-label}
% \end{figure}

% \begin{figure}[h!]
%     \centering
%     \begin{subfigure}{\linewidth}
%         \includegraphics[width=\linewidth]{"experiments/covid_spike/covid_spike_adjacent_residue_5f20.png"}
%         \includegraphics[width=\linewidth]{"experiments/covid_spike/covid_spike_adjacent_residue_10f20.png"}
%         \includegraphics[width=\linewidth]{"experiments/covid_spike/covid_spike_adjacent_residue_15f20.png"}
%     \end{subfigure}
%     \caption{$t=0.25$}
%     \label{fig:enter-label}
% \end{figure}

\begin{figure}[h!]
    \centering
    \begin{subfigure}{\linewidth}
        \includegraphics[width=\linewidth]{"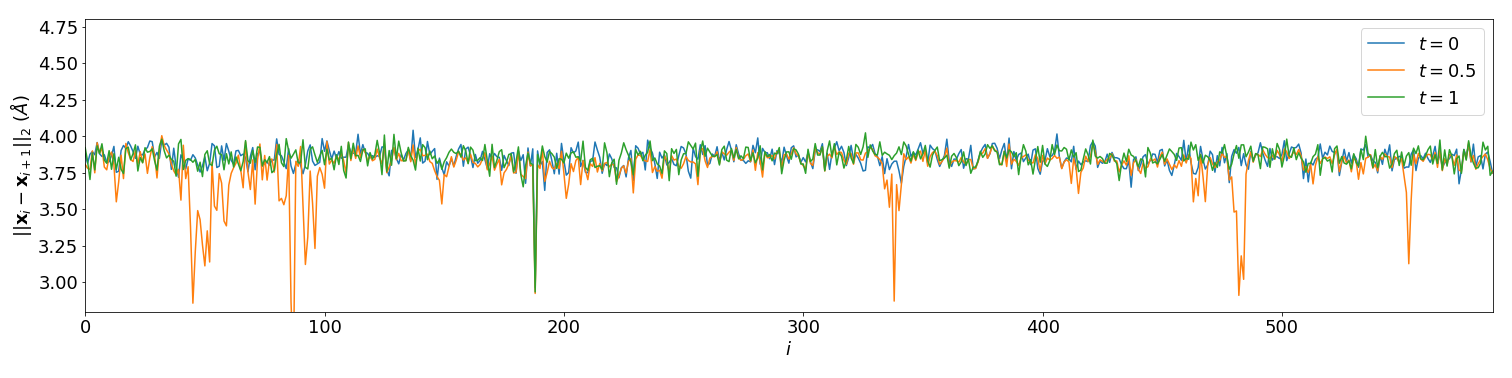"}
        \includegraphics[width=\linewidth]{"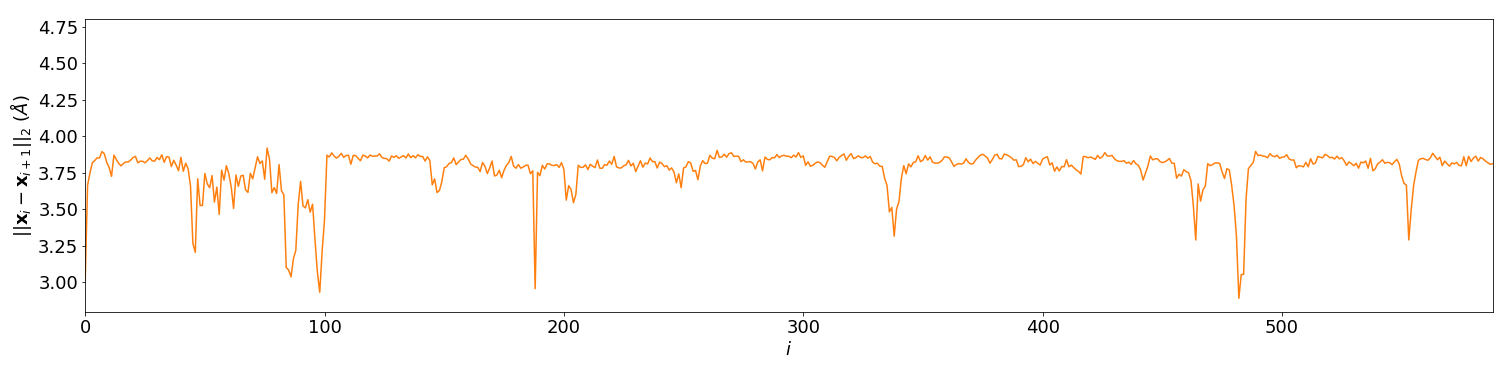"}
    \end{subfigure}
    \caption{The majority of the adjacent $\mathrm{C}_\alpha$ distances for the SARS-CoV-2 helicase nsp 13 $\separation^\GyRaParam$-geodesic midpoint (top) and $\separation^\GyRaParam$-barycentre (bottom) are preserved as predicted by \cref{rem:nice-interpolation}, but there are several zones of degradation. This again indicates a discrepancy between the energy landscape-based Riemannian structure and the actual physical potential, which is stronger in the short-range.}
    \label{fig:pwd-covid}
\end{figure}

\paragraph{SARS-CoV-2 helicase nsp 13.}
The results for SARS-CoV-2 in \cref{fig:pwd-covid} give additional support to the observations made above for the adenylate kinase case. That is, we know that there are larger displacements in this data sets than before. So we would expect that both the $\separation^\GyRaParam$-geodesics and the $\separation^\GyRaParam$-barycentre will show similar, but more intense degradation than in the previous case, which is exactly what happens.

Overall, we highlight that the effects are minor and manageable. In other words, \cref{rem:nice-interpolation} is expected to give good intuition.

% [results are in line with what we have seen in ade [that is. We know that there are larger displacements in this data sets than before. So we would expect that both the geodesics and the barycentre will show similar, but more intense behaviour as in the previous case, which is exactly what happens.]]

\subsection{A note on curvature effects}
\label{app:numerics-protein-geometry-curvature-effects}
Despite the very encouraging results described in \cref{sec:numerics-protein-geometry-geodesics,sec:numerics-protein-geometry-log-low-rank}, \cref{rem:curvature-proteins-low-rank,rem:curvature-proteins-geo} tell us to be cautious to draw any conclusions because of potential curvature effects. To test the severity of the curvature effects in the $\ell^2$-distance we consider the \emph{discrepancy from zero-curvature baselines} of (i) a $\separation^\GyRaParam$-geodesic variation at the begin and end points of the curves from \cref{sec:numerics-protein-geometry-geodesics} and (ii) low rank approximations from \cref{sec:numerics-protein-geometry-log-low-rank}. That is, we use for varying $t$
\begin{equation}
    % \epsilon_{rel}^{\gamma,0} (t) := 
    \frac{\operatorname{RMSD}(\geodesic_{[\eMatC_0],[\eMatB]}(t),\geodesic_{[\eMat],[\eMatB]}(t)) - t \cdot \operatorname{RMSD}(\geodesic_{[\eMatC_0],[\eMatB]}(0),\geodesic_{[\eMat],[\eMatB]}(0)) }{\operatorname{RMSD}(\geodesic_{[\eMatC_0],[\eMatB]}(0),\geodesic_{[\eMat],[\eMatB]}(0))},
    \label{eq:curvature-effect-geo-var-begin}
\end{equation}
where $[\eMatC_0]$ is close to the respective first frame $[\eMat]$,
\begin{equation}
    % \epsilon_{rel}^{\gamma,1}(t) := 
    \frac{\operatorname{RMSD}(\geodesic_{[\eMat],[\eMatC_1]}(t),\geodesic_{[\eMat],[\eMatB]}(t)) - (1-t) \cdot \operatorname{RMSD}(\geodesic_{[\eMat],[\eMatC_1]}(1),\geodesic_{[\eMat],[\eMatB]}(1)) }{\operatorname{RMSD}(\geodesic_{[\eMat],[\eMatC_1]}(1),\geodesic_{[\eMat],[\eMatB]}(1))},
    \label{eq:curvature-effect-geo-var-end}
\end{equation}
where $[\eMatC_1]$ is close to respective final frame $[\eMatB]$, and
\begin{equation}
    % \epsilon_{rel}^{r}(t) := 
    \frac{\operatorname{RMSD}([\eMat_{\lfloor t \numData \rfloor}], \exp^{\separation^\GyRaParam}_{\bar{[\eMat]}}((\Xi_{[\bar{\eMat}]}^r)_{\lfloor t \numData \rfloor})) - \frac{1}{\sqrt{\proteinLen}}\|(\log^{\separation^\GyRaParam}_{[\bar{\eMat}]} [\eMat_{\lfloor t \numData \rfloor}])_{\diamond \bar{\eMat}} -  ((\Xi_{[\bar{\eMat}]}^r)_{\lfloor t \numData \rfloor})_{\diamond \bar{\eMat}}\|_2}{\frac{1}{\sqrt{\proteinLen}}\|(\log^{\separation^\GyRaParam}_{[\bar{\eMat}]} [\eMat_{\lfloor t \numData \rfloor}])_{\diamond \bar{\eMat}} -  ((\Xi_{[\bar{\eMat}]}^r)_{\lfloor t \numData \rfloor})_{\diamond \bar{\eMat}}\|_2},
    \label{eq:curvature-effect-low-rank}
\end{equation}
respectively to test for curvature effects, where $\operatorname{RMSD}(\cdot, \cdot)$ as in \cref{eq:approximation-l2-differences-low-rank}. Note that the right terms in the numerator are the zero-curvature baselines and that a positive value of \cref{eq:curvature-effect-geo-var-begin,eq:curvature-effect-geo-var-end} suggest positive curvature effects and vice versa for negative curvature by \cite[Lemma 1]{bergmann2019recent}. The opposite conclusions are suggested for positive and negative values of \cref{eq:curvature-effect-low-rank} by \cite[Theorem 3.4]{diepeveen2023curvature}.

\begin{figure}[h!]
    \centering
    \begin{subfigure}{0.31\linewidth}
        \includegraphics[width=\linewidth]{"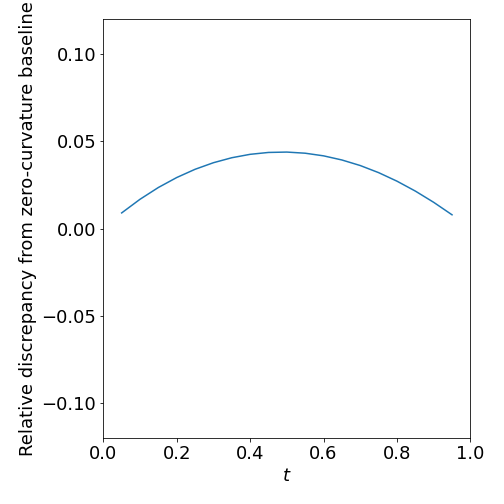"}
        % \caption{Geodesic variation at begin point}
    \end{subfigure}
    \hfill
    \begin{subfigure}{0.31\linewidth}
        \includegraphics[width=\linewidth]{"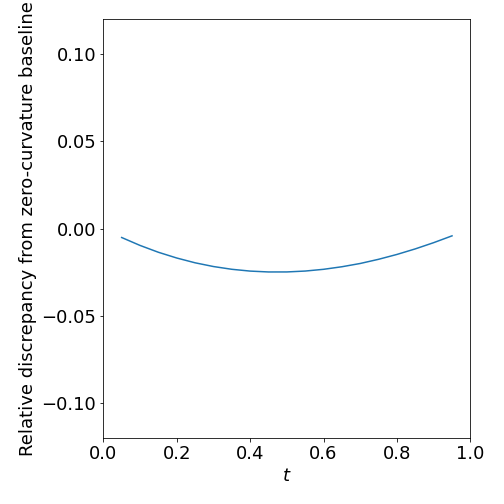"}
        % \caption{Geodesic variation at end point}
    \end{subfigure}
    \hfill
    \begin{subfigure}{0.31\linewidth}
        \includegraphics[width=\linewidth]{"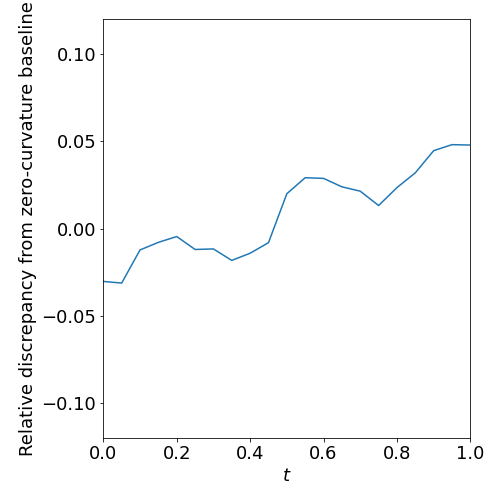"}
        % \caption{Low rank approximation}
    \end{subfigure}
    \caption{The progression of \cref{eq:curvature-effect-geo-var-begin} for geodesic variation at the begin point (left), \cref{eq:curvature-effect-geo-var-end} for geodesic variation at the end point (middle), and \cref{eq:curvature-effect-low-rank} for rank 1 approximation at the $\separation^\GyRaParam$-barycentre (right) tells us that the curvature effects on these data analysis tasks for the adenylate kinase data set are negligible and do not jeopardize stability.
    }
    \label{fig:curvature-effects-4ke}
\end{figure}

\paragraph{Adenylate kinase.}
We take $[\eMatC_0] := [\eMat^{AK}_4]$ and $[\eMatC_1] := [\eMat^{AK}_{99}]$. The progressions of \cref{eq:curvature-effect-geo-var-begin,eq:curvature-effect-geo-var-end,eq:curvature-effect-low-rank} are shown in \cref{fig:curvature-effects-4ke}. The left hand plot suggests overall positive curvature effects for the geodesic variation at the begin point and the middle plot suggests negative curvature effects for the geodesic variation at the end point. The curvature effects for low rank approximation in the right hand plot are in line with these observations, since we expect to see positive curvature effects towards $t=0$ and negative curvature towards $t=1$ -- since the low rank approximation is rank 1. In either case, the effects are minor.

\begin{figure}[h!]
    \centering
    \begin{subfigure}{0.31\linewidth}
        \includegraphics[width=\linewidth]{"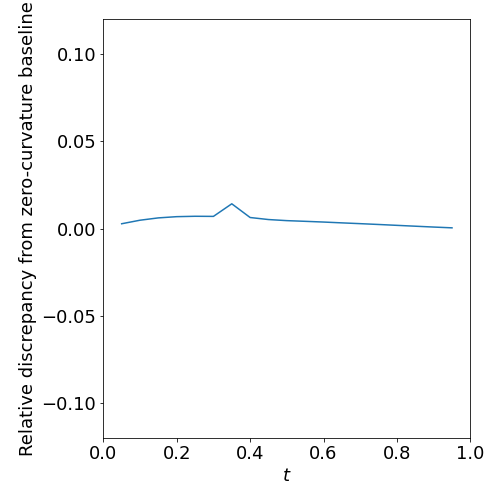"}
        % \caption{Geodesic variation at begin point}
    \end{subfigure}
    \hfill
    \begin{subfigure}{0.31\linewidth}
        \includegraphics[width=\linewidth]{"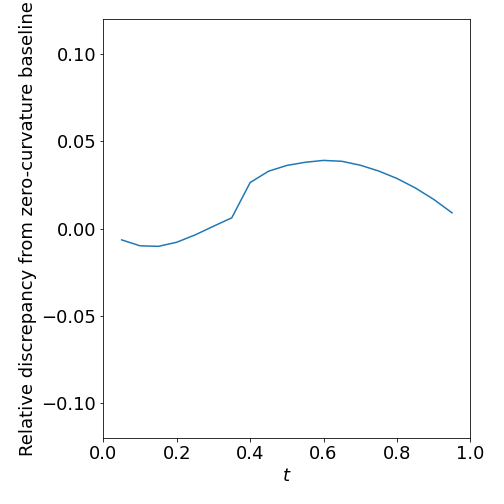"}
        % \caption{Geodesic variation at end point}
    \end{subfigure}
    \hfill
    \begin{subfigure}{0.31\linewidth}
        \includegraphics[width=\linewidth]{"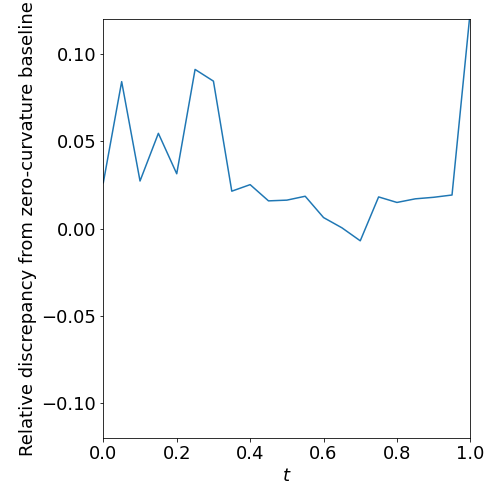"}
        % \caption{Low rank approximation}
    \end{subfigure}
    \caption{The progression of \cref{eq:curvature-effect-geo-var-begin} for geodesic variation at the begin point (left), \cref{eq:curvature-effect-geo-var-end} for geodesic variation at the end point (middle), and \cref{eq:curvature-effect-low-rank} for rank 1 approximation at the $\separation^\GyRaParam$-barycentre (right) tells us that the curvature effects on these data analysis tasks for the SARS-CoV-2 helicase nsp 13 data set are negligible and do not jeopardize stability.
    }
    \label{fig:curvature-effects-covid-spike}
\end{figure}

\paragraph{SARS-CoV-2 helicase nsp 13.}
We take $[\eMatC_0] := [\eMat^{SC}_4]$ and $[\eMatC_1] := [\eMat^{SC}_{197}]$. The progressions of \cref{eq:curvature-effect-geo-var-begin,eq:curvature-effect-geo-var-end,eq:curvature-effect-low-rank} are shown in \cref{fig:curvature-effects-covid-spike}. For this data set we observe slight positive curvature effects in the left hand plot for geodesic variation at the begin point and more pronounced positive curvature effects in the middle plot for variation at the end point. At first glance, these results do not match the negative curvature effects for the low rank approximation in the right hand plot. However, we remind the reader that we cannot expect these observations to be in line if the barycentre does not go through the data set as in the above case (cf. \cref{fig:error-geo-low-rank,fig:error-geo-low-rank-covid}). So since we even need 7 dimensions to approximate the data set well, the seemingly paradoxical results are not alarming. In any case, the most important observation is once again that the curvature effects are minor and do not jeopardize the stability of data analysis tasks.

% \todo[inline]{Change axis labels for the last plots}

\subsection{The $\separation^\GyRaParam$-logarithmic mapping with respect to \cref{eq:candidate-metric-full}}
\label{app:closed-form-separation-log}

\begin{multline}
    (\log_{[\eMat]}^{\separation^\GyRaParam} ([\eMatB]))_{\diamond \eMat} =
    - (\pwdMetricTensor_{\diamond\eMat} + \GyRaParam \MetricTensorCorr_{\diamond\eMat})^\dagger \Bigl(\Bigl[\sum_{\sumIndB\neq \sumIndA} \log \Bigl(\frac{\|\ePoint_\sumIndA - \ePoint_\sumIndB\|_2}{\|\ePointB_\sumIndA - \ePointB_\sumIndB\|_2}\Bigr) \frac{\ePoint_\sumIndA - \ePoint_\sumIndB}{\|\ePoint_\sumIndA - \ePoint_\sumIndB\|_2^2} \\
    + 2 \GyRaParam \log \Bigl(\frac{\det(\sum_{\sumIndB} (\ePoint_\sumIndB - \frac{1}{\proteinLen} \eMat \mathbf{1}) \otimes (\ePoint_\sumIndB - \frac{1}{\proteinLen} \eMat \mathbf{1}))}{\det(\sum_{\sumIndB} (\ePointB_\sumIndB - \frac{1}{\proteinLen} \eMatB \mathbf{1}) \otimes (\ePointB_\sumIndB - \frac{1}{\proteinLen} \eMatB \mathbf{1}))} \Bigr) \bigl(\sum_{\sumIndB} (\ePoint_\sumIndB - \frac{1}{\proteinLen} \eMat \mathbf{1}) \otimes (\ePoint_\sumIndB - \frac{1}{\proteinLen} \eMat \mathbf{1}) \bigr)^{-1} (\ePoint_\sumIndA - \frac{1}{\proteinLen} \eMat \mathbf{1})\Bigr]_{\sumIndA=1}^{\proteinLen} \Bigr) , \quad \eMat \in [\eMat], \eMatB \in [\eMatB].
    \label{eq:Riemannian-grad-separation-sq-log-alpha}
\end{multline}

\subsection{Tangent vectors from the tangent space SVD}
\label{app:tangent-svd-vectors}

% \paragraph{Adenylate kinase}
\begin{figure}[h!]
    \centering
    \begin{subfigure}{0.18\linewidth}
        \includegraphics[width=\linewidth]{"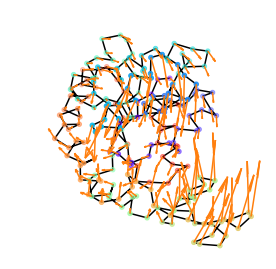"}
        % \caption{}
    \end{subfigure}
    % \hfill
    \begin{subfigure}{0.18\linewidth}
        \includegraphics[width=\linewidth]{"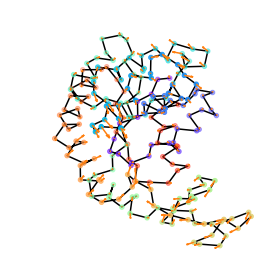"}
        % \caption{}
    \end{subfigure}
    % \hfill
    \begin{subfigure}{0.18\linewidth}
        \includegraphics[width=\linewidth]{"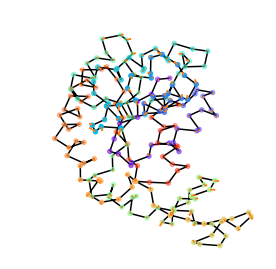"}
        % \caption{$t=0.75$}
    \end{subfigure}
    \caption{Rescaled tangent vectors obtained from tangent space singular value decomposition at the $\separation^\GyRaParam$-barycentre and corresponding to the largest three singular values. The left-most tangent vector -- corresponding to the largest singular value -- captures most of the data.
    }
    \label{fig:4ake-bary-low-rank-tangent-vectors}
\end{figure}

% \paragraph{SARS-CoV-2 helicase nsp 13}

\begin{figure}[h!]
    \centering
    \begin{subfigure}{0.18\linewidth}
        \includegraphics[width=\linewidth]{"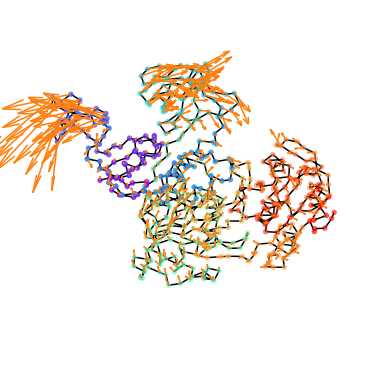"}
        % \caption{}
    \end{subfigure}
    % \hfill
    \begin{subfigure}{0.18\linewidth}
        \includegraphics[width=\linewidth]{"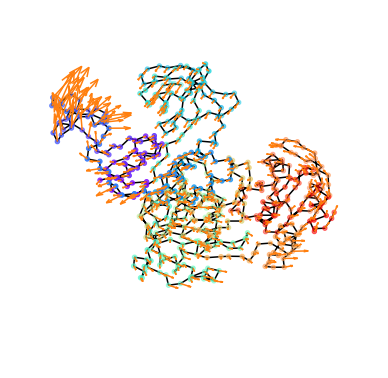"}
        % \caption{}
    \end{subfigure}
    % \hfill
    \begin{subfigure}{0.18\linewidth}
        \includegraphics[width=\linewidth]{"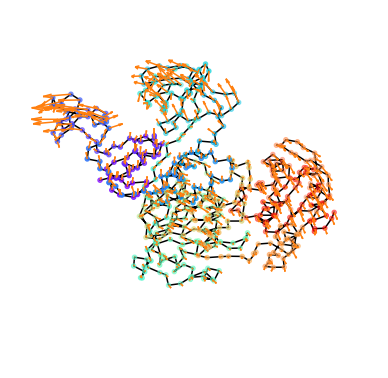"}
        % \caption{$t=0.75$}
    \end{subfigure}
    \begin{subfigure}{0.18\linewidth}
        \includegraphics[width=\linewidth]{"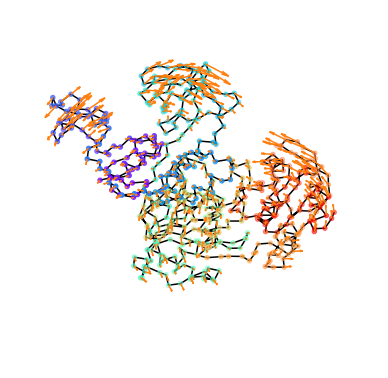"}
        % \caption{$t=0.75$}
    \end{subfigure}
    \begin{subfigure}{0.18\linewidth}
        \includegraphics[width=\linewidth]{"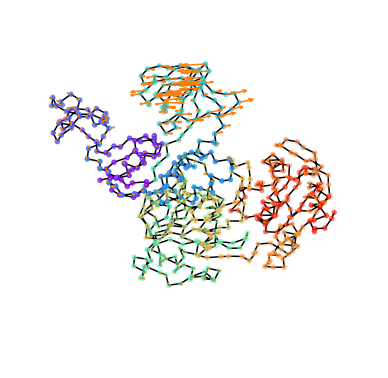"}
        % \caption{$t=0.75$}
    \end{subfigure}
    \\
    \begin{subfigure}{0.18\linewidth}
        \includegraphics[width=\linewidth]{"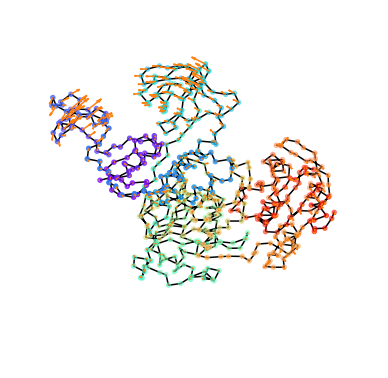"}
        % \caption{$t=0.75$}
    \end{subfigure}
    \begin{subfigure}{0.18\linewidth}
        \includegraphics[width=\linewidth]{"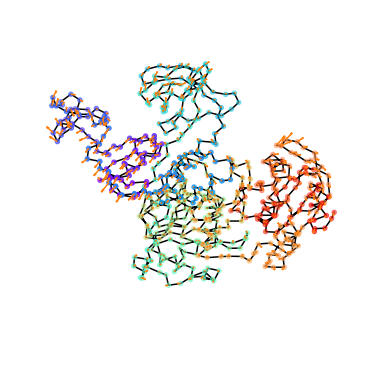"}
        % \caption{$t=0.75$}
    \end{subfigure}
    \begin{subfigure}{0.18\linewidth}
        \includegraphics[width=\linewidth]{"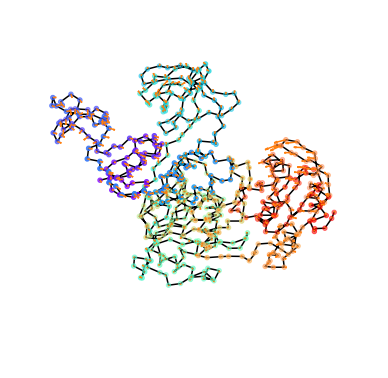"}
        % \caption{$t=0.75$}
    \end{subfigure}
    \begin{subfigure}{0.18\linewidth}
        \includegraphics[width=\linewidth]{"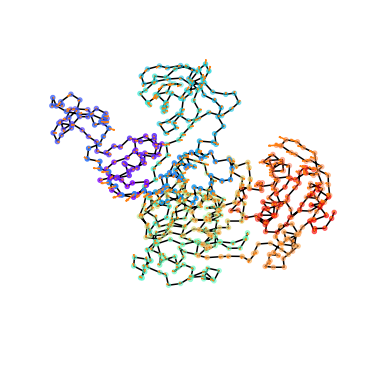"}
        % \caption{$t=0.75$}
    \end{subfigure}
    \caption{Rescaled tangent vectors obtained from tangent space singular value decomposition at the $\separation^\GyRaParam$-barycentre and corresponding to the largest nine singular values. More than one tangent vector is clearly needed to appropriately capture the data set.
    }
    \label{fig:4ake-bary-low-rank-tangent-vectors}
\end{figure}

\end{document}